\documentstyle[preprint,eqsecnum,aps,axodraw]{revtex}
\def \bfgr #1{ \mbox {{\boldmath $#1$}}}
\begin{document}
\draft
\setcounter{page}{0}
\title{Radiative corrections to the neutron $\bbox{\beta
-}$decay \\ within the Standard Model.}
\author{G.  G.  Bunatian \footnote{Email: bunat@cv.jinr.dubna.su}}
\address{\bf Joint Institute for Nuclear Research, 141980, Dubna, Russia}
\date{\today}
\maketitle
\begin{abstract}
Starting with the basic Lagrangian of the Standard Model, the
 radiative corrections to the neutron $\beta-$decay are acquired.
 The electroweak interactions are consistently taken into
consideration amenably to the Weinberg-Salam theory. The effect of
 the strong quark-quark interactions on the neutron $\beta
-$decay is parameterized by introducing the nucleon electromagnetic
form factors and the weak nucleon transition current specified by
the form factors $g_V
\, , \; g_A \, ...$. The radiative corrections to the total decay
probability $W$ and to the asymmetry coefficient of the electron
momentum distribution $A$ are obtained to constitute
$\delta{W}{\approx}8.7{\%} \, , \, \,
\delta{A}{\approx}-2{\%}$. The contribution to the radiative corrections
due to allowance for the nucleon form factors and the nucleon
excited states amounts up to a few per cent to the whole value of
the radiative corrections. The ambiguity in description of the
nucleon compositeness is this surely what causes the uncertainties
${\sim}0.1\%$ in evaluation of the neutron $\beta-$decay
characteristics. For now, this puts bounds to the precision
attainable in obtaining the element $V_{ud}$ of the $CKM$ matrix
and the $g_V \, , \; g_A \,
...$ values from experimental data processing.
\end{abstract} 
\setcounter{page}{1}
\section{Introduction}
\label{sec:level1}

Nowadays, it has been well realized that a thorough and all-round
study of the neutron $\beta -$decay conduces to gain an insight
into
 physical gist of the semiweak processes and into the elementary
particle physics in general.
 That is why for the past decade a great deal of efforts has been
directed to measure with a high accuracy (better than ${\sim}1\%$)
the main characteristics of the $\beta -$decay of free neutrons:
the lifetime $\tau$ \cite{t}, the asymmetry factors (as neutrons
are polarized) of the electron momentum distribution $A$ \cite{a}
and the antineutrino momentum distribution $B$ \cite{ba}, the
recoil proton distribution and the electron-antineutrino
correlation coefficient $a$ \cite{by1}, the coefficient $D$ of
triple correlation of the electron momentum, the antineutrino
momentum and the neutron spin \cite{tk}. Further experiments are
believed to come to fruition before long \cite{f}.

In treating the experimental data, the task is posed to inquire
into the effective 4-fermion interaction \cite{ll,d,com}
\begin{eqnarray}
{\cal L}_{WF}(x)=\frac{G_{F} |V_{ud}|}{\sqrt{2}}({\bar
{\psi}}_{e}(x){\gamma}_{\alpha}(1
- {\gamma}^{5}){\psi}_{\nu}(x)){\times}
\nonumber \\
{\times}{\sum_{{\bf P}_n ,\sigma_n , {\bf P}_p , \sigma_p}}
{\bar{\Psi}}_{p}(P_p, \sigma_p, x)
 \Bigl\{({\gamma}^{\alpha} g_{V}(q^2)+g_{WM}(q^2)\sigma ^{\alpha
\nu} q_{\nu}) - \nonumber \\-({\gamma}^{\alpha}
g_{A}(q^2)+g_{IP}(q^2)q^{\alpha} ){\gamma}^{5}\Bigr\}
{\Psi}_{n}(P_n,
\sigma_n, x) \, , \; \; \; \; \; q=P_p - P_n \, ,
 \label{1}
\end{eqnarray}
the quantities $|V_{ud}| , \; g_V , \; g_A , ... $ herein to be
specified with the same accuracy which has been attained in the
experimental measurements. This effective Lagrangian (\ref{1}) is
generally considered as descending from the Standard Model, the
nowaday elementary particle theory (see, for instance, Ref.
\cite{d}). In the expression (\ref{1}), $\psi_e(x), \;
\psi_{\nu}(x)$ stand for the electron (positron), (anti)neutrino
fields, and $\Psi_N(P_N,
\sigma_N, x), \, N{=}n, p, $ represent the nucleon states with the
momenta $P_N$ and polarizations $\sigma_N$. The system of units
$h{=}c{=}1$ is adapted, and $\gamma^5 , \;
\sigma^{\mu\nu}$ are defined by
$\gamma^5=i\gamma^0\gamma^1\gamma^2\gamma^3 ,
\; {\sigma^{\mu\nu}}{=}{(
\gamma^{\mu}\gamma^{\nu}{-}\gamma^{\nu}\gamma^{\mu})}{/}{2}$.
$G_F$ is the Fermi constant and $|V_{ud}|$ is the
Cabibbo-Kobayashi-Maskawa (CKM) \cite{ckm} quark-mixing matrix
element. By confronting the experimental data with the results of
the appropriate calculations, the $|V_{ud}|, \, g_V , \, g_A ... $
values are to be fixed so strictly that we should be in position to
fathom the principles of the elementary particle theory. In
particular, the $CKM$ unitarity
\begin{equation}
|V_{ud}|^2 +|V_{us}|^2 +|V_{ub}|^2 =1 \label{ckm}
\end{equation}
should be verified as strictly as possible \cite{ckm}.

So far as the transferred momentum $q$ is very small when compared
to the nucleon mass, ${|q|}{/}{M_N}{\sim}{0.0005}$, Eq. (\ref{1})
provides the bulk amplitude ${\cal M}^0$ of the neutron
$\beta-$decay with presuming $M_N{\rightarrow}\infty$, neglecting
the terms with $g_{WM} \, , \; \, g_{IP}$, and
 replacing the functions $g_V(q^2) ,
\; g_A(q^2)$ by their values at $q^2{=}0$ : $g_V(0){=}1 ,
\; g_A(0)$ \cite{ll,d,com}. Finiteness
of the nucleon mass causes the sizable, about $1\%$, corrections to
the calculated decay characteristics \cite{sm} that have been taken
into consideration in experimental data processing
 in Refs.\cite{t,a,ba}.

As we strive to acquire the quantities $|V_{ud}| , \; g_V , \; g_A ,
... $ with an accuracy better than $1\%$, the electromagnetic
corrections are to be allowed for in treating the neutron
$\beta-$decay. Therefore the effective Lagrangian (\ref{1}) is to be
accomplished by the interactions of electrons and nucleons with
electromagnetic field $A$
\begin{eqnarray} {\cal L}_{e\gamma}(x)=-e{\bar
{\psi}}_{e}(x){\gamma}^{\mu}{\psi}_{e}(x)\cdot A_{\mu}(x)
 \; , \qquad \qquad \label{ea}\\ {\cal
L}_{N\gamma}(x)=-e\sum_{N,P_N,\sigma_N}{\bar{\Psi}}_{N}(P_N,\sigma_N,x)
f_{N}^{\mu}(q){\Psi}_{N}(P_N,\sigma_N,x){\cdot}A_{\mu}(x) \,
,\label{ba}
\end{eqnarray}
where $f_{N}^{\mu}(q)$ are the nucleon electromagnetic form
factors. These interactions give rise to the electromagnetic
corrections to the bulk amplitude ${\cal M}^0$.

If the effective Lagrangian
\begin{eqnarray} {\cal L}_{eff}={\cal L}_{WF}+{\cal L}_{e\gamma}
+{\cal L}_{N\gamma} \label{l}
\end{eqnarray}
could consistently describe the radiative $\beta-$decay of neutrons
\begin{eqnarray}
n\Longrightarrow p+e^- +\bar\nu +\gamma \; ,\label{n}
\end{eqnarray}
the actual transition amplitude ${\cal M}$ of order $\alpha$ would
 merely be presented by the set of ordinary Feynman diagrams
originating immediately from the interactions (\ref{1}),
(\ref{ea}), (\ref{ba})

\vspace{1.7cm}

\begin{picture}(17.,5.)
\setlength{\unitlength}{0.80cm}
\multiput(2.5,0.)(6.,0.){3}{\vector(0,1){2.0}}
\put(7.5,1.0){\vector(-1,0){1.35}}
\put(13.5,0.){\vector(-1,-1){1.2}} \thicklines
\multiput(2.5,0.)(6.,0.){3}{\vector(-1,1){2.25}}
\multiput(2.5,0.)(6.,0.){3}{\vector(-1,0){2.5}}
\multiput(5.,0.)(6.,0.){3}{\line(-1,0){2.5}}
\multiput(2.5,2.1)(6.,0.){3}{$\bar{\nu}$}
\multiput(0.,2.2)(6.,0.){3}{$e$}
\multiput(0.,0.2)(6.,0.){3}{$p$}
\multiput(4.8,0.2)(6.,0.){3}{$n$}
\put(12.,-1.4){$\gamma$}
\put(5.9,1.){$\gamma$}
\put(4.,2.){$(a)$}
\put(10.,2.){$(b)$}
\put(16.,2.){$(c)$}
\multiput(2.5,-0.045)(6.,0.){3}{\vector(-1,0){2.5}}
\multiput(5.,-0.045)(6.,0.){3}{\line(-1,0){2.5}}
\end{picture} 

\vspace{0.6cm}

\begin{equation}
.\label{d1}
\end{equation}

\vspace{0.6cm}

\begin{picture}(17.,5.)
\setlength{\unitlength}{0.80cm}
\put(1.5,0.){\circle*{0.35}}
\multiput(2.5,0.)(6.,0.){1}{\circle{0.45}}
\multiput(6.5,0.)(1.,0.){2}{\circle*{0.35}}
\multiput(2.5,0.)(6.,0.){3}{\vector(0,1){2.0}}
\put(1.5,0.){\line(0,1){1.0}}
\put(7.,0.){\oval(1.0,1.0)[b]}
\put(13.,0.){\oval(1.0,1.0)[t]}
\thicklines
\put(14.5,0.){\vector(-1,0){2.3}}
\put(14.5,0.){\vector(-1,-1){1.3}}
\multiput(2.5,0.)(6.,0.){2}{\vector(-1,1){2.25}}
\multiput(2.5,0.)(6.,0.){2}{\vector(-1,0){2.5}}
\multiput(5.,0.)(6.,0.){3}{\line(-1,0){2.5}}
\multiput(2.5,2.1)(6.,0.){3}{$\bar{\nu}$}
\multiput(0.,2.2)(6.,0.){2}{$e$}
\multiput(0.,0.2)(6.,0.){2}{$p$}
\multiput(4.8,0.2)(6.,0.){3}{$n$}
\put(11.85,-0.05){$e$}
\put(13.,-1.67){$p$}
\put(1.18,0.5){$\gamma$}
\put(7.,-0.75){$\gamma$}
\put(13.,0.66){$\gamma$}
\put(4.,2.){$(d)$}
\put(10.,2.){$(e)$}
\put(16.,2.){$(f)$}
\multiput(2.5,-0.043)(6.,0.){2}{\vector(-1,0){2.5}}
\multiput(5.,-0.043)(6.,0.){3}{\line(-1,0){2.5}}
\put(14.5,-0.05){\vector(-1,-1){1.3}}
\put(1.5,0.15){\line(1,0){1}}
\put(1.5,-0.18){\line(1,0){1}}
\put(6.5,0.15){\line(1,0){1}}
\put(6.5,-0.18){\line(1,0){1}}
\end{picture}


\vspace{1.5cm}

where the triplex lines represent various baryonic states, the
blobs depict the form factors $f_N^{\mu}(q)$ in (\ref{ba}) and the
empty circle stands for the matrix element of the interaction
(\ref{1}) with allowance for $q-$dependence. So, upon
straightforward unsophisticated calculating, the amplitude $\cal M$
and, subsequently, the observables $\tau , \, A , \, B , \, a ,
...$ would directly be obtained in terms of the quantities $G_F ,
|V_{ud}| , \, g_V , \, g_A , ...$ residing into ${\cal L}_{WF}$
(\ref{1}). Then, accordingly the aim proclaimed, it would quite
natural appear that these desirable quantities should be
ascertained by confronting the experimental values of $\tau , \, A
, \, B , \, a ,...$ with their values calculated in the aforesaid
way. But, alas, this plain calculation shows up to be contradictory
because the ultra violet (UV)
 divergences (the terms multiple to $\ln{\Lambda}{/}{M_N} , \; \,
\Lambda{\rightarrow}\infty$) inhere in
 the contributions from the one-loop diagrams $(d),(e),(f)$ in
 (\ref{d1}). So far the treatment is solely based upon the
Lagrangian (\ref{l}) itself, there is no way to cope with this
failure. To deal with well-defined quantities in practical
evaluating the observables $\tau , \, A , \, B , \, a , ...$, the
extra UV cut-off $\Lambda{=}M_V{\approx}100 \, \mbox{GeV}$ could be
set up, supplementing the calculation based on the local
interaction (\ref{l}), see, for instance, Refs.
\cite{jap,g1,34,g2,g3,g4,ga,ga1,i,iii}. Yet, this recipe is rather
untenable, and we would never be able to repose full confidence in
the results obtained in this way. Thus, the description of the
radiative decay (\ref{n}) with the effective interaction (\ref{l})
is not self-contained.

Although the 4-fermion local theory is quite sufficient for the
calculations in the lowest order, without the radiative
corrections, it is not satisfactory because of its violation of
unitarity and its nonrenormalizability, which prevents us from
dealing with electroweak high order effects in a convincing way. A
stringent self-contained treatment of the neutron $\beta-$decay
ought to be founded upon the Standard Model of elementary particle
physics. The Standard Model Lagrangian ${\cal L}^{SM}$ \cite{d}
embodies the nowaday knowledge of the strong and electroweak
interactions of the leptons and the quarks,
\begin{equation}
{\cal L}^{SM}={\cal L}^{EW}+{\cal L}^{qq}_{str} \label{l1}.
\end{equation}
There are several review articles and books available which
thoroughly describe the structure of ${\cal L}^{EW} , \; \, {\cal
L}^{qq}_{str}$. In the work presented, we pursue the way paved in
Refs. \cite{d,ao,h1,h2,b}.

In Sec. $\mbox{II}$, we concisely recapitulate the structure of the
basic electroweak Lagrangian ${\cal L}_{EW}$ and the respective
renormalization procedure in view of the current calculation of the
radiative corrections to the neutron $\beta-$decay in the one-loop
approach, with intent to attain an accuracy about $0.1\%$. By
 introducing the nucleon weak transition current and
 electromagnetic form factors, the needful parameterizing of the
effects caused by nucleon compositeness is set forth in Sec.
$\mbox{III}$. In Secs. $\mbox{IV}$-$\mbox{X}$, we acquire
successively, term by term, the total decay amplitude of order
$\alpha$. In particular, the influence of nucleon structure on the
calculated radiative corrections is estimated in Secs. $\mbox{VI}$,
 $\mbox{IX}$. The radiative corrections to the electron momentum
distribution and to neutron lifetime are acquired in Sec.
$\mbox{XI}$. In the last Sec., we fairly well try and compare our
results with the long-known noteworthy assertions of the former
investigations of the radiative corrections to the neutron
$\beta-$decay. We purposely defer this needful discussion till the
final stage of the work to have at our disposal all the desirable
persuasive arguments to be offered for substantiating our
inferences. Upon realizing what is the accuracy actually attainable
in the nowaday calculations, we brief a feasible way to acquire the
quantities $G_F , \; |V_{ud}| , \; g_V , \; g_A , ...$ residing in
Eq. (\ref{1}) as precisely as possible from appropriate
experimental data processing.
\section{Electroweak interactions in
 description of the neutron ${\large \bfgr\beta-}$decay.}
\label{sec:level2}
The basic electroweak Lagrangian to start with,
\begin{equation}
{\cal L}^{EW}(A_{\mu} , Z_{\mu} , W^{\pm}_{\mu} , H , \psi_f , e ,
M_Z , M_W , M_H , m_f , \xi ) \, ,
\label{el}
\end{equation}
is expressed amenably to Refs. \cite{ao,h1,h2,b} in terms of the
bare physical fields and parameters. $A_{\mu} , \; Z_{\mu} ,
\; W^{\pm}_{\mu} , \; H , \; \psi_f $ stand for the electromagnetic,
$Z-$boson, $W^{\pm}-$boson, Higgs-boson and generic fermion fields,
and the quantities ${e}{=}{\sqrt{4\pi\alpha}} , \; M_Z , \; M_W ,
\; M_H , \; m_f $ are the unit of charge and the masses of the
$Z-$boson, $W-$boson, Higgs-boson, and fermions, respectively;
$\xi$ represents generically the gauge parameters. Taking the line
of \cite{ao,h1,h2}, we choose the Feynman gauge, $\xi=1$. The
physical fields $A_{\mu} , \; Z_{\mu} , \; W^{\pm}_{\mu} $ are
related to the isotriplet of vector fields $W_{\mu}^a , \,
{a}{=}{1,2,3}$, and to the isosinglet vector field $B_{\mu}$ by the
equations \cite{ao,h1,h2,b}
\begin{eqnarray}
Z_{\mu} =c_W W^3_{\mu} +s_W B_{\mu} , \; \; \; A_{\mu} =-s_W
W^3_{\mu} +c_W B_{\mu} , \; \; \; W^{\pm}_{\mu}
=\frac{1}{\sqrt{2}}\bigl( W_{\mu}^1 \mp iW^2_{\mu}\bigr) \; .
\label{e2}
\end{eqnarray}
Chosen $e, \, M_{Z,W}, \, m_f$ as input parameters,
\begin{equation}
c_W =\frac{M_W}{M_Z} \; , \; \; \; \; s^2_W =1-c^2_W \label{e3}
\end{equation}
are nothing but merely shorthand notations to simplify formulae. The
gauge coupling constants are given by
\begin{equation}
g_2
=\frac{e}{s_W} , \; \; \; \; \; g_1=\frac{e}{c_W} \,
,
\label{e4}
\end{equation}
 and the masses of physical particles are written as
\begin{equation}
M_W=\frac{1}{2}g_2{\cal V} \; ,
M_Z=\frac{1}{2}\sqrt{g_1^2+g_2^2}{\cal V} \; , m_f=\frac{f_f{\cal
V}}{\sqrt{2}} \, ,\label{e5}
\end{equation}
where ${\cal V}$ is the vacuum expectation value of the Higgs
field, and $f_f$ stand for the Yukawa couplings of fermions to the
Higgs field. ${\cal L}^{EW}$ (\ref{el}) has been constructed in
Refs. \cite{ao,h1,h2} so that the bilinear terms, i.e. the inverse
propagator terms, take eventually the simplest form:
\begin{eqnarray}
{\cal L}^{EW}_{0} =\bar \psi_f (i\gamma^{\mu} \partial_{\mu}
-m_f )\psi_f + W_{\mu}^{+} g^{\mu\nu} (\Box + M_W^2 )W_{\nu}^{-} +
\nonumber \\
\frac{1}{2}Z_{\mu} g^{\mu\nu} (\Box + M_Z^2 )Z_{\nu}
+\frac{1}{2}(\Box + {m_{\gamma}}^2 ) A_{\mu} g^{\mu\nu} A_{\nu}
 \; .\label{l0}
\end{eqnarray}
The propagators of free fields are consequently
\begin{picture}(70,-15)(0,160)
\SetWidth{1}
\SetColor{Blue}
\Photon(-220,130)(-150,130){6}{5}
\Text(-185,147)[t]{$W , Z$}
\Photon(-220,95)(-150,95){3}{10}
\Text(-185,110)[t]{$A\gamma$}
\ArrowLine(-145,52)(-220,57)
\Text(-185,70)[t]{$f{\equiv}e,\nu ,u,d$}
\end{picture}
\begin{eqnarray}
D^{Z,W}_{\alpha\beta}(x)=\delta_{\alpha\beta}
\int\frac{\mbox{d}^4k}{(2\pi)^4}\frac{\exp(-ikx)}{k^2-M^2_{Z,W}+i0}
 \; , \label{dwz} \\
 D^{A\gamma}_{\alpha\beta}(x)=\delta_{\alpha\beta}
\int\frac{\mbox{d}^4k}{(2\pi)^4}\frac{\exp(-ikx)}{k^2-m_{\gamma}^2 +i0}
 \; , \label{d} \\ G^{f}(x)=
\int\frac{\mbox{d}^4p}{(2\pi)^4}\exp(-ipx)\frac{\not p + m_f}
{p^2-m_f^2+i0} \; . \label{gf}
\end{eqnarray}
The fictitious photon mass $m_{\gamma}$ is included in (\ref{l0}),
(\ref{d}) to treat the integrals which involve the photon
propagator $D^{A\gamma}$. It is to mention that in the ensuing
calculation we shall have to deal not only with the infinitesimal
photon mass $m_{\gamma}{=}\lambda{\rightarrow}0$, but also with
$m_{\gamma}{=}M_S$ specified so as $M_N^2{\ll}M_S^2{\ll}M_W^2$.

 To treat thereafter the neutron $\beta-$decay in the one-loop
approach, the electroweak interactions of lepton, quark, $W-$,
$Z-$boson and electromagnetic fields are to be specified
\cite{ao,h1,h2,b}:
\begin{eqnarray}
 {\cal L}^{EW}_{int} ={\cal L}^{WWZ} +{\cal L}^{WWA} +{\cal L}^{Wff}
+{\cal L}^{Zff} +{\cal L}^{Aff} \; ,\label{l2} \\
{\cal L}^{WWZ} =
i\frac{g_2^2}{\sqrt{g_1^2 +g_2^2}}\biggl(
g^{\alpha\gamma}g^{\delta\beta}
-g^{\alpha\delta}g^{\gamma\beta}\biggr) [\partial_{\alpha}
W^{+}_{\beta} W^{-}_{\gamma} Z_{\delta} +\partial_{\alpha}
W_{\beta}^{-} Z_{\gamma} W^{+}_{\delta}
+\partial_{\alpha}Z_{\beta}W^{+}_{\gamma} W^{-}_{\delta}
]=\label{l3}\\
\begin{picture}(300,-10)(0,17)\SetWidth{1}
\SetColor{Blue}
\Vertex(230,0){2}
\Photon(190,0)(230,0){6}{4}
\Photon(230,0)(270,0){6}{4}
\Photon(230,0)(230,35){6}{3.5}
\Text(185,0)[r]{$W$}
\Text(275,0)[l]{$W$}
\Text(220,20)[r]{$Z$}
\end{picture}
=\Gamma_{\mu\nu\lambda}^{WWZ}W^{+\mu}W^{-\nu}Z^{\lambda} \;
,\nonumber
\end{eqnarray}
\\
\begin{eqnarray}
 {\cal L}^{WWA} = i e \biggl( g^{\alpha\gamma}g^{\delta\beta}
-g^{\alpha\delta}g^{\gamma\beta}\biggr) [\partial_{\alpha}
W^{+}_{\beta} W^{-}_{\gamma} A_{\delta} +\partial_{\alpha}
W_{\beta}^{-} A_{\gamma} W^{+}_{\delta}
+\partial_{\alpha}A_{\beta}W^{+}_{\gamma} W^{-}_{\delta} ]=
\label{l4} \\
\begin{picture}(300,-10)(0,17)\SetWidth{1}
\SetColor{Blue}
\Vertex(230,0){2}
\Photon(190,0)(230,0){6}{4}
\Photon(230,0)(270,0){6}{4}
\Photon(230,0)(230,35){3}{5}
\Text(185,0)[r]{$W$}
\Text(275,0)[l]{$W$}
\Text(220,20)[r]{$A$}
\end{picture}
=\Gamma_{\mu\nu\lambda}^{WWA}W^{+\mu}W^{-\nu}A^{\lambda} \;
,\nonumber
\end{eqnarray}\\
\begin{eqnarray}
 {\cal L}^{Wff'} = \frac{g_2}{2\sqrt{2}} \Biggl(\bar\psi_{i+}
 V_{+-} T_i^{+}
\gamma^{\mu} (1-\gamma^5 ) \psi_{i-} W^{+}_{\mu} +\bar\psi_{i-}
V_{-+} T_i^{-} \gamma^{\mu} (1-\gamma^5 )\psi_{i+} W_{\mu}^{-}
\Biggr)=\label{l5}\\
\begin{picture}(300,-10)(0,17)\SetWidth{1}
\SetColor{Blue}
\Vertex(230,0){2}
\Line(190,0)(230,0)
\Line(230,0)(270,0)
\Photon(230,0)(230,35){6}{3}
\Text(185,0)[r]{$f'$}
\Text(275,0)[l]{$f$}
\Text(220,20)[r]{$W$}
\end{picture}
=\bar\psi_f\Gamma_{\mu}^{Wff'}\psi_{f'}W^{\mu\pm} \;
,\nonumber
\end{eqnarray}\\
\begin{eqnarray}
 {\cal L}^{Zff} =\frac{1}{2}\sqrt{g^2_1 +
g^2_2}\Biggl(\bar\psi_{i+}\gamma^{\mu}\bigl(\frac{1-\gamma^5}{2}
-2Q_{i+}\frac{g^2_1}{g^2_1 +g^2_2}\bigr)\psi_{i+} - \nonumber\\
-\bar\psi_{i-}\gamma^{\mu} \bigl(\frac{1-\gamma^5}{2}
+2Q_{i-}\frac{g^2_1}{g^2_1 +g^2_2}\bigr)\psi_{i-}\Biggr) Z_{\mu}
=\bar\psi_f\Gamma_{\mu}^{Zff}\psi_{f}Z^{\mu}
\label{l6} \\
\begin{picture}(300,-10)(20,0)\SetWidth{1}
\SetColor{Blue}
\Vertex(0,0){2}
\Line(-40,0)(20,0)
\Line(0,0)(40,0)
\Photon(0,0)(0,35){6}{3}
\Text(-45,0)[r]{$f$}
\Text(45,0)[l]{$f$}
\Text(-10,20)[r]{$Z$}
\end{picture} \nonumber
\end{eqnarray}
\begin{eqnarray}
\begin{picture}(300,-10)(-15,12)\SetWidth{1}
\SetColor{Blue}
\Vertex(230,0){2}
\Line(190,0)(230,0)
\Line(230,0)(270,0)
\Photon(230,0)(230,30){3}{4}
\Text(185,0)[r]{$e$}
\Text(275,0)[l]{$e$}
\Text(220,20)[r]{$A$}
\end{picture}
 {\cal L}^{Aee} =-e\bar\psi_e \gamma^{\mu} \psi_e A_{\mu} \; ,
\label{l7} \end{eqnarray}
\begin{eqnarray}
\begin{picture}(200,-10)(-5,10)\SetWidth{1}
\SetColor{Blue}
\Vertex(130,0){2}
\Line(90,0)(130,0)
\Line(130,0)(170,0)
\Photon(130,0)(130,25){3}{4}
\Text(85,0)[r]{$q$}
\Text(175,0)[l]{$q$}
\Text(120,20)[r]{$A$}
\end{picture}
 {\cal L}^{Aqq} =e e_q \bar\psi_q \gamma^{\mu} \psi_q A_{\mu}
\equiv e e_q \bar q \gamma^{\mu} q A_{\mu} \; . \label{l8}
\end{eqnarray}

As usual, for leptons $ \psi_{i+} =\psi_{\nu} , \; \; \;
\psi_{i-}
=\psi_e , \; \; \; V_{+-}=1 , \; \; \; \; Q_{i+} =0 , \; \; \;
\; Q_{i-}=-1$ , and in the case of $u,d$ quarks
$\psi_{i+} =\psi_{u}{\equiv}u , \; \; \; \; \psi_{i-}
=\psi_d{\equiv}d , \; \; \; \; V_{+-}=V_{ud} , \; \; \; \; Q_{i+} =e_u
=2/3 , \; \; \; \; Q_{i-} =e_d =-1/3.$  The operator
$T^{+}$ increases, $T^{-}$ decreases weak isospin projection by one
unite:
 $T^{+}\psi_e =\psi_{\nu} \, ,  \; \; \; T^{-}\psi_{\nu} =\psi_{e}
 \, , \; \; \; T^{+}\psi_d =\psi_u \, ,  \; \; \; T^{-}
\psi_u =\psi_d \, , \; \; \;
T^{-}\psi_e =T^{-}\psi_d =T^{+}\psi_{\nu} =T^{+}\psi_u
=0 \, . $
In the interactions (\ref{l3})-(\ref{l8}) and in the analogous
expressions hereupon, the ${\cal N}$products of the field operators
\begin{eqnarray}
W^{+}_{\mu}(x)=\sum_{\bf q}\Bigl(c_{\mu}({\bf q})w^{+}_{\mu}({\bf
q})\mbox{e}^{-iqx}+c_{\mu}^{+}({\bf q})w^{-}_{\mu}({\bf
q})\mbox{e}^{iqx}\Bigr) \; , \label{w}\\
\psi_f(x)=\sum_{{\bf p},r}\Bigl(a_f({\bf p},r)u_f({\bf p},r)
\mbox{e}^{-ipx}+b^{+}_f({\bf p},r)u_f({-\bf p},-r)\mbox{e}^{ipx}\Bigr)
\; , \label{fi}
\end{eqnarray}
and so on, are implied. Here $f$ specifies a sort of fermions and
$r$ stands for other quantum numbers: spin, isospin, their
projections.

In calculating the neutron $\beta-$decay amplitude in the one-loop
approach, we leave out the effects of Higgs-fermion interactions,
since they are of the order of the Higgs coupling to fermions
${\sim}{m_f}{/}{M_W}$ \cite{d,ao,h1,h2,b}. Also only the first
generations of leptons $(e,\nu_e)$ and quarks ($u-,d-$quarks) come
into the forthcoming consideration.

The transition amplitude ${\cal M}$ of the process (\ref{n}), when
calculated in the one-loop approach according to
(\ref{l0})-(\ref{l8}) directly in terms of the bare fields and
parameters, is UV divergent, and renormalization is necessary. The
multiplicative renormalization of the Lagrangian
(\ref{l0})-(\ref{l8}) is performed amenably to the non-minimal
on-mass-shell (OMS) renormalization scheme \cite{ao,h1,h2,d}, with
the renormalization constants and renormalized quantities defined
in such a way that
\begin{eqnarray}
 W^a_{\mu}\Longrightarrow (z^W_2)^{1/2}W^a_{\mu} \; , \; \; \; \;
\; B_{\mu}\Longrightarrow (z^B_2)^{1/2}B_{\mu} ,  \nonumber\\
\psi^{L,R}_f\Longrightarrow (z^f_{L,R})^{1/2}\psi_f^{L,R} , \; \; \;
 \; \; \; \psi^{L,R}_f = \frac{1\mp \gamma^5}{2}\psi_f ,
\label{ct} \\
m_f^2\Longrightarrow m^2_f + \delta{m}^2_f , \; \; \;
M_{W,Z}^2\Longrightarrow M^2_{W,Z} +\delta{M}_{W,Z}^2 ,\nonumber\\
g_2\Longrightarrow z_1^W (z^W_2)^{-3/2} g_2 , \; \; \;
\; \; g_1\Longrightarrow z_1^B (z^B_2)^{-3/2} g_1 \; .\nonumber
\end{eqnarray}
Expanding the renormalization constants
\begin{equation}
z=1+{\delta}z \; , \label{z}
\end{equation}
we obtain
\begin{equation}
{\cal L}^{EW}={\cal L}^{EW}_{tree}+{\cal L}^{EW}_{ct} \; ,\label{z1}
\end{equation}
where the expression for ${\cal L}^{EW}_{tree}$ in terms of
renormalized quantities is identical with the original one,
(\ref{l0})-(\ref{l8}), but now it contains the renormalized
physical parameters and fields. The counter term Lagrangian
\begin{equation}
{\cal L}^{EW}_{ct}(A_{\mu} \, , \; \, Z_{\mu}
\, , \; \, W_{\mu}^{\pm} \, , \; \, H \, , \; \, \psi_f \, , \; \,
e \, , \; \, M_W \, , \; \, M_Z \, , \; \, m_f \, ;
 \; \, {\delta}z_{1,2}^{W,B} \, , \; \, {\delta}z_{L,R}^f \, , \; \,
 \delta M^2_{W,Z} \, , \; \, \delta m^2_f) \label{z3} \\
\end{equation}
is determined by the quantities ${\delta}z_{1,2}^{W,B} \, ,
\, \; {\delta}z_{L,R}^f \, , \, \; \delta M^2_{W,Z} \, , \,
 \; \delta m^2_f$ in (\ref{ct}).
The linear combinations of the field renormalization constants
${\delta}z_{2}^{W,B}$ and the coupling renormalization constants
${\delta}z_{1}^{W,B}$ are introduced \cite{h1,h2}
\begin{eqnarray}
\left( \begin{array}{cc}
\delta{z}_m^{\gamma}\\
\delta{z}_m^Z\\
\end{array} \right)
{=}
\left( \begin{array}{cc}
s^2_W & c^2_W \\ c^2_W & s^2_W\\
\end{array} \right)
{\cdot}
\left( \begin{array}{cc}
\delta{z}_m^{W}\\
\delta{z}_m^B\\
\end{array} \right) , \; \; \;
\delta{z}_m^{\gamma Z}{=}c_W s_W (\delta{z}_m^W -\delta{z}_m^B ){=}
\frac{c_W s_W}{c_W^2-s_W^2}(\delta{z}^Z_m-\delta{z}_m^{\gamma}) \, ,
\label{5}\\
 m=1,2 \, .\nonumber
\end{eqnarray}

Accordingly the OMS renormalization scheme \cite{ao,h1,h2,b}, the
fine structure constant $\alpha{=}{e^2}{/}{4\pi}$
${=}{1}{/}{137.036}$ (defined in the Thomson limit) is used as an
expansion parameter, and all the renormalization constants and the
renormalized quantities in Eqs. (\ref{ct})-(\ref{5}) are fixed on
the mass-shell of gauge bosons, fermions and Higgs bosons. With
this condition, the renormalized masses are identical to the pole
positions of the propagators, i.e. the physical masses. All the
residues in the diagonal propagators are normalized to $1$, and the
  residues in the non-diagonal parts of propagators are chosen to
be equal to $0$ in order to forbid mixing for on-mass-shell
particles, so as no additional renormalization of wave functions is
required, besides what given by Eqs. (\ref{ct}). Thus, the OMS
renormalization scheme does preserve physical meaning of the
original quantities in the electroweak Lagrangian ${\cal L}^{EW}$
(\ref{l0})-(\ref{l8}).

The formulated OMS renormalization conditions \cite{ao,h1,h2} allow
 us to obtain explicitly ${\delta}z_{1,2}^{W,B}$,
 ${\delta}z_{L,R}^f$, $\delta M^2_{W,Z}$, $\delta m^2_f$ (\ref{ct})
in terms of the unrenormalized self-energies of gauge bosons,
${\Sigma}^{W,Z}(M^2_{W,Z})$, ${\Sigma}^A(0)$,
${\Sigma}^{Z\gamma}(0)$, and fermions ${\Sigma}^f(m_f)$, and their
derivatives ${\partial\Sigma^{A,Z,W}(k^2)}{/}{\partial k^2}$,
 ${\partial\Sigma^f(\not p)}{/}{\partial{\not p}} \; ,$ which are
calculated in the one-loop approximation amenably to the Lagrangian
(\ref{l0})-(\ref{l8}). In particular, the fermion self-energies are
given in the usual way by the graphs
\begin{eqnarray}
{\Sigma}^f(\not p)=\begin{picture}(100,25)(0,0)\SetWidth{1}
\SetColor{Blue}
\Vertex(30,0){2}
\Vertex(80,0){2}
\Line(25,0)(85,0)
\PhotonArc(55,0)(25,0,180){4.5}{8}
\Text(55,4)[b]{$f$}
\Text(80,25)[l]{$W,Z,A\gamma$}
\end{picture} \; , \label{fs}
\end{eqnarray}
where the wavy line renders the propagators of $W-,Z-$bosons,
$D^{W,Z}$ (\ref{dwz}), and photons, $D^{A\gamma}$ (\ref{d}), with
the fictitious mass $m_{\gamma}$ which hereafter takes not only the
infinitesimal value $m_{\gamma}{=}{\lambda}{\rightarrow}0$, but
also the value $m_{\gamma}{=}M_S$ specified so as
$M_N^2{\ll}M_S^2{\ll}M_W^2$.

Upon calculating the radiative corrections with the fields, masses
and coupling constants renormalized amenably to the OMS
renormalization scheme, not only the UV divergencies occurring in
the loop expansion (of propagators as well as
 $S-$matrix elements) are absorbed in the infinite parts of the
renormalization constants, ${\delta}z_{1,2}^{W,B}, \;
{\delta}z_{L,R}^f , \; \delta M^2_{W,Z}, \; \delta m^2_f$, but also
the finite parts of the radiative corrections are fixed. These lead
to physically observable consequences.

The essential ingredients to obtain radiative corrections are the
three-particle vertex functions. First we are to acquire the
electroweak radiative corrections to the bare $e\nu W-$vertex
\begin{eqnarray}
\Gamma_{\alpha}^{e\nu
W}=\frac{e}{2\sqrt{2}s_W}\gamma_{\alpha}(1-\gamma^5)=
\begin{picture}(150,0)(0,20)
\SetWidth{1}
\SetColor{Blue}
\Vertex(100,30){2}
\ArrowLine(100,30)(50,30)
\ArrowLine(150,30)(100,30)
\Photon(100,30)(100,0){6}{4}
\Text(48,30)[r]{$p_e,\sigma_e$}
\Text(155,30)[l]{$-p_{\nu},-\sigma_{\nu}$}
\Text(93,15)[r]{$W^{-}$}
\Text(107,15)[l]{$q$}
\Text(75,35)[b]{$e$}
\Text(125,35)[b]{$\nu$}
\end{picture} \;
 \label{go}
\end{eqnarray}
in ${\cal L}^{Wff'} \; $ (\ref{l5}).

The renormalized corrected $e\nu W-$vertex $\hat\Gamma^{e\nu
W}_{\alpha}(p_e,-p_{\nu},q)$ is determined by the matrix element
\begin{eqnarray}
\langle\phi_e^{+}(p_e,\sigma_e)|{\cal
S}^{EW}|\phi_{\nu}(-p_{\nu},-\sigma_{\nu}) , W^{- \,
\alpha}(q)\rangle = \nonumber\\
=i(2\pi )^4\delta(q-p_{\nu}-p_e) \bigl(\bar u_e(p_e,\sigma_e)
\hat\Gamma^{e\nu W}_{\alpha}(p_e,-p_{\nu},q)w^{- \, \alpha}(q)
u_{\nu}(-p_{\nu},-\sigma_{\nu})\bigr) \label{r4}
\end{eqnarray}
of the ${\cal S}^{EW}-$operator
\begin{eqnarray}
{\cal S}^{EW}={\cal T}\exp[i\int\mbox{d}^4x {\cal L}^{EW}_{int}(x)]
\, , \label{r41}
\end{eqnarray}
with ${\cal L}^{EW}_{int}(x)$ given by (\ref{l2}). Here ${\cal T}$
represents ordinary time ordering,
$\phi_{\nu}(-p_{\nu},-\sigma_{\nu})$ stands for a neutrino with the
momentum $-p_{\nu}$ and the polarization $-\sigma_{\nu}$ in an
initial state, and $\phi_e(p_e,\sigma_e)$ stands for an electron
with the momentum $p_e$ and the polarization $\sigma_e$ in a final
state, $u_{e,\nu}$ indicate the Dirac spinors of leptons. In the
transition from the initial to the final state, a $W^--$boson with
the momentum $q=p_e+p_{\nu}$ and the polarization $\alpha$ is
absorbed (or $W^+$ emitted).

Pursuant to the aforecited OMS renormalization scheme
\cite{ao,h1,h2,b}, we obtain in the one-loop order, $O(\alpha)$,
\begin{eqnarray}
\bigl(\bar u_e(p_e,\sigma_e)
\hat\Gamma^{e\nu W}_{\alpha}(p_e,-p_{\nu},q)w^{- \, \alpha}(q)
u_{\nu}(-p_{\nu},-\sigma_{\nu})\bigr) =
\begin{picture}(195,0)(0,20)
\SetWidth{1}
\SetColor{Blue}
\CCirc(90,30){9.5}{Blue}{Red}
\ArrowLine(90,30)(40,30)
\ArrowLine(140,30)(90,30)
\Photon(90,30)(90,0){6}{4}
\Text(38,30)[r]{$p_e,\sigma_e$}
\Text(145,30)[l]{$-p_{\nu},-\sigma_{\nu}$}
\Text(83,15)[r]{$W^{-}$}
\Text(97,15)[l]{$q$}
\Text(65,35)[b]{$e$}
\Text(115,35)[b]{$\nu$}
\end{picture}
\label{r5}\\
 = \begin{picture}(195,40)(0,50)
\SetWidth{1}
\SetColor{Blue}
\Vertex(90,30){2}
\ArrowLine(90,30)(40,30)
\ArrowLine(140,30)(90,30)
\Photon(90,30)(90,0){6}{4}
\Text(38,30)[r]{$p_e,\sigma_e$}
\Text(145,30)[l]{$-p_{\nu},-\sigma_{\nu}$}
\Text(83,15)[r]{$W^{-}$}
\Text(97,15)[l]{$q$}
\Text(65,35)[b]{$e$}
\Text(115,35)[b]{$\nu$}
\end{picture}
+ \begin{picture}(195,40)(0,50)
\SetWidth{1}
\SetColor{Blue}
\Vertex(90,30){2}
\ArrowLine(90,30)(40,30)
\ArrowLine(140,30)(90,30)
\Photon(90,30)(90,0){6}{4}
\Text(38,30)[r]{$p_e,\sigma_e$}
\Text(145,30)[l]{$-p_{\nu},-\sigma_{\nu}$}
\Text(83,15)[r]{$W^{-}$}
\Text(97,15)[l]{$q$}
\Text(65,35)[b]{$e$}
\Text(115,35)[b]{$\nu$}
\PhotonArc(90,30)(36,0,180){6}{11}
\Vertex(126,30){2}
\Vertex(54,30){2}
\Text(120,65)[l]{$Z$}
\end{picture}\nonumber \\
\begin{picture}(195,40)(0,50)
\SetWidth{1}
\SetColor{Blue}
\Vertex(80,-7){2}
\ArrowLine(130,30)(30,30)
\Photon(80,-7)(80,-36){6}{4}
\Text(28,30)[r]{$p_e,\sigma_e$}
\Text(135,30)[l]{$-p_{\nu},-\sigma_{\nu}$}
\Text(73,-20)[r]{$W^{-}$}
\Text(87,-20)[l]{$q$}
\Text(95,35)[b]{$e$}
\PhotonArc(80,30)(36,180,0){6}{11}
\Vertex(116,30){2}
\Vertex(44,30){2}
\Text(115,-3)[l]{$W^-$}
\Text(44,-3)[r]{$A,Z$}
\Text(-13,2)[r]{$+$}
\end{picture}
\begin{picture}(195,40)(0,50)
\SetWidth{1}
\SetColor{Blue}
\Vertex(90,-7){2}
\ArrowLine(140,30)(40,30)
\Photon(90,-7)(90,-36){6}{4}
\Text(38,30)[r]{$p_e,\sigma_e$}
\Text(145,30)[l]{$-p_{\nu},-\sigma_{\nu}$}
\Text(83,-20)[r]{$W^{-}$}
\Text(97,-20)[l]{$q$}
\Text(115,35)[b]{$\nu$}
\PhotonArc(90,30)(36,180,0){6}{11}
\Vertex(126,30){2}
\Vertex(54,30){2}
\Text(125,-3)[l]{$Z$}
\Text(54,-3)[r]{$W^-$}
\Text(0,2)[r]{$+$}
\end{picture}
\nonumber\\
\begin{picture}(400,15)(0,90)
\SetWidth{1.5}
\SetColor{Blue}
\Line(110,0)(220,0)
\Text(195,3)[r]{${\Large\bigotimes^{\large\bf e\bbox{\nu} W}}$}
\Text(100,0)[r]{$+$}
\Text(230,0)[l]{$,$}
\end{picture}\nonumber
\end{eqnarray}

\vspace{3.3cm}

where the last diagram represents the relevant counter term
\begin{eqnarray}
\Gamma^{e\nu W}_{ct \, \alpha}=\Gamma^{e\nu W}_{\alpha}
{\delta}z^{e\nu W} \; , \label{r7}\\ {\delta}z^{e\nu W} =
\Bigl(\frac{1}{2}{\delta}z_L^e+\frac{1}{2}{\delta}z_L^{\nu}+
{\delta}z_1^W-{\delta}z_2^W\Bigr) \; , \label{ze1}
\end{eqnarray}
as one can infer from Eqs. (\ref{l3})-(\ref{l8}),
(\ref{ct})-(\ref{5}). Here ${{\delta}z^{e,\nu}_L}$ render the
renormalization of the electron and neutrino wave functions, and
the difference ${\delta}z^W_1-{\delta}z^W_2$ is expressed through
the $z\gamma-$transition self-energy \cite{h1,h2}
\begin{eqnarray}
{\delta}z^W_1-{\delta}z^W_2{=}\frac{-1}{M_Z^2 s_W
c_W}\Sigma^{Z\gamma}(0)=
 \frac{-\alpha}{4\pi}\frac{2}{s_W^2}\Delta(M_W)
\; . \label{z12}
\end{eqnarray}
Neglecting all the terms of $O({m_e}{/}{M_{Z,W}}) \, , \; \,
O({p_{e,\nu}^2}{/}{M_{W,Z}^2}) \, $ and presuming the fictitious
photon mass in Eq. (\ref{gf}) $m_{\gamma}{=}\lambda{\rightarrow}0$,
we obtain in the one-loop order, $O(\alpha)$,
\begin{eqnarray}
{\delta}z^{e\nu
W}=-\frac{\alpha}{4\pi}\Bigl\{2\ln\frac{\lambda}{m}+\ln\frac{M_Z}{m}
+\frac{9}{4}-\frac{5}{s_W^2}\ln c_W +\frac{1}{s^2_W}
+\frac{10c^2_W+1}{4c_W^2s_W^2}\biggl(\Delta(M_Z)
-\frac{1}{2}\biggr)\Bigr\} \, . \label{ze}
\end{eqnarray}
In (\ref{z12}), (\ref{ze}) and thereafter, the quantities
$\Delta(M_i)$ stand for the UV divergent singular terms for given
masses $M_i$. Within the method of dimensional regularization (see,
for instance, \cite{d,b}), $\Delta(M_i)$ are known to be given as
\begin{equation}
\Delta(M)=\frac{2}{4-D}-\gamma-\ln\frac{M^2}{4\pi{\mu}^2} \, ,
\label{dd}
\end{equation}
where $D \, , \; \gamma \, , \; \mu$ are the space-time dimension,
the Euler constant and the mass scale, respectively. Let us behold
that amenably to the old-established momentum-space cut-off,
$\Delta(M_i)$ could merely be presented as
\begin{equation}
\Delta(M)=\frac{1}{2}+2\ln\frac{\Lambda}{M} \, , \label{dd1}
\end{equation}
with the momentum-space cut-off parameter $\Lambda$
\cite{ll,d,com}. It goes as a matter of course that neither $D \, ,
\; \, \gamma \, , \; \, \mu  \, $, nor
 $\Lambda$ will occur in the corrected renormalized vertexes,
propagators and self-energy parts of fermions and gauge bosons. The
corrected renormalized $e\nu W-$vertex (\ref{r5}) results as
\begin{eqnarray}
\hat\Gamma_{\alpha}^{e\nu W}=
\Gamma^{e\nu W}_{\alpha}\Bigl\{1+\frac{\alpha}{4\pi}\Bigl(
2\ln\frac{m}{\lambda}+\ln\frac{m}{M_Z}-\frac{9}{4}+\frac{3}{s^2_W}
+\frac{6c_W^2-s_W^2}{s^4_W}\ln{c_W}\Bigr)\Bigr\}.\label{i9}
\end{eqnarray}
As seen, the renormalized corrected $e\nu W-$vertexes is multiple
to the bare one, and quarks are not involved in (\ref{i9}), within
the applied one-loop approach. The infrared divergence,
${\sim}{\ln{{\lambda}{/}{m}}}$, occurring in (\ref{i9}) is known to
disappear out of the eventual result for $\beta-$decay probability
\cite{ll,d,com,jfc}.

To acquire the neutron-proton-$W-$boson vertex function
$\hat\Gamma_{\alpha}^{pnW}$ we shall hereafter have to deal with
the renormalized corrected $udW-$vertex $\hat\Gamma_{S
\, \alpha}^{udW}$ for the pure quark transition
$d{\rightarrow}u+W^-$ in the quark system described by the
electroweak Lagrangian (\ref{l2})-(\ref{l8}), with the fictitious
photon mass $m_{\gamma}{=}M_S \, \; (M_N^2{\ll}M_S^2{\ll}M_W^2)$
adopted. In this case, the calculation involves the ``massive
photon" propagator
\begin{equation}
D^{As}_{\alpha\beta}(x)=\delta_{\alpha\beta}
\int\frac{\mbox{d}^4k}{(2\pi)^4}\frac{\exp(-ikx)}{k^2-M_S^2 +i0}
 \; .\label{fss}
\end{equation}
In particular, the wavy line in (\ref{fs}) renders
$D^{As}_{\alpha\beta}(x)$ (\ref{fss}). What is to emphasize is that
this subsidiary mass $M_S$ is negligible as compared to the heavy
boson mass $M_W$, though the nucleon mass $M_N$ is, in turn,
negligible as compared with $M_S$.

In much the same way as in the leptonic case, the corrected
renormalized vertex $\hat\Gamma_{S \, \alpha}^{udW}$ is introduced
by the matrix element
\begin{eqnarray}
\langle\phi_u^{+}(p_u,\sigma_u)|{\cal
S}^{EW}|\phi_{d}(p_{d},\sigma_d) , W^{+ \,
\alpha}(q)\rangle = \nonumber\\
=i(2\pi )^4\delta(q+p_{d}-p_u) \bigl(\bar u_u(p_u,\sigma_u)
\hat\Gamma^{ud W}_{S \, \alpha}(p_u,p_{d},q)w^{+ \, \alpha}(q)
u_{d}(p_{d},\sigma_{d})\bigr)  \label{r8}
\end{eqnarray}
to describe the transition of an initial $d-$quark with the
momentum $p_d$ and polarization $\sigma_d$ into a final $u-$quark
with the momentum $p_u$ and polarization $\sigma_u$, when a
$W^{+}-$boson with the momentum $q{=}p_u-p_d$ and polarization
$\alpha$ is absorbed (or $W^-$ emitted). The quantities $u_{u , d}$
indicate the Dirac spinors of quarks. Following the above expounded
OMS renormalization scheme \cite{ao,h1,h2,b}, we acquire from the
Lagrangian (\ref{l0})-(\ref{l8}),
 with $m_{\gamma}{=}M_S$ assumed, in the one-loop order,
$O(\alpha)$,
\begin{eqnarray}
\bigl(\bar u_u(p_u,\sigma_u)
\hat\Gamma^{ud W}_{S \, \alpha}(p_u,p_{d},q)w^{+ \, \alpha}(q)
u_{d}(p_{d},\sigma_{d})\bigr) = \label{r9} \\
 = \begin{picture}(195,40)(0,25)
\SetWidth{1}
\SetColor{Blue}
\Vertex(90,30){2}
\ArrowLine(90,30)(40,30)
\ArrowLine(140,30)(90,30)
\Photon(90,60)(90,30){6}{4}
\Text(38,30)[r]{$p_u,\sigma_u$}
\Text(145,30)[l]{$p_d,\sigma_d$}
\Text(83,55)[r]{$W^{+}$}
\Text(50,50)[r]{\Large\bf 1}
\Text(97,55)[l]{$q$}
\Text(65,35)[b]{$u$}
\Text(115,35)[b]{$d$}
\end{picture} + \nonumber\\
\begin{picture}(195,40)(0,110)
\SetWidth{1}
\SetColor{Blue}
\Vertex(80,96){2}
\ArrowLine(130,60)(30,60)
\Photon(80,96)(80,136){6}{4}
\Text(28,60)[r]{$p_u,\sigma_u$}
\Text(135,60)[l]{$p_d,\sigma_d$}
\Text(73,129)[r]{$W^{+}$}
\Text(87,129)[l]{$q$}
\Text(120,110)[l]{\Large\bf 2}
\Text(95,65)[b]{$u$}
\PhotonArc(80,60)(36,0,180){6}{11}
\Vertex(116,60){2}
\Vertex(44,60){2}
\Text(115,85)[l]{$W^+$}
\Text(44,85)[r]{$A,Z$}
\Text(-5,80)[r]{$+$}
\end{picture}
\begin{picture}(195,40)(0,110)
\SetWidth{1}
\SetColor{Blue}
\Vertex(80,96){2}
\ArrowLine(130,60)(30,60)
\Photon(80,96)(80,136){6}{4}
\Text(28,60)[r]{$p_u,\sigma_u$}
\Text(135,60)[l]{$p_d,\sigma_d$}
\Text(73,129)[r]{$W^{+}$}
\Text(87,129)[l]{$q$}
\Text(120,110)[l]{\Large\bf 3}
\Text(95,65)[b]{$d$}
\PhotonArc(80,60)(36,0,180){6}{11}
\Vertex(116,60){2}
\Vertex(44,60){2}
\Text(115,85)[l]{$A,Z$}
\Text(44,85)[r]{$W^+$}
\Text(-8,80)[r]{$+$}
\Text(170,80)[l]{$+$}
\end{picture}\nonumber \\
\begin{picture}(195,40)(0,85)
\SetWidth{1}
\SetColor{Blue}
\Vertex(90,30){2}
\ArrowLine(90,30)(40,30)
\ArrowLine(140,30)(90,30)
\Photon(90,60)(90,30){6}{4}
\Text(38,30)[r]{$p_u,\sigma_u$}
\Text(145,30)[l]{$p_d,\sigma_d$}
\Text(83,55)[r]{$W^{+}$}
\Text(97,55)[l]{$q$}
\Text(135,50)[l]{\Large\bf 4}
\Text(65,35)[b]{$u$}
\Text(115,35)[b]{$d$}
\PhotonArc(90,30)(36,180,0){6}{11}
\Vertex(126,30){2}
\Vertex(54,30){2}
\Text(120,0)[l]{$Z$}
\Text(-5,3)[r]{$+$}
\end{picture}
\begin{picture}(195,40)(0,85)
\SetWidth{1}
\SetColor{Blue}
\Vertex(90,30){2}
\ArrowLine(90,30)(40,30)
\ArrowLine(140,30)(90,30)
\Photon(90,60)(90,30){6}{4}
\Text(38,30)[r]{$p_u,\sigma_u$}
\Text(145,30)[l]{$p_d,\sigma_d$}
\Text(83,55)[r]{$W^{+}$}
\Text(97,55)[l]{$q$}
\Text(130,50)[l]{\Large\bf 5}
\Text(65,35)[b]{$u$}
\Text(115,35)[b]{$d$}
\PhotonArc(90,30)(36,180,0){3}{20}
\Vertex(126,30){2}
\Vertex(54,30){2}
\Text(120,0)[l]{As}
\Text(-8,3)[r]{$+$}
\Text(170,3)[l]{$+$}
\end{picture}\nonumber\\
\begin{picture}(400,15)(0,105)
\SetWidth{1.5}
\SetColor{Blue}
\Line(110,0)(220,0)
\Text(195,3)[r]{${\Large\bigotimes^{\large\bf udW}}$}
\Text(200,10)[bl]{\Large\bf 6}
\Text(100,0)[r]{$+$}
\Text(230,0)[l]{$,$}
\end{picture}
\nonumber
\end{eqnarray}
\vspace{3cm}

where the wavy line with the tag $As$ stands for the ``massive
photon" propagator $D^{As}$ (\ref{fss}). The first graph in
(\ref{r9}) depicts the bare $udW-$vertex
\begin{eqnarray}
\Gamma_{\alpha}^{udW}=|V_{ud}|\frac{e}{2\sqrt{2}s_W}\gamma_{\alpha}
(1-\gamma^5) \label{r11}
\end{eqnarray}
originating from ${\cal L}^{Wff'}$ (\ref{l5}), and the last one
accordingly Eqs. (\ref{ct})-(\ref{5}) represents the counter term
\begin{eqnarray}
\hat\Gamma^{udW}_{S \, \alpha \, ct} =
\Bigl(\frac{1}{2}{\delta}{z^u_L}+
\frac{1}{2}{\delta}{z^d_L}+{\delta}{z^W_1}-
{\delta}{z^W_2}\Bigr)\cdot
\Gamma^{udW}_{\alpha} \, , \label{r10}
\end{eqnarray}
where ${\delta}z^{u,d}_L$ render the renormalization of the quark
wave functions, and the difference ${\delta}z_1^W-{\delta}z_2^W$ is
given by (\ref{z12}). Omitting the terms
$O({p_{u,d}^2}{/}{M^2_{W,Z,S}}) \, , O({M^2_S}{/}{M^2_{W,Z}})$ , we
obtain the corrected renormalized vertex
\begin{eqnarray}
\hat\Gamma^{udW}_{S \, \alpha} =
\Gamma^{udW}_{\alpha}\cdot\Gamma(W) \, , \label{i16} \\
\Gamma(W)=
\Biggl\{1+\frac{\alpha}{4\pi}\Bigl(\ln\frac{M_S}{M_Z} +
\frac{3}{s^2_W} +\frac{6c^2_W - s^2_W}{s^4_W}\ln(c_W)\Bigr)\Biggr\}
 \, , \label{i160}
\end{eqnarray}
 multiple to the bare vertex (\ref{r11}). Of course, there occurs
no infrared divergence in $\hat\Gamma^{udW}_{S \, \alpha}$
(\ref{i160}).

So, we have acquired the renormalized corrected $e\nu W-$ and
$udW-$vertices which are needed to calculate the neutron
$\beta-$decay amplitude.
\section{Treatment of nucleon structure in
 describing the neutron ${\large \bfgr\beta-}$decay.}
\label{sec:level3}
Up to now, we have dealt with the pure electroweak interactions $
\, {\cal L}^{EW}_{int} \, $ (\ref{l2})-(\ref{l8}). As the nucleon is a
complex system of strong interacting quarks, the neutron
$\beta-$decay (\ref{n}) can never be reduced to the pure transition
\begin{equation}
d\Longrightarrow u+e^- +\bar\nu +\gamma \, . \label{nq}
\end{equation}

We are to allow for the nucleon compositeness, such as excited
states and form factors associated with the nucleon intrinsic
structure caused by the strong quark-quark interactions. Therefore,
${\cal L}_{int}^{EW}$ (\ref{l1}) is to be completed by ${\cal
L}^{qq}_{str}$ to describe the transition (\ref{nq}) in a system of
strong interacting quarks,
\begin{equation}
{\cal L}_{int}(x)={\cal L}^{EW}_{int}(x)+{\cal L}^{qq}_{str}(x) \,
.
\label{i1}
\end{equation}

Ignored the strong quark-quark interactions ${\cal
L}^{qq}_{str}(x)$, the baryon is a free quark system described (in
terms of quark occupation numbers) by the Heisenberg wave function
$\Phi^q_{0 \, B}(P_B,\sigma_B)$ with the given total momentum
$P_B$, and the spin $\bbox{\sigma}_B$ and polarization
$\sigma_{Bz}$ indicated as $\sigma_{B}$. So far as interactions
vanish at infinity,
\begin{equation}
{\cal L}_{int}(x)\longrightarrow 0 \, , \; \;
\; \text{when} \; \; \; x^0\longrightarrow \mp\infty \, , \label{i01}
\end{equation}
the baryon wave function in the interaction representation is
written in the ordinary form:
\begin{eqnarray}
\Phi^q_B (P_B , \sigma_B
, x^0)={\cal S}_{str}(x^0 , \mp\infty )\Phi^q_B (P_B , \sigma_B ,
\mp\infty )={\cal S}_{str}(x^0 , \mp\infty )\Phi^q_{0B}(P_B ,
\sigma_B) \, , \label{i3}\\
{\cal S}_{str}(x^0 , -\infty )={\cal T}
\exp\Bigl(i\int\limits_{-\infty}^{x^0} \mbox{d}x^0 \int \mbox{d}
{\bf x} {\cal L}^{qq}_{str}(x)\Bigr) \, , \; \; \; \; \; {\cal
S}(t,t')\cdot {\cal S}(t',t_0)={\cal S}(t,t_0) \, . \nonumber
\end{eqnarray}
The operator
\begin{equation}
{\cal S}_{str}(x_1^0 , x_2^0)={\cal T}
\exp\Bigl(i\int\limits_{x_1^0}^{x_2^0} \mbox{d}x^0 \int \mbox{d}
{\bf x} {\cal L}^{qq}_{str}(x)\Bigr) \label{xx}
\end{equation}
transforms a state of the quark system at a time-point $x_1^0$ to a
state at a time-point $x_2^0$ :
\begin{equation}
\Phi^q_B (P_B , \sigma_B
, x_2^0)={\cal S}_{str}(x_2^0 , x_1^0)\Phi^q_B (P_B , \sigma_B ,
x_1^0) \, . \label{i4}
\end{equation}

 The transition amplitude
\begin{eqnarray}
{\Large \cal M}= \begin{picture}(250,25)(0,0)
\SetColor{Blue}
\SetWidth{1}
\CCirc(100,5){15}{Blue}{Red}
\ArrowLine(160,20)(100,5)
\ArrowLine(160,-20)(100,5)
\ArrowLine(100,5)(25,5)
\Photon(100,5)(40,30){3}{8}
\ArrowLine(100,5)(25,-20)
\Text(165,20)[l]{$-p_{\nu}$}
\Text(165,-15)[l]{$P_n$}
\Text(18,5)[r]{$p_e$}
\Text(35,30)[r]{$p_{\gamma}$}
\Text(20,-15)[r]{$P_p$}
\end{picture}\label{im}
\end{eqnarray}
to describe the neutron $\beta-$decay (\ref{n}) is determined by the
 matrix element of ${\cal S}_{int}$
\begin{eqnarray}
{\cal M}\cdot i(2\pi )^4\delta
(P_n-P_p
-p_e -p_{\nu} -p_{\gamma})= \nonumber\\
\langle \Phi_{0p}^{q \, +}(P_p ,\sigma_p ) ,
\phi_e^{+}(p_e ,\sigma -e) , A(p_{\gamma})|{\cal S}_{int}
|\Phi_{0n}^q (P_n ,\sigma_n) ,
\phi_{\nu}(-p_{\nu},-\sigma_e)\rangle \; , \label{i5}\\
{\cal S}_{int}{\equiv}{\cal S}_{int}(\infty ,-\infty)={\cal T}
\exp\Bigl(i\int
\mbox{d}^4x{\cal L}_{int}(x)\Bigr)
= {\cal T} \exp\Bigl(i\int
\mbox{d}^4x[{\cal L}_{int}^{EW}(x) + {\cal L}_{str}^{qq}(x)]\Bigr) \, .
\label{i6}
\end{eqnarray}

For now, there sees no option, but to parameterize the effects of
strong interactions in treating the neutron $\beta-$decay. We do
not intend neither to specify an actual form of ${\cal
L}_{int}^{qq}(x)$, nor to procure an explicit expression of the
baryon wave function $\Phi^q_B(P_B,\sigma_B)$ in
(\ref{i3})-(\ref{i6}), but we posit an appropriate parameterization
of matrix elements of the electroweak interactions ${\cal
L}_{int}^{EW}$ (\ref{l2})-(\ref{l8}) between the baryon wave
functions $\Phi^q_B(P_B,\sigma_B)$. In this respect, by introducing
the ordinary nucleon weak transition current
\begin{equation}
{\cal J}^{\beta}_{np}(k)=\gamma^{\beta}g_V(k^2)+
g_{WM}(k^2)\sigma^{\beta\nu}k_{\nu}-(\gamma^{\beta}g_A(k^2)
+g_{IP}(k^2)k^{\beta})\gamma^5 \, , \label{i8}
\end{equation}
the matrix element of ${\cal L}^{Wff'}$ (\ref{l5})
\begin{eqnarray}
\Lambda_{0 \, \alpha}^{npW}(k){=}
\int \mbox{d}^4y \langle \Phi^{q \, +}_{p}(P_p ,\sigma_p )
|\bar\psi_q(y)\Gamma_{\chi}^{udW}(k){T}^{+}_q\psi_q(y)W^{+ \,
\chi}(y) |\Phi^q_{n}(P_n,\sigma_n),W^{+}_{\alpha}(k)\rangle
\, , \label{j}\\ k{=}P_p-P_n \; ,\nonumber
\end{eqnarray}
is rewritten in terms of the nucleon field operators,
\begin{eqnarray}
\Psi_N(y)=\sum_{P_N,\sigma_N}\biggl(U_N({\bf P}_N,\sigma_N)
 a_N({\bf P}_N,\sigma_N)\exp[-iP_Ny]+\nonumber\\ U_N(-{\bf
 P}_N,-\sigma_N) b^{+}_N({\bf P}_N,\sigma_N)\exp[iP_Ny]\biggr) \, ,
 \label{nn}
\end{eqnarray}
and the nucleon wave functions $\Phi^N_{n,p}(P_{n,p},\sigma_{n,p})$
describing the single-nucleon states with the given
$P_{n,p},\sigma_{n,p}$. What results is
\begin{eqnarray}
\Lambda_{0 \, \alpha}^{npW}(k){=}
\int \mbox{d}^4y\langle \Phi^{N \, +}_{p}(P_p ,\sigma_p )
|\bar\Psi_N(y)\Gamma_{\chi}^{npW}(k){T}^{+}_N\Psi_N(y) W^{+
\, \chi}(y)|\Phi^N_{n}(P_n,\sigma_n),W^{+}_{\alpha}(k)
\rangle{=}\nonumber\\
=(2\pi)^4\delta(P_n-P_p+k)\bar U_p(P_p,\sigma_p)
\Gamma_{\alpha}^{npW}(k)T^{+}_N U_n(P_p,\sigma_n)w^{+}_{\alpha}(k)
 \, , \label{j1a}
\end{eqnarray}
where
\begin{eqnarray}
\Gamma_{\alpha}^{npW}(k)=\frac{e|V_{ud}|}{2\sqrt{2}s_W}{\cal J}_{np
\, \alpha}(k) = \begin{picture}(120,30)(0,0)
\SetColor{Blue}
\SetWidth{1}
\Vertex(50,7){6}
\Photon(50,7)(50,40){6}{4}
\ArrowLine(50,7)(20,7)
\ArrowLine(80,7)(50,7)
\Text(50,-2)[t]{${\cal J}_{np}$}
\Text(17,7)[r]{$p$}
\Text(83,7)[l]{$n$}
\Text(60,30)[l]{$W^{+}$}
\end{picture} , \label{j1b}
\end{eqnarray}
the operator $T^{+}_N$ transforms the neutron into the proton,
$U_{n,p}$ indicate the Dirac spinors of nucleons. So, the matrix
element $\Lambda_0^{npW}(k)$, originally written in terms of the
quark states, results to be expressed through the nucleon states
and the electroweak form factors $g_V \, , \; \, g_A \, , \; \,
g_{WM} \, , \; \, g_{IP}$. Hereafter we shall also have to deal
with the general case of weak transitions between the
single-baryonic states $\Phi_s^B(P_s,\sigma_s)$ including, besides
the neutron and proton,
 various excited states of the nucleon. Alike Eqs. (\ref{j1a}),
 (\ref{j1b}), the matrix elements to describe these processes are
 written in terms of the baryonic field operators $\Psi^B_s(x)$ and
 the appropriate generalized transition currents
\begin{eqnarray}
\Lambda_{0 \, \alpha}^{rsW}=
\int \mbox{d}^4y\langle \Phi^{B \, +}_{r}(P_r ,\sigma_r)
|\bar\Psi_r(y)\Gamma_{\chi}^{rsW}(k){T}^{+}_B\Psi_s(y) W^{+
\, \chi}(y)|\Phi^B_{s}(P_s,\sigma_s),W^{+}_{\alpha}(k)\rangle
 \, , \label{j2a}
\end{eqnarray}
where
\begin{eqnarray}
\Gamma_{\alpha}^{rsW}(k)=\frac{e|V_{ud}|}{2\sqrt{2}s_W}{\cal J}_{rs
\, \alpha}(k) = \begin{picture}(120,30)(0,0)
\SetColor{Blue}
\SetWidth{1}
\Vertex(50,7){6}
\Photon(50,7)(50,40){6}{4}
\ArrowLine(50,7)(20,7)
\ArrowLine(80,7)(50,7)
\Text(50,-2)[t]{${\cal J}_{rs}$}
\Text(17,7)[r]{$r$}
\Text(83,7)[l]{$s$}
\Text(60,30)[l]{$W^{+}$}
\end{picture}  \label{j2b}
\end{eqnarray}
and $T_B^{+}$ increases baryon charge by one unite.

In much the same way, the matrix element of ${\cal L}^{Aqq}$
(\ref{l8}) transforms as follows
\begin{eqnarray}
\int\mbox{d}^4x\langle\Phi_B^{q \, +}(P_B,\sigma_B)|{\cal
L}^{Aqq}(x)|\Phi_{B'}^q(P_{B'},\sigma_{B'}), A^{\alpha}(k)\rangle =
\begin{picture}(120,25)(0,-2)
\SetColor{Blue}
\SetWidth{1}
\Vertex(50,7){6}
\Photon(50,7)(50,30){3}{4}
\ArrowLine(50,7)(20,7)
\ArrowLine(80,7)(50,7)
\Text(48,0)[t]{$f^{BB'}$}
\Text(17,7)[r]{$B'$}
\Text(83,7)[l]{$B$}
\Text(60,25)[l]{$A$}
\end{picture}\nonumber\\
= -e (2\pi)^4\delta(P_{B'}-P_B-k)\biggl(\bar
U_B(P_B,\sigma_B)f^{BB'}_{\alpha}(k)U_{B'}(P_{B'},\sigma_{B'})
\biggr)A^{\alpha}(k) \, , \label{a}
\end{eqnarray}
where the form factors $f^{BB'}_{\alpha}(k)$ to describe the
electromagnetic transitions of baryons $B'{\rightarrow}B$ are of
the usual form \cite{ll,d,com}
\begin{equation}
f^{NN}_{\alpha}(k)=f^{NN}_{1}(k^2)\gamma_{\alpha}+f^{NN}_{2}(k^2)
k^{\beta}\sigma_{\alpha\beta} \label{a1}
\end{equation}
in the case of neutron and proton ($N{=}n,p$) interactions with
electromagnetic field $A^{\alpha}$. At the momentum transferred
$k^2{\lesssim}M_N$, the quantity $g_{WM}$ is given through the
nucleon anomalous magnetic moments,
\begin{equation}
g_{WM}\approx\frac{\mu_n-\mu_p}{2M_p}\approx-\frac{3.7}{2M_p} \, ,
\label{wm}
\end{equation}
the assessment
\begin{equation}
g_{IP}(k^2)\approx\frac{2M_p \, g_A(k^2)}{k^2-m_{\pi}^2}\sim\
\frac{2M_p \, g_A(0)}{m_{\rho}^2-m_{\pi}^2}\sim\frac{8g_A(0)}{2M_p}
 \, , \label{ip}
\end{equation}
is appropriate, and the estimations
\begin{equation}
f^{pp}_{1}(k^2)\approx\frac{-m^2_{\rho}}{k^2-m_{\rho}^2} \, ,
 \; \; f^{pp}_2(k^2)\approx\Bigl(\frac{1.79}{2 M_p}\Bigr)
\frac{-m^2_{\rho}}{k^2-m_{\rho}^2}
 \, , \; \; \; f^{nn}_1=0 \, , \; \;
  f^{nn}_{2}(k^2)=\Bigl(\frac{1.93}{2
 M_n}\Bigr)\frac{m^2_{\rho}}{k^2-m_{\rho}^2} \, \label{ff}
\end{equation}
hold true within the vector-dominant model (see, for instance,
Refs. \cite{ll,d,com}). Here $m_{\pi} \, , \; \, m_{\rho}$ are
conceived to be of the order of the $\pi-$ and $\rho-$meson masses.
Evidently, at $k^2{\ll}M_N^2$, Eqs. (\ref{i8}), (\ref{a1}) are
reduced to
\begin{eqnarray}
 {\cal J}^{\beta}_{np}(0)=\gamma^{\beta}-
\gamma^{\beta}g_A\gamma^5 \, ,\label{j0} \\
\Gamma_{\alpha}^{npW}(0)=\frac{e|V_{ud}|}{2\sqrt{2}s_W}\gamma_{\alpha}
\bigl(1-g_A(0)\gamma^5\bigr) \, , \label{g0} \\
f_{\alpha}^{pp}(0)=\gamma_{\alpha} \, , \; \; \; f^{nn}(0)=0 \, ,
\label{f0}
\end{eqnarray}
and the nucleon is treated as being a point-like particle, except
for the residence of $g_A$ in the nucleon weak transition current
(\ref{j0}).
\section{Transition amplitude.}
\label{sec:level4}
 As dictated by ${\cal L}_{int}$ (\ref{i1}), the transition
 amplitude ${\cal M}$ (\ref{im})-(\ref{i6}) is represented in the
 one-loop order, $O(\alpha)$, by the set of diagrams
\begin{eqnarray}
\begin{picture}(145,30)(0,5)
\SetWidth{1}
\SetColor{Blue}
\Vertex(50,30){2}
\Vertex(50,-10){5}
\Photon(50,-10)(50,30){6}{4}
\ArrowLine(50,30)(15,30)
\ArrowLine(85,30)(50,30)
\ArrowLine(50,-10)(15,-10)
\ArrowLine(85,-10)(50,-10)
\Text(10,30)[r]{$p_e,\sigma_e$}
\Text(90,30)[l]{$-p_{\nu},-\sigma_{\nu}$}
\Text(10,-10)[r]{$P_p,\sigma_p$}
\Text(90,-10)[l]{$P_n,\sigma_n$}
\Text(35,33)[b]{$e$}
\Text(35,-13)[t]{$p$}
\Text(35,10)[r]{\Large\bf 1}
\Text(58,10)[l]{W}
\Text(65,33)[b]{$\nu$}
\Text(65,-13)[t]{$n$}
\end{picture}
\begin{picture}(145,30)(0,5)
\SetWidth{1}
\SetColor{Blue}
\CCirc(75,30){10}{Blue}{Red}
\Vertex(75,-10){5}
\Text(20,10)[r]{$+$}
\Photon(75,-10)(75,30){6}{4}
\ArrowLine(75,30)(40,30)
\ArrowLine(110,30)(75,30)
\ArrowLine(75,-10)(40,-10)
\ArrowLine(110,-10)(75,-10)
\Text(40,33)[b]{$e$}
\Text(40,-13)[t]{$p$}
\Text(110,33)[b]{$\nu$}
\Text(110,-13)[t]{$n$}
\Text(107,10)[l]{\Large\bf 2}
\end{picture}
\begin{picture}(145,30)(0,5)
\SetWidth{1}
\SetColor{Blue}
\Vertex(70,30){2}
\CCirc(70,-10){11}{Blue}{Yellow}
\Vertex(70,-10){4.5}
\Text(10,10)[r]{$+$}
\Text(115,10)[l]{$+$}
\Photon(70,-10)(70,30){6}{4}
\ArrowLine(70,30)(35,30)
\ArrowLine(105,30)(70,30)
\ArrowLine(70,-10)(35,-10)
\ArrowLine(105,-10)(70,-10)
\Text(30,33)[b]{$e$}
\Text(30,-13)[t]{$p$}
\Text(40,10)[r]{\Large\bf 3}
\Text(105,33)[b]{$\nu$}
\Text(105,-13)[t]{$n$}
\end{picture}
\nonumber\\
\begin{picture}(145,30)(50,60)
\SetWidth{1}
\SetColor{Blue}
\Vertex(75,30){2}
\Vertex(75,-10){5}
\Text(39,10)[r]{$+$}
\Text(142,10)[l]{$+$}
\SetWidth{3}
\Photon(75,-10)(75,30){6}{4}
\SetWidth{1}
\ArrowLine(75,30)(40,30)
\ArrowLine(110,30)(75,30)
\ArrowLine(75,-10)(40,-10)
\ArrowLine(110,-10)(75,-10)
\Text(40,33)[b]{$e$}
\Text(40,-13)[t]{$p$}
\Text(110,33)[b]{$\nu$}
\Text(110,-13)[t]{$n$}
\Text(110,10)[l]{\Large\bf 4}
\end{picture}
\begin{picture}(145,30)(30,60)
\SetWidth{1}
\SetColor{Blue}
\Vertex(75,30){2}
\Vertex(75,-10){5}
\Text(147,10)[l]{$+$}
\Photon(75,-10)(75,30){6}{4}
\ArrowLine(75,30)(40,30)
\ArrowLine(110,30)(75,30)
\ArrowLine(75,-10)(40,-10)
\ArrowLine(110,-10)(75,-10)
\Text(40,33)[b]{$e$}
\Text(40,-13)[t]{$p$}
\Text(110,33)[b]{$\nu$}
\Text(110,-13)[t]{$n$}
\Text(110,10)[l]{\Large\bf 5}
\Photon(50,30)(18,15){2.5}{5}
\Text(16,13)[tr]{$\gamma$}
\end{picture}
\begin{picture}(145,30)(0,60)
\SetWidth{1}
\SetColor{Blue}
\Vertex(75,30){2}
\Vertex(75,-10){5}
\Text(120,10)[l]{$+$}
\Photon(75,-10)(75,30){6}{4}
\ArrowLine(75,30)(40,30)
\ArrowLine(110,30)(75,30)
\ArrowLine(75,-10)(40,-10)
\ArrowLine(110,-10)(75,-10)
\Text(40,33)[b]{$e$}
\Text(40,-13)[t]{$p$}
\Text(110,33)[b]{$\nu$}
\Text(110,-13)[t]{$n$}
\Text(103,10)[l]{\Large\bf 6}
\Photon(50,-10)(25,3){2.5}{5}
\Text(23,1)[br]{$\gamma$}
\Line(50,-11.5)(75,-11.5)
\Line(50,-8.5)(75,-8.5)
\Vertex(50,-10){3.5}
\end{picture}
\label{mm}\\
\begin{picture}(145,30)(40,100)
\SetWidth{1}
\SetColor{Blue}
\Vertex(70,30){2}
\Vertex(70,-10){5}
\Text(28,10)[r]{$+$}
\Text(135,10)[l]{$+$}
\Photon(70,-10)(70,30){6}{4}
\ArrowLine(70,30)(35,30)
\ArrowLine(105,30)(70,30)
\ArrowLine(70,-10)(35,-10)
\ArrowLine(105,-10)(70,-10)
\Text(30,33)[b]{$e$}
\Text(30,-13)[t]{$p$}
\Text(105,33)[b]{$\nu$}
\Text(105,-13)[t]{$n$}
\Text(105,10)[l]{\Large\bf 7}
\PhotonArc(70,-10)(27,240,360){2.5}{7}
\Text(53,-37)[r]{$\gamma$}
\Line(70,-11.5)(97,-11.5)
\Line(70,-8.5)(97,-8.5)
\Vertex(97,-10){3.5}
\end{picture}
\begin{picture}(145,30)(40,100)
\SetWidth{1}
\SetColor{Blue}
\Vertex(70,30){2}
\Vertex(70,-10){5}
\Text(145,10)[l]{$+$}
\Photon(70,-10)(70,30){6}{4}
\ArrowLine(70,30)(35,30)
\ArrowLine(105,30)(70,30)
\ArrowLine(70,-10)(35,-10)
\ArrowLine(105,-10)(70,-10)
\Text(30,33)[b]{$e$}
\Text(30,-13)[t]{$p$}
\Text(105,33)[b]{$\nu$}
\Text(105,-13)[t]{$n$}
\Text(105,10)[l]{\Large\bf 8}
\Vertex(70,10){2}
\Photon(70,10)(35,10){2.5}{5}
\Text(33,10)[r]{$\gamma$}
\end{picture}
\begin{picture}(145,30)(0,100)
\SetWidth{1}
\SetColor{Blue}
\ArrowLine(30,30)(0,30)
\ArrowLine(130,30)(100,30)
\ArrowLine(30,-10)(0,-10)
\ArrowLine(130,-10)(100,-10)
\Text(10,33)[b]{$e$}
\Text(10,-13)[t]{$p$}
\Text(120,33)[b]{$\nu$}
\Text(120,-13)[t]{$n$}
\Text(140,0)[l]{{$\bf,$}}
\Text(120,10)[l]{\Large\bf 9}
\SetWidth{4}
\CBox(30,-10)(100,30){Blue}{Red}
\SetWidth{1}
\end{picture} \nonumber
\end{eqnarray}

\vspace{4.5cm}

with the contents heretofore given by (\ref{l2})-(\ref{l8}),
(\ref{fs}), (\ref{g0}), (\ref{r5}), (\ref{r9}), (\ref{j1b})
(\ref{j2b}) and also currently explicated hereafter, as far as
used. At the lowest order in ${\cal L}_{int}^{EW}$
(\ref{l2})-(\ref{l8}), that is without radiative corrections, the
uncorrected Born amplitude ${\cal M}^0$ presented by the first
graph in (\ref{mm}) is determined by
\begin{eqnarray}
{\cal M}^0\cdot i(2\pi)^4\delta (P_n-P_p -p_e -p_{\nu})=
\Bigl(\frac{e}{2\sqrt{2}s_W}\Bigr)^2
|V_{ud}|\int\frac{\mbox{d}^4k}{(2\pi)^4}\int \mbox{d}^4x\times
\nonumber\\
\times\langle \phi^{+}_e (p_e ,\sigma_e )|\bar\psi_e (x)\gamma^{
\alpha}(1-\gamma^5)
\psi_{\nu}(x)|\phi_{\nu}(-p_{\nu},-\sigma_{\nu})\rangle
(i)\frac{g_{\alpha\beta}}{k^2-M_W^2}\times\label{m01}\\
\times\int\mbox{d}^4y e^{-ik(x-y)}
\langle \Phi^{q \, +}_{0p}(P_p , \sigma_p)|{\cal T}
\Bigl\{\bar\psi_q(y)\gamma^{\beta}(1-\gamma^5)T^+_q\psi_q(y)\cdot{\cal
S}_{str}\Bigr\}|\Phi_{0n}^q(P_n , \sigma_n )\rangle \, , \nonumber
\end{eqnarray}
where the strong interactions intrude via ${{\cal
S}_{str}}{\equiv}{{\cal S}_{str}(\infty,-\infty)} \, $ (\ref{xx}).
With allowance for the relations
\begin{eqnarray}
{\cal S}_{str}={\cal S}_{str}(\infty ,y^0){\cal
S}_{str}(y^0,-\infty )
\, , \; \; \; \Phi^q_N (P_N , \sigma_N ,y^0)={\cal S}_{str}(y^0 ,
-\infty )\Phi^q_{0N}(P_N , \sigma_N ) \label{j3} \, ,
\end{eqnarray}
the last integral in (\ref{m01}) is reduced as follows
\begin{eqnarray}
\int \mbox{d}^4 y e^{iky} \langle \Phi^{q \, +}_{0p}(P_p ,\sigma_p )
{\cal S}_{str}(\infty
,y^0)|\bar\psi_q(y)\gamma^{\beta}(1-\gamma^5){
T}^{+}_q\psi_q(y)|{\cal
S}_{str}(y^0,-\infty)\Phi^q_{0n}(P_n,\sigma_n)\rangle
=\nonumber\\
\int \mbox{d}^4 y e^{iky} \langle \Phi^{q \, +}_{p}(P_p ,\sigma_p )
|\bar\psi_q(y)\gamma^{\beta}(1-\gamma^5){
T}^{+}_q\psi_q(y)|\Phi^q_{n}(P_n,\sigma_n)\rangle
 \, . \label{j1}
\end{eqnarray}
Applying to the expressions (\ref{i8})-(\ref{j1b}), the Born
amplitude proves to be
\begin{eqnarray}
{\cal M}^0 =\bar u_e(p_e ,\sigma_e ) \Gamma_{\alpha}^{e\nu
W}u_{\nu}(-p_{\nu},-\sigma_{\nu} ){\times}
\nonumber\\ \times
\int \mbox{d}^4 y e^{iqy} \langle \Phi^{q \, +}_{p}(P_p
,\sigma_p ) |\bar\psi_u(y)
\; \Gamma_{\beta}^{udW}(q) \; \psi_d(y)|\Phi^q_{n}(P_n,\sigma_n)\rangle
\cdot D^W_{\alpha\beta}(q) =\nonumber\\
=\bar u_e(p_e ,\sigma_e )
\Gamma_{\alpha}^{e\nu W}u_{\nu}(-p_{\nu},-\sigma_{\nu})
\cdot \bar U_p(P_p ,\sigma_p )\Gamma_{
\beta}^{npW}(q) U_n(P_n , \sigma_n )\cdot D_{\alpha\beta}^W(q) \,
 ,\label{i7} \\ {\Gamma}_{\alpha}^{e\nu W} =
\frac{e}{2\sqrt{2}s_W}\gamma_{\alpha}(1-\gamma^5)
\, , \; \; \;
\; \Gamma_{\alpha}^{np
W}(q)=|V_{ud}|\frac{e}{2\sqrt{2}s_W}{\cal J}^{\alpha}_{np}(q) \,
,\nonumber \\ q=P_n-P_p-p_e-p_{\nu} \, \nonumber
\end{eqnarray}
As $q^2{\ll}M^2_p{\ll}M^2_W$, the quantities $\Gamma_{\alpha}^{np
W}(q)\, , \; \; {\cal J}^{\alpha}_{np}(q)$ are replaced by
(\ref{g0}), (\ref{j0}), and
\begin{eqnarray}
D_{\alpha\beta}^W (q)=\frac{g_{\alpha\beta}}{q^2-M_W^2}
=\frac{-g_{\alpha\beta}}{M^2_W} \, . \label{i71}
\end{eqnarray}

With allowance for the radiative corrections, the bare, uncorrected
vertexes ${\Gamma}_{\alpha}^{e\nu W}\, , \; \,
\Gamma_{\alpha}^{np W}(q)$ and $W-$propagator $D_{\alpha\beta}^W
(q)$ in ${\cal M}^0$ (\ref{i7}), depicted by the point, blob and
thin wavy line in the graph ${\large\bf 1}$ in Eq. (\ref{mm}), will
give place to the corrected renormalized quantities
${\hat\Gamma}_{\alpha}^{e\nu W}\, , \; \, \hat\Gamma_{\alpha}^{np
W}(q) \, , \; \; \hat D_{\alpha\beta}^W (q)$, what counts is that
the terms presented by the graphs ${\large\bf 2, 3, 4}$ emerge in
${\cal M}$ (\ref{mm}) in the one-loop order, $O(\alpha)$;
${\hat\Gamma}_{\alpha}^{e\nu W}\, , \; \, \hat\Gamma_{\alpha}^{np
W}(q) \, , \; \; \hat D_{\alpha\beta}^W (q)$ are depicted by the
shaded circle, the shaded circle with heavy core, and the heavy
wavy line in the graphs ${\large\bf 2, 3, 4}$, respectively.

The terms presented by the graphs ${\large\bf 5, 6, 7, 8}$ describe
the real $\gamma-$radiation, and the graphs of the type ${\large\bf
9}$, usually called the ``box-diagrams", render generically all the
irreducible four-particle processes.

The contribution of the graph ${\large\bf 2}$ is merely acquired
from (\ref{i7}) by replacement of ${\Gamma}_{\alpha}^{e\nu W}$ in
(\ref{i7}) by ${\hat
\Gamma}_{\alpha}^{e\nu W}$ (\ref{r5}), (\ref{i9}).

The corrected renormalized vertex ${\hat\Gamma}_{\alpha}^{np W}$ in
the graph ${\large\bf 3}$ in (\ref{mm}) describes the
$n{\rightarrow}p$ transition by absorbing a $W^{+ \, \alpha}(q)$
boson with the polarization $\alpha$ and the momentum $q$ (or
emitting $W^{- \, \alpha}(q)$). The contribution of the graph
${\large\bf 3}$ originates from (\ref{i7}) by replacing
${\Gamma_{\alpha}^{np
W}}{\Longrightarrow}{\hat\Gamma_{N\alpha}^{npW}}$. So, the
calculation of $\hat\Gamma_{\alpha}^{np W}(q)$ is in order.
\section{The radiative corrections to the
 \lowercase{pn}w-vertex without involving strong quark-quark
 interactions.}
\label{sec:level5}
 In the third order in the quark part of ${\cal L}^{EW}_{int}$
(\ref{l2}), the vertex $\hat\Gamma_{\alpha}^{np W}(q)$ is defined by
the matrix element which involves besides the electroweak
interactions, ${\cal L}^{Zqq} \, , \; \, {\cal L}^{Wqq} \, , \; \,
{\cal L}^{Aqq} \, , \; \, {\cal L}^{ZWW} \, , \; \, {\cal L}^{AWW} $
(\ref{l3})-(\ref{l8}), the strong quark-quark interactions
 ${\cal L}_{str}^{qq}$ as well, via ${\cal S}_{str}{\equiv}{\cal
 S}_{str}(\infty ,-\infty)$ (\ref{xx}) :
\begin{eqnarray}
i(2\pi)^4\delta (P_n-P_p+q) \; \bigl( \bar U_p(P_p ,
\sigma_p )\hat\Gamma_{\alpha}^{pnW}(P_n,P_p,q)
 w^{+ \; \alpha}(q) U_n(P_n , \sigma_n ) \bigr)=i\Lambda^{npW}_{0
 \, \alpha}(q)+
\nonumber\\
(-i)\int\mbox{d}^4 x_1\int\mbox{d}^4 x_2\int\mbox{d}^4 x_3
\langle\Phi^{q \, +}_{0\, p}(P_p , \sigma_p)|{\cal T}\Bigl\{\Bigl(
{\cal L}^{Wqq}(x_1){\cal L}^{Wqq}(x_2){\cal L}^{Wqq}(x_3)+
\label{i10}\\ {\cal L}^{Wqq}(x_1){\cal L}^{Zqq}(x_2){\cal
L}^{ZWW}(x_3)+ {\cal L}^{Wqq}(x_1){\cal L}^{Aqq}(x_2){\cal
L}^{AWW}(x_3)+
\nonumber\\ {\cal L}^{Wqq}(x_1){\cal L}^{Zqq}(x_2){\cal
L}^{Zqq}(x_3)+
 {\cal L}^{Wqq}(x_1){\cal L}^{Aqq}(x_2){\cal L}^{Aqq}(x_3)\Bigr) \,
 \cdot {\cal S}_{str} \Bigr\} |\Phi^q_{0n} (P_n,\sigma_n),
 W^{+}_{\alpha}(q)\rangle \equiv
\nonumber\\
\equiv i\Lambda^{npW}_{0 \, \alpha}(q)+\Lambda_{\alpha}^{WWW}(q) +
 \Lambda_{\alpha}^{WWZ}(q)+
\Lambda_{\alpha}^{WAW}(q)+\Lambda_{\alpha}^{WZZ}(q)
+\Lambda_{\alpha}^{WAA}(q)
 \, .\nonumber
\end{eqnarray}
The processes of different kinds contribute to
$\hat\Gamma_{\alpha}^{pnW}(P_n,P_p,q)$ (\ref{i10}).

All the terms but last in the integrand in (\ref{i10}) prove to
incorporate the propagators of heavy gauge bosons
$D^{W,Z}_{\alpha\beta}$ (\ref{dwz}). So, in the r.h.s. of
(\ref{i10}), $\Lambda^{WWW} \, , \; \, \Lambda^{WWZ} \, , \; \,
\Lambda^{WAW} \, , \; \, \Lambda^{WZZ}$ render the processes where
the quark-quark electroweak interactions are due to the heavy gauge
bosons exchange that corresponds to large momenta transferred,
$q^2{\sim}M_{Z.W}^2{\gg}M_N^2$, and therefore the short-range,
${\sim}1{/}M_{W,Z}$, quark-quark electroweak interactions cause
these processes. By emitting or absorbing a virtual heavy gauge
boson, large momenta $q^2{\sim}M_{W,Z}^2$ is transferred to the
quarks constituting the nucleon. As quark momenta inside the
nucleon are relatively small, $q^2{\lesssim}M_N^2$, quarks possess
 large momenta, $q^2{\sim}M_{Z,W}^2{\gg}M_N^2$, in the intermediate
states between emission and absorption of heavy gauge bosons in the
vertexes ${\cal L}^{Wqq}(x_1) \, , \; \, {\cal L}^{Zqq}(x_2)$ in
(\ref{i10}). What is the underlying inherent principle of the
Standard Model to emphasize at this very stage is that the strong
quark-quark interactions die out when quarks possess the large
momenta $q^2{\gg}M_N^2$. Consequently, given the fact that quarks
 have got such a large momenta, the strong quark-quark interactions
die out, i.e. ${\cal L}_{str}^{qq}$ vanishes, in these intermediate
states, and we deal with free quarks \cite{d,com,ao,h1,h2,b}. In
this respect, on rewriting (with allowance for Eqs.
(\ref{l3})-(\ref{l8}), (\ref{i3})-(\ref{i4}), (\ref{j3})) the
quantities $\Lambda^{WWZ}
\, , \; \,
\Lambda^{WAW}$ in the form
\begin{eqnarray}
\Lambda_{\alpha}^{WWZ}(q)
=-\int\mbox{d}^4 x_1\int\mbox{d}^4 x_2\int\mbox{d}^4 x_3
\langle\Phi^{q \, +}_{p}(P_p , \sigma_p )|{\cal T}\Bigl\{\bigl(\bar
q(x_1)\Gamma^{Wqq}_{\delta}(x_1) T^{+}_q q(x_1)\bigr){\cdot}{\cal
S}_{str}(x^0_1 , x^0_2){\times}\nonumber\\
\bigl(\bar
q(x_2)\Gamma^{Zqq}_{\beta}(x_2)q(x_2)\bigr)
\Gamma^{WWZ}_{\chi\nu\lambda}(x_3)W^{+ \, \chi}(x_3)\Bigr\}
|\Phi^q_{n} (P_n,\sigma_n),
W^{+}_{\alpha}(q)\rangle{\cdot}D^Z_{\beta\lambda}(x_1-x_3){\cdot}D^W_{\nu\delta}
(x_3-x_2) \, , \label{j12}\\
\Lambda_{\alpha}^{WAW}(q)
=-\int\mbox{d}^4 x_1\int\mbox{d}^4 x_2\int\mbox{d}^4 x_3
\langle\Phi^{q \, +}_{p}(P_p , \sigma_p )|{\cal T}\Bigl\{\bigl(\bar
q(x_1)\Gamma^{Wqq}_{\delta}(x_1) T^{+}_q q(x_1)\bigr){\cdot}{\cal
S}_{str}(x^0_1 , x^0_2){\times}\nonumber\\
\bigl(\bar q(x_2) e_q
\gamma^{\beta}q(x_2)\bigr)
\Gamma^{WAW}_{\chi\nu\lambda}(x_3)W^{+ \, \chi}(x_3)\Bigr\}
|\Phi^q_{n} (P_n,\sigma_n),
W^{+}_{\alpha}(q)\rangle{\cdot}D^A_{\beta\lambda}(x_1-x_3){\cdot}D^W_{\nu\delta}
(x_3-x_2) \, ,\label{j13}
\end{eqnarray}
we presume $ \; {\cal S}_{str}(x^0_1 , x^0_2){=}1 \, $ herein, so
far as ${\cal L}^{qq}_{str}(x){=}0$ at $x_1^0{\leq}x^0{\leq}x_2^0$
in (\ref{xx}). Then, without involving the strong quark-quark
interactions, the sum $\Lambda^{WWZ}+\Lambda^{WAW}$ transforms to
the matrix element of the ${\cal T}-$product of quark field
operators presented by the diagrams ${\large\bf 2}$ and ${\large\bf
3}$ in (\ref{r9}) between the neutron and proton wave functions
(\ref{i3}),
\begin{eqnarray}
\Lambda_{\alpha}^{WWZ}(q)+\Lambda_{\alpha}^{WAW}(q)=
 \qquad \qquad \qquad \qquad \nonumber \\
=i\int \mbox{d}^4 y e^{iqy}
\langle \Phi^{q \, +}_{p}(P_p,\sigma_p ) |\bar\psi_u(y)
\, \Gamma_{\chi}^{udW}(y) W^{+ \, \chi}(y)
\psi_d(y)|\Phi^q_{n}(P_n,\sigma_n), W^{+}_{\alpha}(q) \rangle{\cdot}
\Gamma(WZA) \, . \label{vza}
\end{eqnarray}
Here, the bare vertex $\Gamma_{\alpha}^{udW}$ is given by
(\ref{r11}) and
\begin{eqnarray}
\Gamma(WZA){=}\frac{3\alpha}{4\pi}\Bigl\{\frac{1}{2s^2_W}
[(1-2e_us_W^2)+(1+2e_ds^2_W)]\bigl(\Delta(M_Z)-\frac{1}{2}\bigr)
{+}(e_u-e_d)\bigl(\Delta(M_W)-\frac{1}{2}\bigr)\Bigr\}{+}\nonumber\\
{+}\frac{\alpha}{4\pi}\Bigl(4(e_u-e_d)+\Bigl(
4+6\frac{c_W^2}{s_W^2}\ln c_W
\Bigr) [\frac{1}{s^2_W}+(e_d-e_u) s_W^2]\Bigr) \, , \label{wza}
\end{eqnarray}
accordingly a direct evaluation of the contribution from the
diagrams ${\large\bf 2}$ and ${\large\bf 3}$ in (\ref{r9}). With
making use of Eqs. (\ref{i8})-(\ref{j1b}), (\ref{j0}), (\ref{g0}),
the expression (\ref{vza}) results as
\begin{eqnarray}
\Lambda_{\alpha}^{WWZ}(q)+\Lambda_{\alpha}^{WAW}(q)=\nonumber\\
{=}i(2\pi)^4 \delta(P_n-P_p+q)\Bigl(\bar
U_p(P_p,\sigma_n)\Gamma^{npW}_{\alpha}(q) w^{+ \,
\alpha}(q) U_n(P_n,\sigma_n)\Bigr)\Gamma(WZA) \, .
\label{vza1}
\end{eqnarray}
Certainly, $\Gamma^{npW}_{\beta}(0)$ (\ref{g0}) resides herein at
$q^2{\ll}M_N^2$.

Recalling Eqs. (\ref{ct})-(\ref{fs}), we acquire in much the same
way
\begin{eqnarray}
\Lambda_{\alpha}^{WWW}(q)=
i\int \mbox{d}^4 y e^{iqy}
\langle \Phi^{q \, +}_{p}(P_p,\sigma_p ) |\bar\psi_u(y)
\, \Gamma_{\chi}^{udW} W^{+ \, \chi}(y)  \,
\psi_d(y)|\Phi^q_{n}(P_n,\sigma_n), W^{+}_{\alpha}(q)
\rangle{\times}\nonumber\\
\times\Bigl({\delta}z^W_1-{\delta}z^W_2+
\frac{1}{2}{\delta}z^u_L(M_W)+
\frac{1}{2}{\delta}z^d_L(M_W)\Bigr)=\nonumber\\
i(2\pi)^4 \delta(P_n-P_p+q)\Bigl(\bar
U_p(P_p,\sigma_n)\Gamma^{npW}_{\alpha}(q) w^{+}_{\alpha}(q)
U_n(P_n,\sigma_n)\Bigr){\times}\nonumber\\
\times \Bigl({\delta}z^W_1-{\delta}z^W_2+
\frac{1}{2}{\delta}z^u_L(M_W)+
\frac{1}{2}{\delta}z^d_L(M_W)\Bigr) \, , \label{j14}
\end{eqnarray}
where the difference ${\delta}z^W_1-{\delta}z^W_2$ is given by
(\ref{z12}), and the quantities
\begin{equation}
\frac{1}{2}{\delta}z^u_L(M_W)=\frac{1}{2}{\delta}z^d_L(M_W)=
\frac{-1}{4s^2_W}\Bigl(\Delta(M_W)-\frac{1}{2}\Bigr)\frac{\alpha}{4\pi}
\label{wz}
\end{equation}
specify renormalization of the $u-, d-$quark wave functions caused
by the quark self-energies (\ref{fs}) with a virtual $W-$boson.

Amenably to Eqs. (\ref{i4}), (\ref{j3}), (\ref{ct})-(\ref{fs}), the
quantity $\Lambda^{WZZ}$ is presented likewise $\Lambda^{WWZ}$,
 $\Lambda^{WAW}$, $\Lambda^{WWW}$ (\ref{j12})-(\ref{wz}) in the
 form
\begin{eqnarray}
\Lambda_{\alpha}^{WZZ}(q)=
-\int\mbox{d}^4 x_1\int\mbox{d}^4 x_2\int\mbox{d}^4 x_3
\langle\Phi^{q \, +}_{p}(P_p , \sigma_p )|
{\cal T}\Bigl\{\Bigl(\bar
q(x_2)\Gamma_{\chi}^{Zqq}(x_2)q(x_2)\Bigr){\cal
S}_{str}(x^0_2,x^0_1)\nonumber\\
\Bigl(\bar q(x_1)\Gamma^{Wqq}_{\mu}(x_1) W^{+\, \mu}(x_1)
 T^{+}_q(x_1) q(x_1)\Bigr){\cal
 S}_{str}(x^0_1,x^0_3)\times\nonumber\\ \times \bar
 q(x_3)\Gamma_{\beta}^{Zqq}(x_3)q(x_3)
\Bigr\} |\Phi^q_{n} (P_n , \sigma_n), W^{+}_{\alpha}(q)
\rangle D^{Z}_{\chi\beta}(x_2-x_3) \, , \label{j15}
\end{eqnarray}
where we can presume the strong quark-quark interactions die out in
the intermediate states, $${\cal S}_{str}(x^0_1,x^0_3){=}{\cal
S}_{str}(x^0_1,x^0_3){=}1 \; ,
$$ alike in Eqs. (\ref{j12}), (\ref{j13}). Then, in much the same
way as $\Lambda^{WWZ}
\, , \; \, \Lambda^{WAW} \, , \; \, \Lambda^{WWW}$ have transformed
to (\ref{vza}), (\ref{vza1}), (\ref{j14}), the quantity
 $\Lambda^{WZZ}$ (\ref{j15}) transforms as follows
\begin{eqnarray}
\Lambda_{\alpha}^{WZZ}(q)=
i\int \mbox{d}^4 y e^{iqy}
\langle \Phi^{q \, +}_{p}(P_p,\sigma_p ) |\bar\psi_u(y)
\, \Gamma_{\chi}^{udW} W^{+ \, \chi}(y) \,
\psi_d(y)|\Phi^q_{n}(P_n,\sigma_n), W^{+}_{\alpha}(q)
\rangle{\times}\nonumber\\
\times\Bigl\{\Gamma(WZ)+\frac{1}{2}{\delta}z_L^u(M_Z)
+\frac{1}{2}{\delta}z_L^d(M_Z)\Bigr\}=\nonumber\\ i(2\pi)^4
\delta(P_n-P_p+q)\Bigl(\bar U_p(P_p,\sigma_p)\Gamma^{npW}_{\alpha}(q)
w^{+}_{\alpha}(q) U_n(P_n,\sigma_n)\Bigr)\times\nonumber\\
\times\Bigl\{\Gamma(WZ)+\frac{1}{2}{\delta}z_L^u(M_Z)
+\frac{1}{2}{\delta}z_L^d(M_Z)\Bigr\}
 \, , \label{vz}
\end{eqnarray}
where the value of $\Gamma(WZ)$ is presented by the diagram
${\large\bf 4}$ in (\ref{r9}) what counts is
\begin{eqnarray}
\Gamma(WZ)=-\frac{\alpha}{4\pi}\frac{1}{4s^2_W c^2_W}[1+2s^2_W
(e_d-e_u)-4e_de_us^4_W]\Bigl(\Delta(M_Z)-\frac{1}{2}\Bigr)
\, , \label{gwz}
\end{eqnarray}
and, amenably to Eqs. (\ref{ct})-(\ref{5}), the renormalization
constants of the $u- ,d-$quark wave functions
\begin{eqnarray}
\frac{1}{2}{\delta}z^d_L(M_Z)=
-\frac{\alpha}{4\pi}\frac{1}{8c^2_Ws^2_W}(1+2e_ds^2_w)^2
\Bigl(\Delta(M_Z)-\frac{1}{2}\Bigr) \, , \label{zz1}\\
\frac{1}{2}{\delta}z^u_L(M_Z)=
-\frac{\alpha}{4\pi}\frac{1}{8c^2_Ws^2_W}(1-2e_us^2_w)^2
\Bigl(\Delta(M_Z)-\frac{1}{2}\Bigr)  \label{zz2}
\end{eqnarray}
 are caused by the self-energies (\ref{fs}) with a virtual
 $Z-$boson.

For the consistent treatment of the issue of strong interactions,
we rewrite the last term $\Lambda^{WAA}$ in (\ref{i10}) as follows
\begin{eqnarray}
\Lambda_{\alpha}^{WAA} = \Lambda^{WAA}_{s \, \alpha}
+ \Lambda^{WAA}_{l \, \alpha}=\nonumber\\
=-e^2\int\mbox{d}^4x_1\int\mbox{d}^4x_2\int\mbox{d}^4x_3
\langle\Phi^{q \, +}_{0p}(P_p ,\sigma_p )|{\cal T}\Bigl\{\Bigl({\cal
L}^{Wqq}(x_1)\Bigl(\bar q(x_2)e_q\gamma^{\nu}q(x_2)\times\nonumber\\
\biggl(D^{As}_{\mu\nu}(x_2-x_3) +
D^{Al}_{\mu\nu}(x_2-x_3)\biggr)\times\label{i11}\\
\bar q(x_3)e_q\gamma^{\mu}q(x_3)\Bigr)\cdot {\cal
S}_{str}\Bigr\}|\Phi^q_{0n}(P_n ,\sigma_n ) ,
W^{+}_{\alpha}(q)\rangle\ \, , \nonumber
\end{eqnarray}
 the propagator $D^{A\lambda}(x_2-x_3)$ (\ref{d}) of a virtual
photon is split herein into two parts
\begin{eqnarray}
D^{A\lambda}_{\mu\nu}(x_2-x_3){=}g_{\mu\nu}
\int\frac{\mbox{d}^4k}{(2\pi)^4}
\Bigl(\frac{1}{k^2-M^2_S+i0} +
\frac{-M_S^2}{(k^2-\lambda^2+i0)(k^2-M^2_S+i0)}\Bigr)e^{-ik(x_2-x_3)}
{=}\label{i12} \\
=D^{As}_{\mu\nu}(x_2-x_3) + D^{Al}_{\mu\nu}(x_2-x_3) \, ,
\qquad \qquad \qquad \qquad \qquad \quad \nonumber
\end{eqnarray}
with introducing the subsidiary matching parameter $M_S$, chosen so
that $M_p^2{\ll}M_S^2{\ll}M_W^2$ \cite{sr1,h}. The quantity
$D^{As}(x)$, involving only the integration over large momenta
$k^2{\gtrsim}M_S^2$, is natural to be treated as the propagator of
 a ``massive photon" with the mass $M_S$.

The corrected renormalized vertex $\hat\Gamma^{npW}_{\alpha}$ in Eq.
(\ref{i10}) is written as the sum
\begin{equation}
\hat\Gamma_{\alpha}^{npW} = \hat\Gamma_{s \, \alpha}^{npW} +
 \hat\Gamma_{l \, \alpha}^{npW} \, , \label{i13}
\end{equation}
where the quantities $\hat\Gamma_{s \, \alpha}^{npW}$ and
$\hat\Gamma_{l \, \alpha}^{npW}$ are determined as follows
\begin{eqnarray}
i(2\pi)^4 \delta(P_n-P_p+q)\Bigl(\bar
U_p(P_p,\sigma_p)\hat\Gamma^{npW}_{s \, \beta}(q) w^{+ \,
\beta}(q) U_n(P_n,\sigma_n)\Bigr)=\nonumber\\
\Lambda_{0 \, \beta}^{npW}+\Lambda_{\beta}^{WWZ}+\Lambda_{\beta}^{WAW}+
\Lambda_{\beta}^{WWW}+\Lambda_{\beta}^{WZZ}
+\Lambda_{s \, \beta}^{WAA} \, , \label{i14}\\ i(2\pi)^4
\delta(P_n-P_p+q)\Bigl(\bar U_p(P_p,\sigma_p)\hat\Gamma^{npW}_{l \,
\beta}(q) w^{+ \,
\beta}(q) U_n(P_n,\sigma_n)\Bigr)=
\Lambda_{l \, \beta}^{WAA} \, .\label{i14a}
\end{eqnarray}
So, $\hat\Gamma_s^{np W}$ is due to the electroweak quark-quark
interactions mediated by $W-,Z-$bosons and ``massive photons",
whereas $\hat\Gamma_l^{np W}$ is due to ``soft photons". The
quantity $\Lambda_s^{WAA}$ in (\ref{i11}), which involves the
propagator $D^{As}$ of a ``massive photon" (\ref{fss}),
 (\ref{i12}), describes the processes where quarks interact
exchanging virtual ``massive photons". Consequently, the large
momenta, $k^2{\sim}M_S^2{\gg}M_p^2$, are transferred to the quark
system by the electromagnetic interactions thereby. Therefore
quarks possess the large momenta in the intermediate states between
emission and absorption of a ``massive photon", alike in the
processes described by $\Lambda^{WWZ} \, , \;
\, \Lambda^{WAW} \, ,
\; \, \Lambda^{WWW} \, \; \Lambda^{WZZ}$, where the quark-quark
electroweak interactions are mediated by $W- ,Z-$bosons. In this
respect, the strong quark-quark interactions in these intermediate
states can be ignored in treating $\Lambda^{WAA}_s$. Consequently,
 $\Lambda^{WAA}_s$ in (\ref{i11}) can be written in much the same
 way as $\Lambda^{WZZ}$ (\ref{j15}) in the form
\begin{eqnarray}
\Lambda^{WAA}_{s \, \alpha}(q)
=-e^2\int\mbox{d}^4 x_1\int\mbox{d}^4 x_2\int\mbox{d}^4 x_3
\langle\Phi^{q \, +}_{p}(P_p , \sigma_p )|
{\cal T}\Bigl\{\Bigl(\bar q(x_2)e_q\gamma_{\chi}q(x_2)\Bigr){\cal
S}_{str}(x^0_2,x^0_1)\times\nonumber\\
\times\Bigl(\bar q(x_1)\Gamma^{Wqq}_{\mu}(x_1) W^{+\, \mu}(x_1)
 T^{+}_q(x_1) q(x_1)\Bigr){\cal S}_{str}(x^0_1,x^0_3)\times \qquad
 \nonumber\\
\times \Bigl(\bar q(x_3)e_q\gamma_{\beta}q(x_3)\Bigr)
\Bigr\} |\Phi^q_{n} (P_n , \sigma_n), W^{+}_{\alpha}(q)
\rangle\ D^{As}_{\chi\beta}(x_2-x_3) \, , \qquad \label{j16}
\end{eqnarray}
with accepting $ \, {\cal S}(x^0_1,x^0_2){=}{\cal
S}(x^0_1,x^0_3){=}1$ (\ref{xx}) herein. So, we are again to treat,
alike in Eqs. (\ref{vza}), (\ref{vz}), (\ref{j14}), the matrix
element of the ${\cal T}-$product of the pure quark field operators
presented by the graphs ${\large\bf 5}$ in the expression
(\ref{r9}) between the neutron and proton wave functions
(\ref{i3}). In much the same way as in calculating $\Lambda^{WZZ}
\, , \; \, \Lambda^{WWW} \, , \; \, \Lambda^{WWZ} \, , \; \,
\Lambda^{WAW}$ (\ref{vza}), (\ref{j14}), (\ref{vz}), we acquire
\begin{eqnarray}
\Lambda^{WAA}_{s \, \alpha}(q)=
i\int \mbox{d}^4 y e^{iqy}
\langle \Phi^{q \, +}_{p}(P_p,\sigma_p ) |\bar\psi_u(y)
\, \Gamma_{\chi}^{udW} W^{+ \, \chi} \,
\psi_d(y)|\Phi^q_{n}(P_n,\sigma_n), W^{+}_{\alpha}(q)
\rangle{\times}\nonumber\\
\times\Bigl\{\Gamma(WAS)+\frac{1}{2}{\delta}z_L^u(M_S)
+\frac{1}{2}{\delta}z_L^d(M_S)\Bigr\}=\nonumber\\ i(2\pi)^4
\delta(P_n-P_p-q)\Bigl(\bar U_p(P_p,\sigma_p)\Gamma^{npW}_{\alpha}(q)
w^{+}_{\alpha}(q) U_n(P_n,\sigma_n)\Bigr)\times\nonumber\\
\times\Bigl\{\Gamma(WAS)+\frac{1}{2}{\delta}z_L^u(M_S)
+\frac{1}{2}{\delta}z_L^d(M_S)\Bigr\}
 \, . \label{va}
\end{eqnarray}
Here
\begin{equation}
\Gamma(WAS)=\frac{\alpha}{4\pi}e_u e_d
\Bigl(\Delta(M_S)-\frac{1}{2}\Bigr) \, , \label{wa}
\end{equation}
and
\begin{equation}
\frac{1}{2}{\delta}z^u_L(M_S)+\frac{1}{2}{\delta}z^d_L(M_S)=
-\frac{\alpha}{4\pi}\frac{e_u^2+e_d^2}{2}
\Bigl(\Delta(M_S)-\frac{1}{2}\Bigr) \, \label{zas}
\end{equation}
provides the renormalization (\ref{ct}) of the $u- ,d-$quark wave
functions caused by the $u- ,d-$quark self-energies (\ref{fs})
where the wavy line stands for the ``massive photon" propagator
$D^{As}$ (\ref{fss}), (\ref{i12}).

Summarizing the results (\ref{vza1}), (\ref{j14}), (\ref{vz}),
(\ref{va}), the quantity $\hat\Gamma^{npW}_{s \, \alpha}$ in Eqs.
(\ref{i13}), (\ref{i14}) proves to be
\begin{equation}
\hat\Gamma^{npW}_{s \, \alpha}= \Gamma^{npW}_{\alpha}
\cdot \Gamma^{W} \, , \label{i16w}
\end{equation}
where $\Gamma^{W}$ and $\Gamma^{npW}_{\alpha}$ are given by (\ref{i160})
and (\ref{j1b}), (\ref{g0}), (\ref{j0}), (\ref{i7}).
\section{Effect of strong interactions on the
{\lowercase{np}}{W}-vertex.}
\label{sec:level6}
As only the momenta $k^2{\lesssim}M_s^2$ contribute into $D^{Al}$
(\ref{i12}), only these comparatively small momenta are transformed
to quarks by emitting or absorbing virtual photons in the processes
described by the quantity $\Lambda^{WAA}_l$ in (\ref{i10}),
(\ref{i11}), (\ref{i14a}). Possessing the comparatively small
momenta, $k^2{\lesssim}M_s^2$, quarks can be considered to
constitute the baryon in the intermediate state between emitting
and absorbing a virtual ``soft photon". Then, with allowance for
Eqs. (\ref{i4}), (\ref{j3}), $\Lambda^{WAA}_l$ can be transformed
as follows
\begin{eqnarray}
\Lambda_{l \, \alpha}^{WAA}(q)=
-e^2\int\mbox{d}^4 x_1\int\mbox{d}^4 x_2\int\mbox{d}^4 x_3
\langle\Phi^{q \, +}_{p}(P_p , \sigma_p )|
{\cal T}\Bigl\{\Bigl(\bar q(x_2)e_q\gamma_{\chi}q(x_2)\Bigr){\cal
S}_{str}(x^0_2,x^0_1)\times\nonumber\\
\times\Bigl(\bar q(x_1)\Gamma^{Wqq}_{\mu}(x_1) W^{+ \, \mu}(x_1)
 T^{+}_q(x_1) q(x_1)\Bigr){\cal
 S}_{str}(x^0_1,x^0_3)\times\nonumber\\
\times \Bigl(\bar q(x_3)e_q\gamma_{\beta}q(x_3)\Bigr)
\Bigr\} |\Phi^q_{n} (P_n , \sigma_n), W^{+}_{\alpha}(q)
\rangle\ D_{\chi\beta}^{Al}(x_2-x_3)=\label{j17}\\
=-e^2\int\mbox{d}^4 x_1\int\mbox{d}^4 x_2\int\mbox{d}^4 x_3
\sum_{r,s}{\cal T}\Bigl\{
\langle\Phi^{q \, +}_{p}(P_p , \sigma_p )
|\Bigl(\bar
q(x_2)e_q\gamma_{\chi}q(x_2)\Bigr)|\Phi_r^q(P_r,\sigma_r)\rangle
\times \nonumber\\
\times\langle\Phi^{q \, +}_r(P_r,\sigma_r)
|\Bigl(\bar q(x_1)\Gamma^{Wqq}_{\mu}(x_1) W^{+\, \mu}(x_1)
 T^{+}_q(x_1) q(x_1)\Bigr)|\Phi^q_s(P_s,\sigma_s),
 W^{+}_{\alpha}(q)\rangle\times\nonumber\\ \times\langle\Phi^{q \,
 +}_s(P_s,\sigma_s)|
\Bigl(\bar q(x_3)e_q\gamma_{\beta}q(x_3)\Bigr)
|\Phi^q_{n} (P_n , \sigma_n)
\rangle \Bigr\} D_{\chi\beta}^{Al}(x_2-x_3)
 \, . \nonumber
\end{eqnarray}
Here the sum runs over the intermediate quark states with
relatively small momenta $P_{r,s}^2{\lesssim}M_S^2$ described by
the baryonic wave functions $\Phi^q_{r,s}(P_{r,s},\sigma_{r,s})$
(\ref{i3}). Of course, the proton and neutron intermediate states
are included therein too. The matrix elements of the ${\cal
T}-$products of quark operators between
$\Phi^q_{r,s}(P_{r,s},\sigma_{r,s})$ (\ref{i3}) are defined by Eqs.
(\ref{j})-(\ref{a1}) in terms of the matrix elements of the ${\cal
T}-$products of the baryon field operators $\Psi_r$ between the
baryon wave functions $\Phi^B_{r,s}$, with the baryon form factors
$\Gamma^{rsW}_{\beta} \, , \; \, f^{rs}_{\alpha}$ (\ref{j1b}),
(\ref{j2b}), (\ref{a}) introduced thereby. Defined ordinarily the
baryon field propagator
\begin{eqnarray}
{\cal G}_{rs}(x-y)=-i\langle 0|\Psi_r(x)\bar\Psi_s(y)|0\rangle =
\delta_{rs}\frac{1}{(2\pi)^4}\int\mbox{d}^4p \, {\cal
G}_r(p) \, \mbox{e}^{-p(x-y)} \, , \label{gb}
\end{eqnarray}
and the baryon self-energy (\ref{fs}) with the virtual ``soft
photon" (\ref{i12})
\begin{eqnarray}
\Sigma_{N \, l}(P_N) = -e^2
\sum_{r}\int\frac{\mbox{d}^4k}{i(2\pi)^4} \, f_{\alpha}^{Nr}(k)
D_{\alpha\beta}^{Al}(k) \, f_{\beta}^{rN}(k) \, {\cal G}_r(P_N-k)
=\nonumber\\
=-e^2 \sum_{r}\int\frac{\mbox{d}^4k}{i(2\pi)^4} \,
f^{Nr}_{\alpha}(k) \, {\cal
G}_r(P_N-k)f^{rN}_{\beta}(k)\frac{-M_S^2
g^{\alpha\beta}}{(k^2-\lambda^2+i0)(k^2-M^2_S+i0)}= \label{j18}\\
= \begin{picture}(300,70)(0,0)
\SetColor{Blue}
\SetWidth{1}
\Vertex(100,15){5}
\Vertex(200,15){5}
\PhotonArc(150,15)(50,0,180){3}{19}
\Line(100,16.5)(200,16.5)
\Line(60,15)(240,15)
\Line(100,13.5)(200,13.5)
\Text(58,15)[r]{$N$}
\Text(242,15)[l]{$N$}
\Text(100,7.5)[t]{$f^{Nr}$}
\Text(200,7.5)[t]{$f^{ \, rN}$}
\Text(150,7.5)[t]{${\cal G}_r$}
\Text(195,45)[bl]{$Al$}
\Text(150,19)[b]{$r$}
\end{picture}{\large\bfgr ,} \nonumber
\end{eqnarray}
the corrected renormalized vertex $\hat\Gamma^{npW}_{l \, \alpha}$
in Eq. (\ref{i14a}) proves to be
\begin{eqnarray}
\hat\Gamma^{npW}_{l \, \alpha}=\Gamma_{\alpha}(WAl)+
\Gamma^{npW}_{\alpha}\cdot[\frac{1}{2}{\delta}z^p +
\frac{1}{2}{\delta}z^n]=\nonumber\\
\begin{picture}(140,70)(100,0)
\SetColor{Blue}
\SetWidth{1}
\Vertex(100,15){5}
\Vertex(200,15){5}
\PhotonArc(150,15)(50,0,180){3}{19}
\Line(100,16.5)(200,16.5)
\Line(85,15)(215,15)
\Line(100,13.5)(200,13.5)
\Text(84,15)[r]{$p$}
\Text(216,15)[l]{$n$}
\Text(100,7.5)[t]{$f^{ps}$}
\Text(200,7.5)[t]{$f^{ \, rn}$}
\Text(195,45)[bl]{$Al$}
\Text(175,19)[b]{$r$}
\Text(125,19)[b]{$s$}
\Vertex(150,15){5}
\Photon(150,15)(150,-15){6}{4}
\Text(157,-7)[l]{$W^+$}
\Text(150,24)[b]{${\cal J}_{sr}$}
\end{picture}
\begin{picture}(140,30)(-40,-8)
\SetColor{Blue}
\SetWidth{1}
\Vertex(50,7){6}
\Photon(50,7)(50,40){6}{4}
\Line(50,7)(-15,7)
\Line(80,7)(50,7)
\Text(50,-2)[t]{${\cal J}_{np}$}
\Text(-16,7)[r]{$p$}
\Text(-44,7)[l]{$+$}
\Text(83,7)[l]{$n$}
\Text(60,30)[l]{$W^{+}$}
\Text(22,7)[r]{${\bf\bigotimes}$}
\Text(19,16)[b]{${\bf\frac{1}{2}{\bbox{\delta}}{z^p}}$}
\end{picture}
\begin{picture}(140,30)(-10,-8)
\SetColor{Blue}
\SetWidth{1}
\Vertex(50,7){6}
\Photon(50,7)(50,40){6}{4}
\Line(50,7)(20,7)
\Line(110,7)(50,7)
\Text(50,-2)[t]{${\cal J}_{np}$}
\Text(17,7)[r]{$p$}
\Text(-8,7)[l]{$+$}
\Text(111,7)[l]{$n$}
\Text(60,30)[l]{$W^{+}$}
\Text(82,7)[l]{${\bf\bigotimes}$}
\Text(85,0)[t]{${\bf\bbox{\frac{\bf 1}{\bf 2}\delta{\bf z^n}}}$}
\end{picture}
 \, , \label{j19}
\end{eqnarray}
where the first graph represents the quantity
\begin{eqnarray} \bar
U_p(P_p,\sigma_p)\Gamma_{\alpha}(WAl)U_n(P_n,\sigma_n) w^{+
\; \alpha}(q)=\nonumber\\
\frac{e^3|V_{ud}|}{2\sqrt{2} s_W}\bar U_p(P_p,\sigma_p)\sum_{r,s}
\int\frac{\mbox{d}^4k}{i(2\pi)^4}f^{pr}_{\mu}(k){\cal
G}_r(P_p-k)\times\nonumber\\
\times {\cal J}^{\alpha}_{rs}(q)\bigl(T^{+}_{B}\bigr)_{rs}{\cal
G}_s(P_n-k)f^{sn}_{\nu}(k)D^{Al}_{\mu\nu}(k)
U_n(P_n,\sigma_n)w_{\alpha}^+(q) \, , \label{wal}
\end{eqnarray}
and the finite renormalization constants
$\frac{1}{2}{\delta}z^{n,p}$ of the neutron and proton wave
functions come from
\begin{eqnarray}
{\delta}{z^N} = -\frac{\partial \Sigma_{Nl}(P)}{{\partial}{\not
P}}\mid
_{\not P =M_N} \, . \label{j20}
\end{eqnarray}
In (\ref{j18})-(\ref{wal}), the wavy lines tagged by $Al$
 represent the ``soft photon" propagator $D^{Al}$ (\ref{i12}), the
triplex lines generically render various baryonic states (
including the nucleon), and the blobs stand for the
$NB\gamma-,BN\gamma-,BB'\gamma-,BB'W-$vertices with the appropriate
form factors (\ref{j1b})-(\ref{ff}). Apparently, as only the
integration over the momenta $k^2{\lesssim}M_S^2$ contributes to
(\ref{j18})-(\ref{j20}), no UV divergence emerges therein.

The prevailing part of (\ref{j18})-(\ref{j20}) is obtained by
retaining in the sum over $r,s$ only the single nucleon
intermediate states $r,s=N$ with the propagator
\begin{equation}
G_N(P_N)=\frac{{\not P}_N +M_N}{P_N^2-M_N^2+i0}
\, , \label{gn}
\end{equation}
and also presuming (\ref{j0})-(\ref{f0}). Then, the quantity
$\Gamma_{\alpha}(WAl)$ (\ref{wal}) evidently vanishes, as
$f^{nn}{=}0$ is utilized, and we arrive at
\begin{eqnarray}
\hat\Gamma^{pnW}_{l \, \alpha} =
\Biggl(\frac{1}{2}{\delta}{z_0^p}+\frac{1}{2}{\delta}{z_0^n}\Biggr)
 {\Gamma}_{\alpha}^{npW} \, , \label{i17}
\end{eqnarray}
with the finite renormalization constants (\ref{j20}) of the neutron
and proton wave functions
\begin{equation}
{\delta}{z^{p}_0}=-\frac{\alpha}{4\pi}\Bigl(2\ln\frac{M_S}{M_p} +
\frac{9}{2}-4\ln\frac{M_p}{\lambda}\Bigr) \, , \; \; \; \;
{\delta}{z^n_0}=0 \, . \label{i18}
\end{equation}

To estimate the effect of nucleon structure on ${\delta}z^N$, we
first retain only the single nucleon intermediate state with
 $G_N$ (\ref{gn}) in (\ref{j18})-(\ref{j20}), yet specify the
nucleon form factors into (\ref{j18})-(\ref{j20}) by Eqs.
(\ref{a1}), (\ref{ff}) which are plausible at the momenta $k^2$
transferred by a virtual ``soft photon". Then, after a due
calculation, laborious but rather plain, we arrive at the
estimation
\begin{eqnarray}
{\delta}{\tilde z}^p =
-\frac{\alpha}{2\pi}\Bigl\{-2\ln\frac{M_p}{\lambda}+\frac{9}{4}-J(r)+
\frac{r}{2}\frac{\partial}{\partial{r}}J(r)\Bigl\}
+\frac{\alpha}{2\pi}\frac{1.79^2}{2}\Bigl\{I(0)-I(r)-
\frac{r}{2}\frac{\partial}{\partial{r}}I(r)\Bigl\} \, , \label{s14}\\
{\delta}{\tilde z}^n =
\frac{\alpha}{2\pi}\frac{1.93^2}{2}\Bigl\{I(0)-I(r)-
\frac{r}{2}\frac{\partial}{\partial{r}}I(r)\Bigl\} \, , \label{s15}
\end{eqnarray}
where $r{=}m_{\rho}{/}{M_p}$ is to set, and
\begin{eqnarray}
J(r)=\int_{0}^{1}\frac{d x}{x^2 +r^2
(1-x)}[r^2(\frac{x^2}{2}-x)+x(2x-2+x^2 )] \, , \nonumber\\
I(r)=\int_{0}^{1}\frac{x d x}{8(x^2+r^2 (1-x))}[r^4 (x+6)+2r^2 (3x^2
-6x- 8)-8x^2 (x-3)] .\nonumber
\end{eqnarray}

For the intermediate states in (\ref{j18}) with $r{\neq}N$, the
quantities $f^{pr}_{\alpha} \, , \; \, f^{rn}_{\alpha}$ describe
the transitions between these nucleon excited states $r$ and the
proton and neutron states $p, n$ , respectively. These intermediate
states are naturally to be treated as the well-known excited states
of the proton, such as the $\Delta_{33}-$isobar, Roper-resonance,
and so on. To realize the effect of the exited states on
${\delta}z^N$ (\ref{j20}), (\ref{j18}), we consider the
contribution into (\ref{j20}) due to an intermediate
$\Delta_{33}-$isobar, the simplest proton excited state, the
internal structure of which is much the same as the structure of
the nucleon ground state. In the nucleon as well as in the
$\Delta_{33}-$resonance, all three quarks occupy the state
$1S_{1/2}$. Therefore, the amplitude $f_{\alpha}^{p\Delta_{33}}$ in
(\ref{j18}), (\ref{j20}) does not differ substantially from
$f^{pp}_{\alpha}$. Also along these lines, the very distinction of
${\cal G}_{\Delta_{33}}$ (\ref{gb}) from $G_p$ (\ref{gn}), which is
of vital importance for the current estimation, actually results in
replacing $M_p{\longrightarrow}M_{\Delta}$ (
 see, for instance, Refs. \cite{23}). What is of crucial value in
 evaluating (\ref{j18}), (\ref{j20}) with $r{\neq}N$ is that
\begin{eqnarray}
M_r^2 - M^2_p \sim M_p^2 \; , \; \; \; d=\frac{M_{\Delta_{33}}^2 -
M^2_p}{M_{\Delta_{33}}}{\approx}\frac{1}{2} \, . \label{s16}
\end{eqnarray}
Then, by assuming the form factors (\ref{j0})-(\ref{f0}), the
direct estimation of the contribution to (\ref{j20}) from the term
with the $\Delta_{33}$ intermediate state gives
\begin{eqnarray}
{\delta}z^p_{\Delta} =
-\frac{\alpha}{2\pi}\bigl\{J_{\Delta}(M_S{/}M_{\Delta}) -
J_{\Delta}(0)\bigr\} \, , \label{s17}\\ J_{\Delta}(r)=\int_{0}^{1}
\frac{\mbox{d}x}{x^2+d x(1-x)+r^2(1-x)}\bigl\{\bigl(x-\frac{x^2}{2}
\bigr)[2x+d (1-2x)-r^2]-2x(1-x^2)\bigr\} \, .\nonumber
\end{eqnarray}
The relations $M_S^2{\gg}M_N^2 , m_{\rho}^2 , M^2_{\Delta}-M_N^2$
were utilized in obtaining (\ref{i18})-(\ref{s15}), (\ref{s17}).
Let us behold that ${\delta}z^p_{\Delta} \, ,\; \,
{\delta}{\tilde{z}}^n $ are free of the infrared divergencies,
unlike ${\delta}z^p \, , \; \, {\delta}{\tilde{z}}^p$.

Now it is only a matter of straightforward numerical evaluation to
become convinced that the difference
\begin{eqnarray}
[({\delta}z^p_{\Delta}+{\delta}{\tilde{z}}^p+{\delta}{\tilde{z}}^n)
-{\delta}z_0^p ]\lesssim 0.1\cdot{\delta}z_0^p  \; \label{s18}
\end{eqnarray}
constitutes less than ${\sim}10\%$ to the main quantity
${\delta}z^p_0 $ (\ref{i18}).

Except for the $\Delta_{33}-$isobar, the structure of the nucleon
excited states and the structure of the ground state of the nucleon
are disparate. Therefore, the values of $f_{\alpha}^{pr}$ with
$r{\neq}p,\Delta_{33}$ are anyway substantially smaller than the
$f_{\alpha}^{pp}$ value. Consequently, the contribution of these
excited states into (\ref{j18}), (\ref{j20}) is still far smaller
than (\ref{s17}).

The quantity (\ref{wal}) is exclusively caused by the small form
factors $f^{nn} \, , \; \, f^{sn}$ (\ref{a}), (\ref{ff}). It
incorporates also two baryonic intermediate states. In this
respect, the contribution of (\ref{wal}) into (\ref{j19}) is
realized to be still far smaller than (\ref{s14}), (\ref{s15}),
(\ref{s17}). All the more so, we may abandon the contribution of
simultaneous allowance for the nucleon form factors and the nucleon
excited states.

Thus, with an accuracy better than ${\sim}10\%$, Eq. (\ref{i17})
holds true, the quantity (\ref{wal}) is negligible, and the
renormalization constants of the neutron and proton wave functions
are given by (\ref{i18}). As the whole radiative corrections
constitute a few per cent to the uncorrected $\beta-$decay
probability, we commit an error ${\lesssim}0.1\%$ but never more,
making use of (\ref{i17}), (\ref{i18}) in the further calculations.

Finally, adding (\ref{i16w}) and (\ref{i17}), the corrected
renormalized $npW-$vertex proves (with the aforesaid accuracy) to
be multiple to the uncorrected vertex (\ref{j1b}):
\begin{eqnarray}
\hat\Gamma^{npW}_{\alpha}(P_n , P_p ,q) =
\hat\Gamma^{npW}_{s \, \alpha}(P_n , P_p ,q) +
\hat\Gamma^{npW}_{l \, \alpha}(P_n , P_p ,q) =\nonumber\\
= \Gamma^{npW}_{\alpha}(q)
\Biggl\{1+\frac{\alpha}{4\pi}\Bigl(\ln\frac{M_p}{M_Z}
- 2\ln\frac{\lambda}{M_p} - \frac{9}{4} + \frac{3}{s^2_W} +
\frac{6c^2_W - s^2_W}{s^4_W}\ln(c_W)\Bigr)\Biggr\} \, . \label{i21}
\end{eqnarray}
This quantity is just what is depicted by the shaded circle with
heavy core in the graph ${\large\bf 3}$ in the amplitude
(\ref{mm}).
\section{The radiative corrections to the $W-$boson propagator.}
\label{sec:level7}
Next, the propagator $D^W(q)$ (\ref{dwz}) of the bare $W-$boson in
(\ref{i7}) gives place to the corrected regularized $W-$boson
propagator $\hat{D}^W(q) \, $ \cite{ao,h1,h2,b,ms} \, ,
\begin{eqnarray}
D^W(q)=\frac{1}{q^2-M^2_W+i0}\Longrightarrow \hat D^W(q)
=\frac{1}{q^2-M_W^2
+\hat\Sigma(q^2)}\approx
\nonumber \\
\approx\Bigl(-\frac{1}{M_W^2}\Bigr)\frac{1} {1-\frac{\hat\Sigma
(0)}{M_W^2}}\, , \; \; \; \text{for} \;  q^2{\ll}M^2_W \,
,\label{i22}
\end{eqnarray}
as represented by the graph ${\large\bf 4}$ in the expression
${\cal M}$ (\ref{mm}) where the heavy wavy line stands for $\hat
D^W$. The renormalized $W-$boson self-energy $\hat\Sigma(0)$ is
rather not amenable to a precise reliable evaluation because it
includes light quarks contributions in the momentum region where
 strong interaction effects cannot be ignored \cite{h2}.
Fortunately, one can acquire from the analysis of the $\mu-$meson
decay \cite{h2,ms} that
\begin{eqnarray}
\frac{G_{\mu}}{\sqrt{2}}=\frac{\alpha\pi(1+{\delta}_v)}
{2M^2_W s^2_W \Bigl(1-\frac{\hat\Sigma^W(0)}{M^2_W}\Bigr)} \; , \;
\; \;  G_{\mu}=1.1663\cdot10^{-5}
\, \mbox{GeV}^{-2} \, , \; \; \; \; \delta_{\it v}\approx
0.006 ,\ . \label{r2}
\end{eqnarray}
 The estimation ${\hat\Sigma^W(0)}{/}{M^2_W}{\approx}0.066$ was
ascertained in Refs. \cite{h1,h2}.

It is expedient to redefine ${\cal M}_0$ as the sum of the
amplitudes ${\large\bf 1}$ and ${\large\bf 4}$ in expression ${\cal
M}$ (\ref{mm}), writing hereupon ${\cal M}_0$ as
\begin{eqnarray}
{\cal M}^0=\Biggl(\frac{e}{2\sqrt{2}s_W}\Biggr)^2
\hat{D}^W(q) |V_{ud}| \bigl(\bar u_e
(p_e)\gamma_{\alpha}(1-\gamma^5)u_{\nu}(-p_{\nu})\bigr)\cdot\bigl(\bar
U_p(P_p)\gamma^{\alpha}(1-\gamma^5 g_A)U_n(P_n)\bigr) \,
. \label{i23}
\end{eqnarray}
Accordingly (\ref{i22}), (\ref{r2}), the coefficient in (\ref{i23})
reads
\begin{equation}
\Biggl(\frac{e}{2\sqrt{2}s_W}\Biggr)^2
\hat{D}^W(q)=-\frac{G_{\mu}}{\sqrt{2}}(1-{\delta}_v)=-
\frac{G}{\sqrt{2}} \, .
\label{r3}
\end{equation}

The contributions from all the diagrams in (\ref{mm}) but
${\large\bf 4}$ are themselves of the order ${\alpha}{/}{4\pi}$,
even without allowance for replacing $D^W{\rightarrow}{\hat D}^W$.
Therefore, in treating the $\alpha$-order radiative corrections
caused by the processes depicted by these graphs, it stands to
reason to set
\begin{equation}
\Biggl(\frac{e}{2\sqrt{2}s_W}\Biggr)^2\frac{1}{M^2_W}=\frac{G}{\sqrt{2}}
  \, , \label{r55}
\end{equation}
which is put to use henceforward.
\section{The radiative corrections due to the irreducible
{\lowercase{npe}}{$\bbox{\nu}-$}vertex (the ``box diagrams").}
\label{sec:level8}
By now, we have considered the terms in ${\cal M}$ (\ref{mm}) which
stem from the Born amplitude ${\cal M}_0$ (\ref{i7}) by replacing
the vertices $\Gamma_{\alpha}^{e\nu W} \, , \; \,
\Gamma_{\alpha}^{npW}$ and the $W-$boson propagator $D^W$ with the
corrected renormalized quantities $\hat\Gamma_{\alpha}^{e\nu W} \,
, \; \, \hat\Gamma_{\alpha}^{npW} \, , \; \, \hat{D}^W$. Besides
these terms, which are due to the aforesaid modification of the
separate blocks in the graph ${\large\bf 1}$ (\ref{mm}), the total
amplitude ${\cal M}$ (\ref{im}), (\ref{i5}) incorporates also the
part represented by the graphs ${\large\bf 9}$ in (\ref{mm}) which
are of the second order both in the lepton and quark electroweak
interactions (\ref{l5})-(\ref{l8}). The matrix element
\begin{eqnarray}
i(2\pi)^4\delta (P_n -P_p -p_e -p_{\nu}) {\cal M}_{2\gamma}
=\nonumber\\
\int\mbox{d}^4x_1\int\mbox{d}^4x_2\int\mbox{d}^4x_3\int\mbox{d}^4x_4
\langle\Phi_{0p}^{q \, +}(P_p ,\sigma_p) , \phi^{+}_e(p_e,\sigma_e)|
\nonumber\\
{\cal T}\Biggl\{\Bigl({\cal L}^{Zqq}(x_1){\cal L}^{Wqq}(x_2){\cal
L}^{We\nu}(x_3){\cal L}^{Zee}(x_4)+{\cal L}^{Zqq}(x_1){\cal
L}^{Wqq}(x_2){\cal L}^{We\nu}(x_3){\cal
L}^{Z\nu\nu}(x_4)+\nonumber\\ {\cal L}^{Aqq}(x_1){\cal
L}^{Aee}(x_2){\cal L}^{Wqq}(x_3){\cal
L}^{We\nu}(x_4)\Bigr)\cdot{\cal S}_{str}\Biggr\}
|\Phi_{0n}^q(P_n,\sigma_n) ,
\phi_{\nu}(-p_{\nu},-\sigma_{\nu})\rangle=
\nonumber\\
=\Lambda^{ZW} + \Lambda^{AW} \, , \label{i24}
\end{eqnarray}
defines this part of the amplitude ${\cal M}_{2\gamma}$, usually
referred to as the contribution from the ``box-type" diagrams. It
comprises the terms of different nature, the strong quark-quark
interactions ${\cal L}_{str}^{qq}$ entangled herein through ${\cal
S}_{str}{\equiv}{\cal S}_{str}(\infty , -\infty)$. The second term,
$\Lambda^{AW}$, in r.h.s. of (\ref{i24}) involves the interactions
of quarks ${\cal L}^{Aqq}$ (\ref{l7}) and electrons ${\cal
L}^{Aee}$ (\ref{l8}) with electromagnetic field. Inasmuch as
$\Lambda^{AW}$ describes the processes in which a photon is
exchanged between an electron and a quark, the expression of
$\Lambda^{AW}$ includes the virtual photon propagator
$D^{A\lambda}$ (\ref{d}). Then, by disparting $D^{A\lambda}$ into
the ``massive" $D^{As}$ and ``soft" $D^{Al}$ photon propagators,
pursuant to Eq. (\ref{i12}), $\Lambda^{AW}$ is split into two parts
corresponding to large, $k^2{\gtrsim}M_S^2$, and comparatively
small, $k^2{\lesssim}M_S^2$, momenta transferred from leptons to
quarks by
 a virtual photon, much in the same way as in the case of Eq.
 (\ref{i11}). So, with allowance for Eqs. (\ref{i3}), (\ref{i4}),
 (\ref{j3}), the quantity $\Lambda^{AW}$ in (\ref{i24}) is written
 in the form
\begin{eqnarray}
\Lambda^{AW}
=\int\mbox{d}^4x_1\int\mbox{d}^4x_2\int\mbox{d}^4x_3\int\mbox{d}^4x_4
\langle\Phi_{p}^{q+}(P_p ,\sigma_p) , \phi^{+}_e(p_e,\sigma_e)|
\nonumber\\
{\cal T}\Biggl\{ e \bigl(\bar\psi_q(x_1)
\gamma^{\alpha}e_q\psi_q(x_1)\bigr) {\cal S}_{str}(x^0_1 , x^0_2)
\bigl(\bar\psi_u(x_2)
\Gamma_{\mu}^{udW}(x_2) T^{+}_q \psi_d(x_2)\bigr)\times\label{i25}\\
\times \bigl(\bar\psi_e(x_3)\Gamma_{\rho}^{e\nu W}(x_3)
\psi_{\nu}(x_3)\bigr)
(-e)\bigl(\bar\psi_e(x_4)
\gamma^{\beta}\psi_e(x_4)\bigr)
\Biggr\} \nonumber\\
|\Phi^q_{n}(P_n ,\sigma_n)
,\phi_{\nu}(-p_{\nu},-\sigma_{\nu})
\rangle\Biggl(D_{\alpha\beta}^{As}(x_1-x_4)
 + D_{\alpha\beta}^{Al}(x_1-x_4)\Biggr)
 D^W_{\mu\rho}(x_2-x_3)=\Lambda^{AW s} + \Lambda^{AW l} \, .
 \nonumber
\end{eqnarray}

 In (\ref{i24}), the term $\Lambda^{ZW}$ including the electroweak
interactions of heavy bosons with quarks and leptons, ${\cal
L}^{Zqq} , {\cal L}^{Wqq} , {\cal L}^{Zee} , {\cal L}^{Z\nu\nu} ,
{\cal L}^{We\nu} $, is due to the $Z-$boson exchange between quarks
and leptons. It contains the propagators $D^{W,Z}$ (\ref{dwz}) of
virtual heavy gauge bosons. This case evidently corresponds to the
large momenta, $ \, q^2{\gtrsim}M^2_{S}\gg M_p^2$ , transferred
from leptons to quarks. Recalling Eqs. (\ref{i3}), (\ref{i4}),
(\ref{j3}), we find out
\begin{eqnarray}
\Lambda^{ZW}
=\int\mbox{d}^4x_1\int\mbox{d}^4x_2\int\mbox{d}^4x_3\int\mbox{d}^4x_4
\langle\Phi_{p}^{q+}(P_p ,\sigma_p) , \phi^{+}_e(p_e,\sigma_{e})|
\nonumber\\
{\cal T}\Biggl\{\bigl(\bar\psi_q(x_1)
\Gamma_{\mu}^{Zqq}(x_1)\psi_q(x_1)\bigr) {\cal S}_{str}(x^0_1 , x^0_2)
\bigl(\bar\psi_u(x_2)
\Gamma_{\lambda}^{udW}(x_2) T^{+}_q \psi_d(x_2)\bigr)\times\label{j23}\\
\times \bigl(\bar\psi_e(x_3)\Gamma_{\alpha}^{e\nu W}(x_3)
\psi_{\nu}(x_3)\bigr)
[\bigl(\bar\psi_e(x_4)
\Gamma_{\beta}^{Zee}(x_4) \psi_e(x_4)\bigr)+\bigl(\bar\psi_{\nu}(x_4)
\Gamma_{\beta}^{Z\nu\nu}(x_4) \psi_{\nu}(x_4)\bigr)]
\Biggr\} \nonumber\\
|\Phi^q_{n}(P_n ,\sigma_n)
,\phi_{\nu}(-p_{\nu},-\sigma_{\nu})\rangle
D_{\mu\beta}^{Z}(x_1-x_4) D^W_{\lambda\alpha}(x_2-x_3) \, .
\nonumber
\end{eqnarray}

The amplitude ${\cal M}_{2\gamma}$ is written as the sum
\begin{equation}
{\cal M}_{2\gamma}={\cal M}_{2\gamma s}+{\cal M}_{2\gamma l} \, ,
\label{sms}
\end{equation}
where the quantities ${\cal M}_{2\gamma s} \, , \; \, {\cal
M}_{2\gamma l}$ are defined as follows
\begin{eqnarray}
i(2\pi)^4\delta (P_n -P_p -p_e
-p_{\nu}) {\cal M}_{2\gamma s}=\Lambda^{ZW} +
\Lambda^{AWs} \, ,\label{sas}\\
i(2\pi)^4\delta (P_n -P_p -p_e
-p_{\nu}){\cal M}_{2\gamma l}=\Lambda^{AWl} \, . \label{sal}
\end{eqnarray}

Quark momenta inside the nucleon are known to be relatively small,
$k^2{\lesssim}M_p^2$. Large momenta, $k^2{\gtrsim}M^2_S{\gg}M_p^2
\, , \; \, k^2{\gtrsim}M^2_{Z,W}{\gg}M_p^2$, are transferred by
virtual gauge bosons and ``massive" photons to the quark system in
the processes described by $\Lambda^{ZW} \, , \; \, \Lambda^{AWs}$
(\ref{sas}). Therefore, quarks have got the large momenta
$k^2{\gg}M_p^2$ in the intermediate states between
 emission and absorption of gauge bosons and ``massive" photons at
 the time-points $x^0_1$ and $x^0_2$ in such processes. At this
 point, we invoke again the Standard Model assumption that the
 strong quark-quark interactions ${\cal L}_{str}^{qq}$ vanish
 provided quarks possess the momenta $k^2{\gg}M_p^2$. Consequently,
 the operator ${\cal S}_{str}(x^0_1 , x^0_2)$ (\ref{xx}) in
 $\Lambda^{ZW} \, , \; \, \Lambda^{AWs}$ turns out to be unit,
 ${\cal S}_{str}(x^0_1 , x^0_2){=}1 $. Then, by straightforward
 calculating $\Lambda^{ZW} \, ,
\; \, \Lambda^{AWs}$ (\ref{i25}), (\ref{j23}), we obtain
(\ref{sas})
\begin{eqnarray}
\Lambda^{ZW}+\Lambda^{AWs}=i(2\pi)^4{\cal M}_{2\gamma s} \; \delta
(P_n-P_p-p_e-p_{\nu})=\nonumber\\
=\int\mbox{d}^4 x\langle\Phi_p^{q
\, +}(P_p ,\sigma_p) , \phi_e(p_e,\sigma_e)
|\Bigl(\bar\psi_e(x)\bar\psi_q(x)
 \; \hat\Gamma^{e\nu ud} \;
 \psi_q(x)\psi_{\nu}(x)\Bigr)|\Phi^q_n(P_n ,\sigma_n)
 ,\phi_{\nu}(-p_{\nu},-\sigma_{\nu})\rangle \, , \label{j25}
\end{eqnarray}
with the operator $\Bigl(\bar\psi_e(x)\bar\psi_q(x)
 \; \hat\Gamma^{e\nu ud} \; \psi_q(x)\psi_{\nu}(x)\Bigr)$ to
describe the pure electroweak transitions of leptons and quarks
presented by the set of diagrams
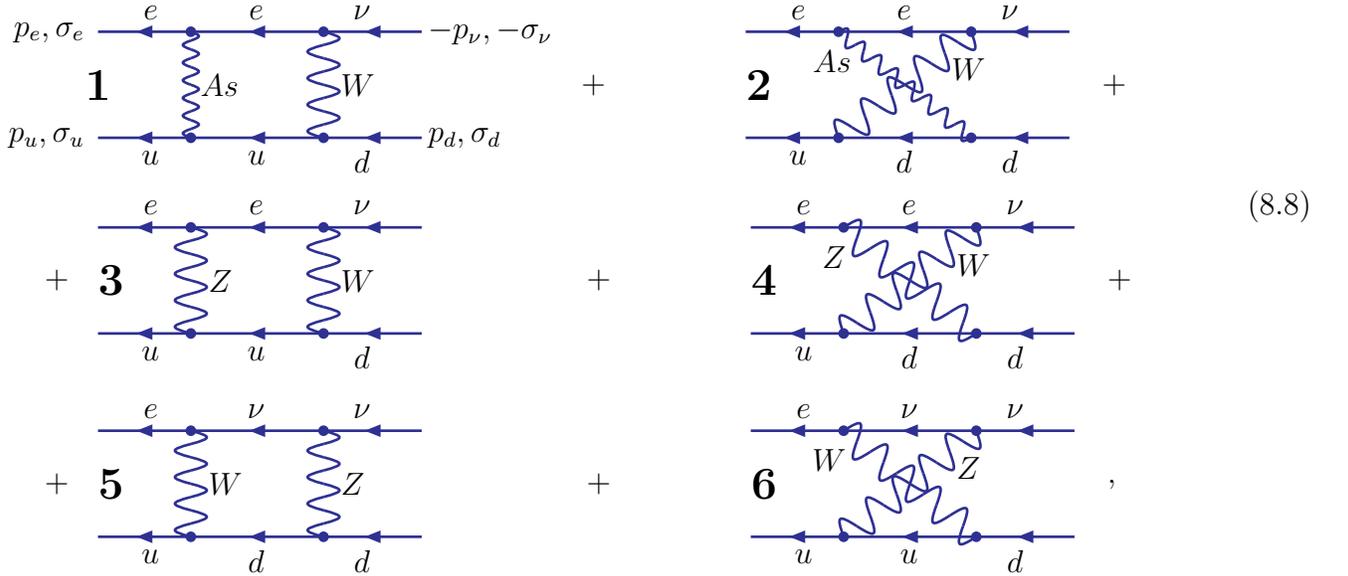
\begin{eqnarray}
\begin{picture}(210,35)(0,0)
\SetWidth{1}
\SetColor{Blue}
\Vertex(50,30){2}
\Vertex(50,-10){2}
\Vertex(100,30){2}
\Vertex(100,-10){2}
\Text(54,10)[l]{$As$}
\Text(107,10)[l]{$W$}
\Photon(100,-10)(100,30){6}{4}
\ArrowLine(50,30)(15,30)
\ArrowLine(100,30)(50,30)
\ArrowLine(50,-10)(15,-10)
\ArrowLine(100,-10)(50,-10)
\ArrowLine(137,30)(100,30)
\ArrowLine(137,-10)(100,-10)
\Photon(50,30)(50,-10){3}{6}
\Text(10,30)[r]{$p_e,\sigma_e$}
\Text(140,30)[l]{$-p_{\nu},-\sigma_{\nu}$}
\Text(10,-10)[r]{$p_u,\sigma_u$}
\Text(140,-10)[l]{$p_d,\sigma_d$}
\Text(35,35)[b]{$e$}
\Text(35,-15)[t]{$u$}
\Text(75,35)[b]{$e$}
\Text(75,-15)[t]{$u$}
\Text(20,10)[r]{\Large\bf 1}
\Text(115,35)[b]{$\nu$}
\Text(115,-15)[t]{$d$}
\end{picture}
\begin{picture}(210,35)(-35,0)
\SetWidth{1}
\SetColor{Blue}
\Vertex(50,30){2}
\Vertex(50,-10){2}
\Vertex(100,30){2}
\Vertex(100,-10){2}
\Text(55,23)[tr]{$As$}
\Text(93,20)[tl]{$W$}
\Photon(50,-10)(100,30){6}{5}
\ArrowLine(50,30)(15,30)
\ArrowLine(100,30)(50,30)
\ArrowLine(50,-10)(15,-10)
\ArrowLine(100,-10)(50,-10)
\ArrowLine(137,30)(100,30)
\ArrowLine(137,-10)(100,-10)
\Photon(50,30)(100,-10){3}{8}
\Text(35,35)[b]{$e$}
\Text(35,-15)[t]{$u$}
\Text(75,35)[b]{$e$}
\Text(75,-15)[t]{$d$}
\Text(25,10)[r]{\Large\bf 2}
\Text(-47,10)[l]{$+$}
\Text(150,10)[l]{$+$}
\Text(115,35)[b]{$\nu$}
\Text(115,-15)[t]{$d$}
\end{picture}\nonumber \\
\begin{picture}(210,35)(0,35)
\SetWidth{1}
\SetColor{Blue}
\Vertex(50,30){2}
\Vertex(50,-10){2}
\Vertex(100,30){2}
\Vertex(100,-10){2}
\Text(57,10)[l]{$Z$}
\Text(107,10)[l]{$W$}
\Photon(100,-10)(100,30){6}{4}
\ArrowLine(50,30)(15,30)
\ArrowLine(100,30)(50,30)
\ArrowLine(50,-10)(15,-10)
\ArrowLine(100,-10)(50,-10)
\ArrowLine(137,30)(100,30)
\ArrowLine(137,-10)(100,-10)
\Photon(50,30)(50,-10){6}{4}
\Text(35,35)[b]{$e$}
\Text(35,-15)[t]{$u$}
\Text(75,35)[b]{$e$}
\Text(75,-15)[t]{$u$}
\Text(25,10)[r]{\Large\bf 3}
\Text(-5,10)[l]{$+$}
\Text(115,35)[b]{$\nu$}
\Text(115,-15)[t]{$d$}
\end{picture}
\begin{picture}(210,35)(-37,35)
\SetWidth{1}
\SetColor{Blue}
\Vertex(50,30){2}
\Vertex(50,-10){2}
\Vertex(100,30){2}
\Vertex(100,-10){2}
\Text(51,23)[tr]{$Z$}
\Text(93,20)[tl]{$W$}
\Photon(50,-10)(100,30){6}{5}
\ArrowLine(50,30)(15,30)
\ArrowLine(100,30)(50,30)
\ArrowLine(50,-10)(15,-10)
\ArrowLine(100,-10)(50,-10)
\ArrowLine(137,30)(100,30)
\ArrowLine(137,-10)(100,-10)
\Photon(50,30)(100,-10){6}{5}
\Text(35,35)[b]{$e$}
\Text(35,-15)[t]{$u$}
\Text(75,35)[b]{$e$}
\Text(75,-15)[t]{$d$}
\Text(25,10)[r]{\Large\bf 4}
\Text(-47,10)[l]{$+$}
\Text(150,10)[l]{$+$}
\Text(115,35)[b]{$\nu$}
\Text(115,-15)[t]{$d$}
\end{picture}
  \label{j26}\\
\begin{picture}(210,35)(0,70)
\SetWidth{1}
\SetColor{Blue}
\Vertex(50,30){2}
\Vertex(50,-10){2}
\Vertex(100,30){2}
\Vertex(100,-10){2}
\Text(57,10)[l]{$W$}
\Text(107,10)[l]{$Z$}
\Photon(100,-10)(100,30){6}{4}
\ArrowLine(50,30)(15,30)
\ArrowLine(100,30)(50,30)
\ArrowLine(50,-10)(15,-10)
\ArrowLine(100,-10)(50,-10)
\ArrowLine(137,30)(100,30)
\ArrowLine(137,-10)(100,-10)
\Photon(50,30)(50,-10){6}{4}
\Text(35,35)[b]{$e$}
\Text(35,-15)[t]{$u$}
\Text(75,35)[b]{$\nu$}
\Text(75,-15)[t]{$d$}
\Text(25,10)[r]{\Large\bf 5}
\Text(-5,10)[l]{$+$}
\Text(115,35)[b]{$\nu$}
\Text(115,-15)[t]{$d$}
\end{picture}
\begin{picture}(210,35)(-37,70)
\SetWidth{1}
\SetColor{Blue}
\Vertex(50,30){2}
\Vertex(50,-10){2}
\Vertex(100,30){2}
\Vertex(100,-10){2}
\Text(51,23)[tr]{$W$}
\Text(93,20)[tl]{$Z$}
\Photon(50,-10)(100,30){6}{5}
\ArrowLine(50,30)(15,30)
\ArrowLine(100,30)(50,30)
\ArrowLine(50,-10)(15,-10)
\ArrowLine(100,-10)(50,-10)
\ArrowLine(137,30)(100,30)
\ArrowLine(137,-10)(100,-10)
\Photon(50,30)(100,-10){6}{5}
\Text(35,35)[b]{$e$}
\Text(35,-15)[t]{$u$}
\Text(75,35)[b]{$\nu$}
\Text(75,-15)[t]{$u$}
\Text(25,10)[r]{\Large\bf 6}
\Text(-47,10)[l]{$+$}
\Text(150,10)[l]{$,$}
\Text(115,35)[b]{$\nu$}
\Text(115,-15)[t]{$d$}
\end{picture}\nonumber
\end{eqnarray}

\vspace{3cm}

where, in particular, the wavy line with the tag $As$ depicts
 the ``massive photon" propagator $D^{As}$ (\ref{fss}),
(\ref{i12}). Recalling Eqs. (\ref{m01}), (\ref{i7}), (\ref{i71}),
(\ref{j1b}), (\ref{g0}), we eventually find the amplitude
\begin{eqnarray}
{\cal M}_{2\gamma s}= - {\cal M}^0 \frac{\alpha}{4\pi}
\Biggl\{\Bigl(1+\frac{5
c^4_W}{s^4_W}\Bigr)\ln{(c_W)}- 6\ln\frac{M_W}{M_S}\Biggr\}
\,  \label{i27}
\end{eqnarray}
 being multiple to the Born amplitude ${\cal M}_0$ (\ref{i7}). It
is to emphasize once again the relations
\begin{eqnarray}
m_f{\ll}M_p \, , \; \; \; M_n{-}M_p{\ll}M_N \, , \; \; \;
|p_f|^2{\ll}M_N^2 \, , \; \, f{\equiv}e,\nu ,u,d
\, , \; \, |{\bf P}_N|^2{\ll}M_N^2 \, , \; \, M^2_p{\ll}M^2_S{\ll}M^2_W
\, , \nonumber\\ \frac{M_n{-}M_p}{M_p}\ln\frac{M_n{-}M_p}{M_p}{\sim}0
 \, , \; \frac{M_S}{M_W}\ln\frac{M_W}{M_S}{\sim}0 \qquad \qquad
 \qquad \qquad \qquad \label{j27}
\end{eqnarray}
were used in obtaining (\ref{j25})-(\ref{i27}), as well as far and
wide over the work.

The second term in (\ref{i27}) is due to the contributions of the
first and second diagrams in (\ref{j26}). In view of the discussion
given in the last section, it is to take cognizance that if we had
a neutral initial particle instead of a $d-$quark and a final
particle with the charge $+1$ instead of an $u-$quark, the
contribution of the second diagram in (\ref{j26}) would apparently
vanish and the coefficient in front of $\ln{M_W}{/}{M_S}$ would be
equal to $8$ instead of $6$.
\section{The irreducible
{\lowercase{npe}}{$\bbox{\nu}-$}vertex with allowance for nucleon
 compositeness.}
\label{sec:level9}
Unlike the case of $\Lambda^{AWs}$, in the processes described by
$\Lambda^{AWl} \, $ (\ref{sal}), (\ref{i25}), quarks and leptons
exchange a virtual $W-$boson and a virtual ``soft photon"
 (\ref{i12}). The amplitude ${\cal M}_{2\gamma l}$ (\ref{sal})
includes the ``soft photon" propagator $D^{Al}$ (\ref{i12}). This
case corresponds to the comparatively small momenta,
$k^2{\lesssim}M_S^2$, transferred from leptons to quarks.
Therefore, the intermediate quark system, between quark
interactions with a $W-$boson and a ``soft photon", possesses
 the relatively small momenta, and we deal with the intermediate
baryonic states $B$, the ground or excited states of the nucleon.
With allowance for (\ref{i3}), (\ref{j3}), $\Lambda^{AWl} \, $
(\ref{sal}) is written as the sum over these baryonic states
\begin{eqnarray}
\Lambda_l^{AW}=i(2\pi)^4 {\cal M}_{2\gamma l}
\delta (P_n-P_p-p_e-p_{\nu})= \nonumber \\
\int\mbox{d}^4x_1\int\mbox{d}^4x_2\int\mbox{d}^4x_3
\int\mbox{d}^4x_4 \sum_{B}
\langle\Phi_{p}^{q \, +}(P_p ,\sigma_p) , \phi^{+}_e(p_e,\sigma_{e})
|\nonumber\\
{\cal T}\Biggl\{\bigl(\bar\psi_e(x_4)
\Gamma_{\rho}^{e\nu W}\psi_{\nu}(x_4)\bigr)(-e)\bigl(\bar\psi_e(x_2)
\gamma^{\beta} \psi_{e}(x_2)\bigr)
\Bigl(\bar\psi_u(x_3)\Gamma_{\mu}^{duW}\psi_d(x_3)\Bigr)|
\Phi_{B}^q(P_B ,\sigma_B)\rangle\nonumber\\
\times\langle\Phi_{B}^{q \, +}(P_B ,\sigma_B)|
\Bigl(\bar\psi_q(x_1)\gamma^{\alpha} e e_q \psi_q(x_1)
\Bigr)\Biggr\}|\nonumber\\
 \Phi_{n}^q(P_n ,\sigma_n) ,
 \psi_{\nu}(-p_{\nu},-\sigma_{\nu})\rangle D^W_{\mu\rho}(x_3-x_4)
 D_{\alpha\beta}^{Al}(x_1-x_2) \, .\label{j28}
\end{eqnarray}
Recalling Eqs. (\ref{j})-(\ref{a1}), (\ref{gb}), the amplitude
${\cal M}_{2\gamma l}$ is presented as the sum of the contributions
of two diagrams
\begin{eqnarray}
\begin{picture}(210,55)(0,-10)
\SetWidth{1}
\SetColor{Blue}
\Vertex(50,30){2}
\Vertex(50,-10){5}
\Vertex(100,30){2}
\Vertex(100,-10){5}
\Text(48,-17)[t]{$f^{pB}$}
\Text(102,-17)[t]{${\cal J}_{Bn}$}
\Text(57,10)[l]{$Al$}
\Text(107,10)[l]{$W$}
\Photon(100,-10)(100,30){6}{4}
\ArrowLine(50,30)(15,30)
\ArrowLine(100,30)(50,30)
\ArrowLine(50,-10)(15,-10)
\ArrowLine(100,-10)(50,-10)
\Line(100,-11.5)(50,-11.5)
\Line(100,-8.5)(50,-8.5)
\ArrowLine(137,30)(100,30)
\ArrowLine(137,-10)(100,-10)
\Photon(50,30)(50,-10){3}{6}
\Text(35,35)[b]{$e$}
\Text(15,-15)[t]{$p$}
\Text(75,35)[b]{$e$}
\Text(75,-15)[t]{${\cal G}_{B}$}
\Text(115,35)[b]{$\nu$}
\Text(135,-15)[t]{$n$}
\end{picture}
\begin{picture}(210,55)(-37,-10)
\SetWidth{1}
\SetColor{Blue}
\Vertex(50,30){2}
\Vertex(50,-10){5}
\Vertex(100,30){2}
\Vertex(100,-10){5}
\Text(48,-17)[t]{${\cal J}_{pB}$}
\Text(102,-17)[t]{$f^{Bn}$}
\Text(51,23)[tr]{$Al$}
\Text(93,20)[tl]{$W$}
\Photon(50,-10)(100,30){6}{5}
\ArrowLine(50,30)(15,30)
\ArrowLine(100,30)(50,30)
\ArrowLine(50,-10)(15,-10)
\ArrowLine(100,-10)(50,-10)
\Line(100,-11.5)(50,-11.5)
\Line(100,-8.5)(50,-8.5)
\ArrowLine(137,30)(100,30)
\ArrowLine(137,-10)(100,-10)
\Photon(50,30)(100,-10){3}{7}
\Text(35,35)[b]{$e$}
\Text(15,-15)[t]{$p$}
\Text(75,35)[b]{$e$}
\Text(75,-15)[t]{${\cal G}_B$}
\Text(-47,10)[l]{$+$}
\Text(160,10)[l]{$=$}
\Text(115,35)[b]{$\nu$}
\Text(135,-15)[t]{$n$}
\end{picture}  \label{j29}\\
.\nonumber\\
=\int\frac{\mbox{d}
k^4}{(2\pi)^4 i}\bigl(\bar u_e(p_e)(-e)\gamma^{\beta}G_e(p_e-k)
\Gamma^{e\nu W \, \alpha} u_{\nu}(-p_{\nu})\bigr)
\frac{-M^2_S}{(k^2-\lambda^2+i0)(k^2-M^2_S+i0)}\times \nonumber \\
\frac{1}{k^2-M_W^2+i0}
\sum_{B}\Biggl\{\bigl(\bar U_p(P_p) e f_{\beta}^{pB}(k){\cal G}_B(P_p+k)
 \Gamma_{\alpha}^{BnW}(k) U_n(P_n)\bigr)+\qquad \qquad
\qquad \qquad \nonumber\\
+\bigl(\bar U_p(P_p) \Gamma_{\alpha}^{pBW}(k){\cal G}_B(P_p-k) e
f_{\beta}^{Bn}(k) U_n(P_n)\bigr)\Biggr\} \; ,
\qquad \qquad \qquad \qquad \nonumber
\end{eqnarray}
where the wavy line tagged by $Al$ stands for the ``soft photon"
propagator $D^{Al}$ (\ref{i12}) and the triplex line represents
generically the propagator of a quark system in the intermediate
states. The forthcoming estimations will
 prove that omitting all the nucleon excited states and describing
the nucleon form factors and nucleon transition current by Eqs.
 (\ref{j0})-(\ref{f0}), we commit no more than a few
per cent error in evaluating ${\cal M}_{2\gamma l}$, in much the
same way as in evaluating $\hat\Gamma^{npW}_{l \, \alpha}$
(\ref{i17}). In this approach, liable for providing the dominant
part of
 ${\cal M}_{2\gamma l}$, the contribution of the second term in
 (\ref{j29}) disappears, as $f^{nn}{=}0$ is adopted, and the
 contribution of the first term gets simplified, utilizing
 (\ref{gn}), (\ref{j0})-(\ref{f0}). Then, with allowance for
 (\ref{j27}), straightforward calculation gives
\begin{eqnarray}
\begin{picture}(210,55)(0,-24)
\SetWidth{1}
\SetColor{Blue}
\Vertex(50,30){2}
\Vertex(50,-10){2}
\Vertex(100,30){2}
\Vertex(100,-10){4.5}
\Text(100,-16)[t]{${\cal J}_{pn}(0)$}
\Text(57,10)[l]{$Al$}
\Text(107,10)[l]{$W$}
\Photon(100,-10)(100,30){6}{4}
\ArrowLine(50,30)(15,30)
\ArrowLine(100,30)(50,30)
\ArrowLine(50,-10)(15,-10)
\ArrowLine(100,-10)(50,-10)
\ArrowLine(137,30)(100,30)
\ArrowLine(137,-10)(100,-10)
\Text(8,10)[r]{${\cal M}_{2\gamma l}^0 =$}
\Text(150,10)[l]{$=$}
\Photon(50,30)(50,-10){3}{6}
\Text(35,35)[b]{$e$}
\Text(15,-15)[t]{$p$}
\Text(75,35)[b]{$e$}
\Text(75,-15)[t]{$p$}
\Text(115,35)[b]{$\nu$}
\Text(135,-15)[t]{$n$}
\end{picture}
 \label{0}\\
= \Bigl(\frac{e}{2\sqrt{2}s_W}\Bigr)^2\frac{|V_{ud}|}{(2\pi)^4}
\frac{-e^2}{M_W^2}
\Biggl\{\frac{1}{2} I_1(2M_p\varepsilon , \lambda)
{\cal P}_0^{\alpha\beta} h^0_{\beta\alpha} -
\frac{1}{2}I_1(2M_pk_{\delta} , \lambda) {\cal P}_1^{\beta\delta\alpha}
h^0_{\beta\alpha} -\nonumber\\
-{\cal P}_1^{\beta\delta\alpha}
h^{1\nu}_{\beta\alpha}\bigl(I_1(k_{\delta}k_{\nu},\lambda) -
I_1(k_{\delta}k_{\nu},M_S)\bigr)\Biggr\} \, , \qquad \qquad
\qquad \qquad \nonumber
\end{eqnarray}
where
\begin{eqnarray}
I_1({\cal C},\mu)=\int\mbox{d}^4k \, \varphi (k,P_p,p_e)\frac{{\cal
C}} {k^2-\mu^2+i0} \; , \; \; \nonumber\\ \varphi (k,P_p,p_e)=
\frac{i}{[(p_e-k)^2-m^2+i0][(P_p+k)^2-M_p^2+i0]} \; , \; \;
\label{vph}\\
{\cal P}_0^{\beta\alpha}=\bar{u}_e(p_e)\Bigl(\gamma^{\beta}\bigl(
\frac{\hat{p}_e+m}{\varepsilon}\bigr)\gamma^{\alpha}(1-\gamma^5)
\Bigr)u_{\nu}(-p_{\nu}) \; , \; \label{if0} \\
{\cal
P}_1^{\beta\delta\alpha}=\bar{u}_e(p_e)\gamma^{\beta}\gamma^{\delta}
\gamma^{\alpha}(1-\gamma^5)u_{\nu}(-p_{\nu}) \; , \; \nonumber\\
h^0_{\alpha\beta}={\bar U}_p(P_p)\gamma_{\beta}\Bigl(\frac{\hat
P_p}{M_p}+1\Bigr)\gamma_{\alpha}(1-\gamma^5 g_A)U_n(P_n) \, ,
\nonumber\\
 h^{1\nu}_{\beta\alpha}=\bar{U}_p(P_p)\gamma_{\beta}\gamma^{\nu}
\gamma_{\alpha}(1-g_A\gamma^5)U_n(P_n) \, , \; \; \;
\hat{p}{\equiv}p_{\alpha}\gamma^{\alpha}
\, , \; \;\nonumber
\end{eqnarray}
and $\varepsilon{=}\sqrt{m^2+{\bf p}_e^2} \, , \; \; v{=}{|{\bf
p}_e|}{/}{\varepsilon}$ are the electron energy and velocity.

Now, the point is to acquire what comes out of allowance for the
nucleon compositeness: form factors and excited states. In what
follows, we shall treat concisely these two effects separately, one
after other.
 We start with retaining only the pure single proton intermediate
 state, $B{=}p$, in the first term in (\ref{j29}) and approximating
 thereby the nucleon form factors by Eq. (\ref{i8}),
 (\ref{a1})-(\ref{ff}). Then, with allowance for (\ref{j27}), we
 obtain the respective contribution to (\ref{j29})
\begin{eqnarray}
\begin{picture}(210,55)(90,-22)
\SetWidth{1}
\SetColor{Blue}
\Vertex(50,30){2}
\Vertex(50,-10){5}
\Vertex(100,30){2}
\Vertex(100,-10){5}
\Text(50,-17)[t]{$f^{pp}$}
\Text(100,-17)[t]{${\cal J}_{pn}$}
\Text(57,10)[l]{$Al$}
\Text(107,10)[l]{$W$}
\Photon(100,-10)(100,30){6}{4}
\ArrowLine(50,30)(15,30)
\ArrowLine(100,30)(50,30)
\ArrowLine(50,-10)(15,-10)
\ArrowLine(100,-10)(50,-10)
\ArrowLine(137,30)(100,30)
\ArrowLine(137,-10)(100,-10)
\Text(8,10)[r]{${\cal M}_{2\gamma l}^1 =$}
\Text(150,10)[l]
{{\Large$= \Bigl(\frac{e}{2\sqrt{2}s_W}\Bigr)^2\frac{|V_{ud}|}{(2\pi)^4}
\frac{-1}{M_W^2}\times$}}
\Photon(50,30)(50,-10){3}{6}
\Text(35,35)[b]{$e$}
\Text(15,-15)[t]{$p$}
\Text(75,35)[b]{$e$}
\Text(75,-15)[t]{$p$}
\Text(115,35)[b]{$\nu$}
\Text(135,-15)[t]{$n$}
\end{picture}\nonumber\\
\times\Biggl\{\frac{1}{2} I_1(2M_p\varepsilon , \lambda)
{\cal P}_0^{\alpha\beta} h^0_{\beta\alpha}
-\Bigl(I_1(2k_{\delta}M_p,\lambda)-{\delta}_{0\delta}I_1(2k_{\delta}M_p,
m_{\rho})\Bigr)\frac{1}{2}{\cal
P}_1^{\beta\delta\alpha}h^0_{\beta\alpha} -\nonumber\\
-\Biggl(\bigl(I_1(k_{\delta}k_{\nu},\lambda)
-I_1(k_{\delta}k_{\nu},M_S)\bigr)
- \bigl(I_1(k_{\delta}k_{\nu},m_{\rho})-I_1(k_{\delta}k_{\nu},M_S)\bigr)
\Biggr){\cal P}_1^{\beta\delta\alpha} h^{1\nu}_{\beta\alpha}
+ \; \; \label{m11}\\
 +[I_1({k_{\delta}k_{\nu}k_{\rho}}{/}{M_p},\lambda) -
 I_1({k_{\delta}k_{\nu}k_{\rho}}{/}{M_p},m_{\rho})] {\cal
 P}_1^{\beta\delta\alpha} M_p h_{\alpha\beta}^{2\nu\rho}\Biggr\} \;
 , \; \nonumber
\end{eqnarray}
 where, in addition to (\ref{if0}), we have defined
\begin{eqnarray}
 h^{2\mu\nu}_{\beta\alpha}=\bar{U}_p(P_p)\Biggl(
\gamma_{\beta}\gamma^{\mu}
\bigl(g_{WM}\sigma_{\alpha}^{\nu}-g_{IP}{\delta}_{\alpha\nu}\gamma^5
\bigr)
+\frac{1.79}{2M_p}\sigma_{\beta}^{\mu}\gamma^{\nu}\gamma_{\alpha}
(1-g_A\gamma^5)\Biggr) U_n(P_n)\sim \frac{1}{M_p} \, . \;
\;\label{if1}
\end{eqnarray}
In (\ref{0}), (\ref{m11}), the terms involving $h^0_{\beta\alpha}
 \, \; \, h^1_{\beta\alpha}$ (\ref{if0}) are associated with the
 electric form factor, whereas $h^{2\nu\mu}_{\beta\alpha}$
 (\ref{if1}) is due to the magnetic form factors and electroweak
 form factors (\ref{i8})-(\ref{ff}). Hereafter, the calculation of
 the $\alpha-$order total decay probability and electron momentum
 distribution will call for the real part of ${\cal M}_{2\gamma}$,
 as ${\cal M}^0$ is real, and integrating over the antineutrino and
 proton momenta is performed, see Sec. XI below. All the integrals
 $I_1$ but $I_1(2M_p\varepsilon , \lambda)$ are real, and their
 expressions prove to be rather plain,
\begin{eqnarray}
I_1(2M_pk_{\delta},\lambda)=\frac{p_{e \, \delta} I_1}{\varepsilon}
+ {\delta}_{0\delta}I_{10} \, , \; \; \; I_1=\frac{\pi^2}{v}\ln(x)
\, , \; \; \nonumber\\
I_{10}=\pi^2[2\ln\bigl(\frac{m}{M_p}\bigr)-\frac{1}{v}\ln(x)] \, ,
\; \; \; x=\frac{1-v}{1+v} \, , \; \; \label{11i}\\
I_1(k_{\delta}k_{\nu},\lambda)-I_1(k_{\delta}k_{\nu},M_S)=-g_{\delta\nu}
(I_2-{\delta}_{\delta 0}I_{20}) \, , \; \; \; \nonumber\\
I_2=\frac{\pi^2}{4}\bigl(\frac{3}{2}+2 \ln\frac{M_S}{M_p}\Bigr) \,
, \; \; \; I_{20}=\frac{\pi^2}{2} \, . \nonumber
\end{eqnarray}
Following the method of \cite{jfc}, the careful calculation of
$\mbox{Re}I_1(2M_p\varepsilon,\lambda)$ was carried out in Ref.
\cite{iii} with the result
\begin{eqnarray}
\mbox{Re}I_1(2M_p\varepsilon,\lambda)={\cal I}(P_p,p_e,\varepsilon )
=\nonumber\\
=-\frac{\pi^2}{v} [\ln(x) \ln({\lambda}{/}{m}) -
\frac{1}{4}(\ln{(x)})^2+F(1/x-1)-\frac{M_p{\pi}^2}{A}{\cdot}
\frac{v \, \varepsilon}{t_2-t_1} \, ] \, , \label{1ij}
\end{eqnarray}
where
\begin{eqnarray}
t_{1,2}=-\frac{m^2 -M_p^2\pm 2 \cdot \sqrt{(P_p \, p_{e})^2 -M_p^2
m^2}}{m^2 +M_p^2 +2(P_p \, p_{e})} \, , \; \; \; \;
4A{=}m^2+M_p^2+2p_eP_p \, , \nonumber
\end{eqnarray}
and $F$ is the Spence-function \cite{sf}. This quantity (\ref{1ij})
determines the first, most important term in the amplitudes
(\ref{0}) and (\ref{m11}). Let us behold that the ``Coulomb
correction" is incorporated therein in the natural way, via the
last term in ${\cal I}(P_p,p_e,\varepsilon )$ (\ref{1ij}).

The second term in (\ref{m11}) comes out of the second term in
(\ref{0}) by subtracting $I_1(2M_pk_{\delta},m_{\rho})$ from
$I_1({2M_pk_{\delta},\lambda})$. For the mass
${\mu}{\gtrsim}m_{\rho}$, the estimation is obtained
\begin{eqnarray}
I_1 (2M_pk_{\alpha},\mu)\approx -\pi^2\delta_{0\alpha} [\frac{(r^2
-4)^{{3}{/}{2}}}{12r}\ln \Bigl(\frac{r\sqrt{r^2
-4}}{2}+\frac{r^2}{2}-1\Bigr)+
\Bigl(1- \frac{r^2}{6}\Bigr)\ln{r}
+\frac{1}{6}] \; , \label{a37}
\end{eqnarray}
where ${r}{=}{\frac{\mu}{M_p}}$. At ${\mu}{=}{m_{\rho}} \, , \; \,
r{\approx}1$, we have got
\begin{equation}
I_1 (2M_pk_{\alpha},m_{\rho})\approx -\pi^2{\delta}_{0\alpha} \, .
\label{a38}
\end{equation}
This value is to be compared to
\begin{equation}
{\delta}_{0\alpha}I_{10}\approx \pi^2
\ln\frac{m}{M_p}\delta_{0\alpha}\approx
-15\pi^2{\delta}_{0\alpha}
\, \label{a39}
\end{equation}
in $I_1(2M_pk_{\alpha},\lambda )$. As seen, $I_1
(2M_pk_{\alpha},m_{\rho})$ can be omitted in (\ref{m11}) with an
error smaller than $6\%$.

Taking into consideration (\ref{j27}), the differences which
determine the third terms in (\ref{0}) and in (\ref{m11}) are
reduced to
\begin{eqnarray}
I_1(k_{\delta}k_{\nu},\lambda)-I_1(k_{\delta}k_{\nu},M_S)=
-g_{\alpha\beta}\frac{\pi^2}{2}[
(\frac{3}{4}+{\ln}\frac{M_S}{M_p} )-\delta_{0\alpha}] \; , \;
\label{id1}\\
I_1(k_{\delta}k_{\nu},m_{\rho})-I_1(k_{\delta}k_{\nu},M_S)=
\frac{\pi^2}{10}{\delta}_{\delta\nu}{\delta}_{0\delta}-\frac{\pi^2}{2}
g_{\delta\nu}\Bigl(\ln\frac{M_S}{M_p}-\frac{3}{4}-I(m_{\rho})\Bigr)
\, , \;  \label{id2}\\
 I(\mu) = \int_0^1 x \mbox{d}x \int_0^1 \mbox{d}y \, {\ln}[x^2
(y-1)^2 + r_{\mu}^2 (1-x)] \, , \; \; \; r_{\mu}=\frac{\mu}{M_p}
\,.
\end{eqnarray}

Next, it is only a matter of straightforward numerical evaluation
to become convinced that the quantity (\ref{id2}) makes up no more
than ${\sim}10\%$ to (\ref{id1}). So, with this accuracy, the third
term in ${\cal M}^1_{2\gamma l}$ (\ref{m11}) is seen to coincide
with the third term in ${\cal M}^0_{2\gamma l}$ (\ref{0}).

In the last term in (\ref{m11}), the factor ${\cal
P}_1^{\beta\delta\alpha} M_p h_{\beta\alpha}^{2\nu\rho}$ is
realized to be of the same order, as the factors ${\cal
P}_1^{\beta\delta\alpha} h_{\beta\alpha}^{0}$ and ${\cal
P}_1^{\beta\delta\alpha} h_{\beta\alpha}^{1\nu}$ in (\ref{0}),
(\ref{m11}). Upon a labor-consuming but rather unsophisticated
evaluation of the corresponding integrals
$I_1({k_{\delta}k_{\nu}k_{\rho}}{/}{M_p},\mu)$, we arrive at the
estimation of the difference
\begin{eqnarray}
I_1({k_{\delta}k_{\nu}k_{\rho}}{/}{M_p},\lambda)-
I_1({k_{\delta}k_{\nu}k_{\rho}}{/}{M_p},m_{\rho})\approx\nonumber\\
\approx-\frac{\pi^2
r_{\mu}^2}{6}\Bigl(-1+2r^2\int^1_{0} \frac{\mbox{d}z z^2}{z^2
-z(2-r^2)+1}\Bigr)\approx-\frac{\pi^2}{10} \, , \; \; \;
 r=\frac{m_{\rho}}{M_p} \, , \label{id3}
\end{eqnarray}
which constitutes ${\lesssim}1{\%}$ to the integrals
$I_1(2M_p\varepsilon,\lambda) \, , \; \, I_1(2M_pk_{\delta},\lambda)
\, , \; \, I_1(k_{\delta}k_{\nu},\lambda)
  - I_1(k_{\delta}k_{\nu},\lambda),$ determining ${\cal M}_{2\gamma
l}^0$ (\ref{0}). So, the last term in ${\cal M}^1_{2\gamma l}$
(\ref{m11}) is seen to constitute no more than ${\sim}1\%$ to
${\cal M}^0_{2\gamma l}$ (\ref{0}) and can be abandoned with this
accuracy.

Thus, we have realized the difference ${\cal M}^1_{2\gamma l}
-{\cal M}^0_{2\gamma l}$ caused by allowance for the nucleon
form factors (\ref{i8})-(\ref{ip}) amounts to less than
${\sim}10\%$ to the dominant quantity ${\cal M}^0_{2\gamma l}$
(\ref{0}). Consequently, with committing an error less than
${\sim}10\%$, the form factors (\ref{i8})-(\ref{ip}) can be
replaced by (\ref{j0})-(\ref{f0}) so that ${\cal M}^1_{2\gamma l}$
reduces to ${\cal M}^0_{2\gamma l}$. All the more so, we can
neglect, at least with the same accuracy, the contribution from the
second term in (\ref{j29}) which is due to nothing but the neutron
form factor $f_2^{nn}{\sim}{(M_n-M_p)}{/}{M_p} \sim 0 \, $
(\ref{ff}) exclusively, even in the simplest case corresponding to
the pure neutron intermediate state, $B{=}n$.

Now, we are to consider the terms with $B{\neq}N$ in the sum in
(\ref{j29}) which present the processes involving the virtual
excited states of the nucleon, depicted by the triplex lines in the
diagrams (\ref{j29}). These intermediate states are naturally to be
treated as the well-known nucleon excited states, such as the
$\Delta_{33}-$isobar, the Roper-resonance and so on, with the
propagators ${\cal G}_B$ (\ref{gb}) (depending on the masses $M_B
\, , \; \, M_N{<}M_B{\ll}M_S$) instead of the nucleon propagator
$G_N$ (\ref{gn}). For the current estimation, it is of a drastic
value that the quantities $m^2 \, , \, \; (M_n{-}M_p)^2$ are
actually negligible as compared to the differences $M_B^2{-}M_N^2$
,
\begin{equation}
\frac{(M_n-M_p)^2}{M_B^2-M_N^2}\sim 0 \; , \; \; \; \;
\frac{m^2}{M_B^2-M_N^2}\sim 0 \; . \label{dm1}
\end{equation}
Indeed, even in the case of the $\Delta_{33}-$isobar, the lowest
nucleon excited state, we have got $M_{\Delta}-M_p{\approx}300 \,
\mbox{MeV}$. All the more so, Eqs. (\ref{dm1}) hold true for any
other nucleon excited state $B{\neq}\Delta_{33}$. Moreover, the
important relation is obviously valid
\begin{equation}
M_B^2-M_N^2\sim M_N^2 \, . \label{dm2}
\end{equation}
In the processes involving these intermediate states $B{\neq}N$,
the quantities (\ref{i8})-(\ref{ip}) describe the weak and
electromagnetic transitions between the excited and ground states
of the nucleon. For purpose of the current estimation, we take up
the processes with a $\Delta_{33}-$isobar, $B{=}\Delta_{33}$, the
simplest exited state of the nucleon, the internal structure of
which is much the same as that of the nucleon ground state. In the
nucleon as well as in the $\Delta_{33}-$isobar, all three quarks
occupy the same state $1S_{1{/}2}$. Therefore, it is plausible in
the current estimation to presume the amplitudes
$f_{\mu}^{N\Delta}(k)$,
 ${\cal J}_{\alpha}^{n\Delta}(k)$ do not differ substantially from
$f_{\mu}^{NN}(k)$, ${\cal J}_{\alpha}^{np}(k)$
(\ref{i8})-(\ref{ff}). Also along these lines, as $G_p$ gives place
to ${\cal G}_B$ in the amplitude ${\cal M}^0$ (\ref{0}), the very
modification which is of vital importance for the qualitative
assessment actually consists in replacing
\begin{equation}
M_p\Longrightarrow M_{\Delta} \,  \label{dm3}
\end{equation}
in the proton propagator. Then the respective contribution into the
amplitude (\ref{j29}) reduces to
\begin{eqnarray}
\begin{picture}(210,55)(40,-15)
\SetWidth{1}
\SetColor{Blue}
\Vertex(50,30){2}
\Vertex(50,-10){3}
\Vertex(100,30){2}
\Vertex(100,-10){5}
\Text(102,-15.5)[t]{${\cal J}_{np}(0)$}
\Text(57,10)[l]{$Al$}
\Text(107,10)[l]{$W$}
\Photon(100,-10)(100,30){6}{4}
\ArrowLine(50,30)(15,30)
\ArrowLine(100,30)(50,30)
\ArrowLine(50,-10)(15,-10)
\SetWidth{1.5}
\ArrowLine(100,-10)(50,-10)
\SetWidth{1}
\ArrowLine(137,30)(100,30)
\ArrowLine(137,-10)(100,-10)
\Text(8,10)[r]{${\cal M}_{2\gamma l}^{\Delta} =$}
\Text(150,10)[l]{$=$}
\Photon(50,30)(50,-10){3}{6}
\Text(35,35)[b]{$e$}
\Text(15,-15)[t]{$p$}
\Text(75,35)[b]{$e$}
\Text(69,-13)[t]{$M_{\Delta}$}
\Text(115,35)[b]{$\nu$}
\Text(135,-15)[t]{$n$}
\end{picture}\nonumber \\
=\int\frac{\mbox{d}
k^4}{(2\pi)^4 i}\bigl(\bar u_e(p_e)(-e)\gamma^{\beta}G_e(p_e-k)
\Gamma^{e\nu W \, \alpha} u_{\nu}(-p_{\nu})\bigr)\times \qquad
 \qquad \qquad \label{dm5} \\
\frac{-M^2_S}{(k^2-\lambda^2+i0)(k^2-M^2_S+i0)(k^2-M_W^2+i0)}
\Biggl\{\bigl(\bar U_p(P_p) e \gamma_{\beta} {\cal G}_{\Delta}(P_p+k)
 \Gamma_{\alpha}^{pnW}(k) U_n(P_n)\bigr)\Biggr\} \, .\nonumber
\end{eqnarray}
The estimation of ${\cal M}_{2\gamma l}^{\Delta}$ (\ref{dm5}) is
procured by replacing
\begin{equation}
I_1\Longrightarrow I_{1\Delta} \label{dm6}
\end{equation}
in ${\cal M}^0_{2\gamma l}$ (\ref{0}), where $I_{1\Delta}$ comes
out of $I_1$ (\ref{f0}) with changing the proton mass $M_p$ by the
$\Delta_{33}-$isobar mass $M_{\Delta}$ in the function
$\varphi(k,P_p,p_e)$ (\ref{vph}). What is of crucial importance for
the current evaluation is that
\begin{equation}
(P_{\Delta}+k)^2-M_p^2=M_{\Delta}^2-M_p^2\sim M_p^2\gg (M_n-M_p)^2
\, \label{dm4}
\end{equation}
at $k{=}0$ in the denominators of the integrands in $I_{1\Delta}$,
instead of zero in the integrands of $I_1$ (\ref{f0}), i.e. with
$M_p$ in place of $M_{\Delta}$. In particular, that is why there
occurs no infrared divergence in the integral
$I_{1\Delta}(2M_{p}\varepsilon , \lambda)$, as opposed to
$I_{1}(2M_{p}\varepsilon , \lambda)$. As ${\cal M}_{2\gamma
l}^{\Delta}$ is expressed in terms of $I_{1\Delta}$ alike ${\cal
M}_{2\gamma l}^{0}$ is expressed in terms of $I_{1}$, the integrals
$I_{1\Delta}({\cal C},\mu)$ are to be evaluated with $\mu =
\lambda , M_S$ and confronted to the respective integrals
$I_1({\cal C},\mu)$ in order to assess the ${\cal M}_{2\gamma
l}^{\Delta}$ (\ref{dm5}) value as compared with the value of ${\cal
M}_{2\gamma l}^{0}$ (\ref{0}). The most important integrals
 $I_{1\Delta}(2M_{p}\varepsilon , \mu)$ which determine the
 dominant part of ${\cal M}_{2\gamma l}^{\Delta}$ (as
 $I_{1}(2M_{p}\varepsilon , \mu)$ do in the case of ${\cal
 M}_{2\gamma l}^{0}$) are given by
\begin{equation}
I_{1\Delta}(2M_p\varepsilon , \mu)\approx
2M_p\varepsilon\pi^2\int\limits_{0}^{1}\mbox{d}x x
\int\limits_{0}^{1}dy\frac{1}{y^2 x^2 M_p^2 +y x
(M_{\Delta}^2-M_p^2) +{\mu}^2 (1-x) +x^2 m^2} \, .\label{dm7}
\end{equation}
With allowance for Eqs. (\ref{dm1})-(\ref{dm4}), we acquire the
estimation of the integral $I_{1\Delta}(2M_{p}\varepsilon ,
\lambda)$ in ${\cal M}_{2\gamma l}^{\Delta}$ (\ref{dm5})
\begin{equation}
I_{1\Delta}(2M_p\varepsilon,\lambda)\approx \frac{4\pi^2
M_p\varepsilon}{(M_{\Delta}^2-M_p^2)}\ln
\Bigl(\frac{(M_{\Delta}^2-M_p^2)}{m M_p}\Bigr)\sim 0 \, .\label{dm8}
\end{equation}
instead of the integral $I_{1}(2M_{p}\varepsilon ,
\lambda)$  (\ref{1ij}), mostly determining the
evaluation of ${\cal M}_{2\gamma l}^{0}$ (\ref{0}). Likewise, the
estimation of (\ref{dm7}) at $\mu{=}M_S$ gives
\begin{equation}
I_{1\Delta}(2M_p\varepsilon,M_S )\approx \frac{4\pi^2
\varepsilon}{M_S}\ln \Bigl(\frac{(M_{\Delta}^2-M_p^2)}{m M_p}\Bigr)\sim 0
\, .\label{dm9}
\end{equation}

In much the same way, it is straightforward to become convinced that
all the remaining integrals $I_{1\Delta}({\cal C},\mu)$ in ${\cal
M}_{2\gamma l}^{\Delta}$ (\ref{dm5}) prove also to be negligible as
compared to the respective integrals $I_1({\cal C},\mu)$ in ${\cal
M}_{2\gamma l}^{0}$ (\ref{0}), and consequently ${\cal M}_{2\gamma
l}^{\Delta}$ (\ref{dm5}) results to be rather negligible as compared
with ${\cal M}_{2\gamma l}^{0}$ (\ref{0}).

Except for the aforesaid $\Delta_{33}-$resonance case, the
structure of excited states of the nucleon differs drastically from
the structure of the nucleon ground state. Therefore, the values of
all the amplitudes ${\cal J}_{nB}^{\alpha}
\, , \; \, f^{pB}_{\mu}$ with $B{\neq}N$ and $B{\neq}\Delta_{33}$ are
substantially smaller than ${\cal J}_{pn}^{\mu}{\sim}{\cal
J}_{\Delta n}^{\mu} \, , \; \, f^{pp}_{\mu}{\sim}f^{\Delta p}$.
Given this fact, it stands to reason that the contribution to
${\cal M}_{2\gamma l}$ (\ref{j29}) due to these intermediate states
can not exceed anyway the contribution from the intermediate
$\Delta_{33}-$isobar state considered above. So, all the
corrections to ${\cal M}_{2\gamma l}^{0}$ (\ref{0}) caused by the
terms involving the intermediate excited states with $B{\neq}p$ in
Eq. (\ref{j29}) prove to be negligible, as a matter of fact. All
the more so, we can abandon the contributions to ${\cal M}_{2\gamma
l}$ (\ref{j29}) which are due to simultaneous allowance for the
excited states, $B{\neq}p$, and the form factors $f^{BB'}_{\mu} \,
,
\; \, {\cal J}_{rs}^{\mu}$ (\ref{j2b}), (\ref{a}), respecting the
above estimations associated with Eqs. (\ref{a37})-(\ref{id3}) and
the relevant discussion thereat.

Thus, summing up, we have ascertained the amplitude ${\cal
M}_{2\gamma l}$ (\ref{j29}) can be reduced to ${\cal M}_{2\gamma
l}^{0}$ (\ref{0}) with the accuracy better than ${\sim}10\%$. On
substituting (\ref{if0}), (\ref{1ij}) in (\ref{0}), ${\cal
M}_{2\gamma l}$ is finally put into the explicit form:
\begin{eqnarray}
{\cal M}_{2\gamma
l}=\Bigl(\frac{e}{2\sqrt{2}s_W}\Bigr)^2|V_{ud}|\frac{1}{M_W^2}
\cdot\frac{\alpha}{4\pi} \,
\Biggl\{\Biggl(\bar
u_e(p_e)\gamma^{\beta}(\not
p_e+m)\gamma^{\alpha}(1-\gamma^5)u_{\nu}(-p_{\nu})
\frac{1}{2\varepsilon M_p v}\times\nonumber\\
\times [\ln{(x)}\ln{\frac{\lambda}{m}}-\frac{1}{4}(\ln{(x)})^2+F(1/x-1)
-\frac{M_p{\pi}^2}{A}{\cdot}\frac{v \, \varepsilon}{t_2-t_1} \, ]
- \nonumber\\
-\bar u_e(p_e)\gamma^{\beta}\gamma^{\delta}\gamma^{\alpha}
(1-\gamma^5)u_{\nu}(-p_{\nu})\frac{1}{2M_p} \, [-\frac{p_{e \,
\delta}}{v\varepsilon}\ln{(x)}+
\delta_{0\delta}\bigl(\frac{1}{v}\ln{(x)}-2\ln{\frac{m}{M_p}}\bigr)]
\Biggr)\times \; \label{i29} \\
\times\bigl(\bar U_p(P_p)\gamma_{\beta}(\not P_p+M_p)\gamma_{\alpha}
(1-\gamma^5g_A)U_n(P_n)\bigr)
- \bigl(\bar u_e(p_e)\gamma^{\beta}\gamma^{\delta}\gamma^{\alpha}
(1-\gamma^5)u_{\nu}(-p_{\nu})\bigr)\times\nonumber\\
\times \bigl(\bar
U_p(P_p)\gamma_{\beta}
\gamma^{\nu}\gamma_{\alpha}(1-g_A\gamma^5)U_n(P_n)\bigr)
 g_{\delta\nu}\Bigl(\frac{3}{8}+\frac{1}{2}\bigl(
\ln{\frac{M_W}{M_p}} - \frac{M_W^2}{M_W^2-M_S^2}\ln{\frac{M_W}{M_S}}
 \bigr) - \delta_{0\delta}\frac{1}{2}\Bigr)\Biggr\} \, . \nonumber
\end{eqnarray}

What is the inherent feature of ${\cal M}_{2\gamma l}$ (\ref{j29}),
(\ref{i29}) to be emphasized is that this amplitude shows up to be
not multiple to the uncorrected Born amplitude ${\cal M}^{0}$
(\ref{i7}), even though ${\cal M}_{2\gamma l}$ (\ref{i29}) has
ensued from the general expression (\ref{j29}) on leaving aside the
effects of nucleon structure. In this regard, ${\cal M}_{2\gamma
l}$ on principle differs from the above considered quantities
${\cal M}_{2\gamma s} \, $ (\ref{i27}), $ \,
\hat{\Gamma}_{\alpha}^{npW}
\, $ (\ref{i21}), $ \, \hat{\Gamma}^{e\nu W}\, $ (\ref{i9}), which
all have turned out to be proportional to the corresponding
uncorrected quantities ${\cal M}^0 \, , \; \,
{\Gamma}_{\alpha}^{npW} \,
 , \; \, {\Gamma}^{e\nu W}$ (\ref{i7}), (\ref{g0}), (\ref{j1b}).
\section{Real $\Large\bbox\gamma{-}$radiation.}
\label{sec:level10}
In the first $\alpha-$order, the real $\gamma-$emission
accompanying the neutron $\beta-$decay
 is presented by the diagrams ${\large\bf 5-8}$ in the amplitude
${\cal M}$ (\ref{mm}). The
 triplex lines in the graphs ${\large\bf 6,7}$ represent the
 conceivable excited states of the nucleon. As $m,
 M_n-M_p{\ll}M_p{\ll}M_W$, the contributions from the diagrams
 ${\large\bf 6-8}$ are negligible as compared to the one coming out
 of the diagram ${\large\bf 5}$, which renders the common
 bremsstrahlung of a final electron. The corresponding amplitude of
 the
  real $\gamma-$radiation with the momentum $k$ and the
 polarization $\bbox{\epsilon}^{(r)}$
\begin{eqnarray}
{\cal M}_{1\gamma}^{(r)}(k) =
\Bigl(\frac{e}{2\sqrt{2}s_W}\Bigr)^2|V_{ud}|\bigl(\frac{-1}{M_W^2}\Bigr)
e \epsilon_a^{(r)} \bigl(\bar u_e(p_e)\gamma^a\frac{(\not p_e+\not
k+m)}{( p_e +k)^2-m^2}
\gamma^{\lambda}(1-\gamma^5)u_{\nu}(-p_{\nu})\bigr)\times\label{i30}\\
\times \bigl(\bar U_p(P_p)\gamma_{\lambda}(1-g_A\gamma^5)U_n(P_n)\bigr)
 \, , \qquad \qquad (a , r =1 , 2 , 3 ) \; , \nonumber
\end{eqnarray}
is seen to be not proportional to the uncorrected quantity ${\cal
M}^0$ (\ref{i7}), alike ${\cal M}_{2\gamma l}$, yet against ${\cal
M}_{2\gamma s} \, , \; \, \hat{\Gamma}_{\alpha}^{npW} \, ,
\; \,  \hat{\Gamma}^{e\nu W}\, $ which all are multiple to
 the uncorrected quantities ${\cal M}^0 \, , \; \,
 {\Gamma}_{\alpha}^{npW} \, ,
\; \, {\Gamma}^{e\nu W}\, $.
\section{The radiative corrections to
the electron momentum distribution and the neutron lifetime.}
\label{sec:level11}
With allowance for the radiative corrections of order $\alpha$, the
absolute square of the transition amplitude ${\cal M}$ (\ref{mm})
proves expedient to be written in the form
\begin{eqnarray}
|{\cal M}|^2 = |{\cal M}^R +{\cal M}_{2\gamma l} +{\cal
M}^{(r)}_{1\gamma}|^2\approx |{\cal M}^R|^2 + |{\cal
M}^{(r)}_{1\gamma}|^2 + 2\mbox{Re}[{\cal M}^0{\cal M}_{2\gamma l}]
,\label{abc}
\end{eqnarray}
where
\begin{eqnarray}
{\cal M}^R = \bigl(\bar u_e(p_e)\hat \Gamma_{\alpha}^{e\nu\mu}
u_{\nu}(-p_{\nu})\bigr){\cdot}\bigl(\bar U_p(P_p) \hat
\Gamma_{\beta}^{npW} U_n(P_n)\bigr)
\hat{D}^W_{\alpha\beta}(p_{\nu}+p_e)+{\cal M}_{2\gamma
s}\approx\nonumber\\
\approx{\cal
M}^0\Bigl\{1-\frac{\alpha}{4\pi}\Bigr(2\ln{\frac{M_Z}{M_p}}+
4\ln{\frac{\lambda}{m}}+\frac{9}{2}-\ln{\frac{M_p}{m}}
 - \frac{6}{s_W^2} - 6\ln{\frac{M_Z}{M_S}}-
\frac{3+4c^2_W}{s^4_W}\ln{(c_W)}
\Bigr)\Bigr\}\label{i31}
\end{eqnarray}
comprises all the terms proportional to the Born amplitude ${\cal
M}^0$ (\ref{i7}).

As a final state after the neutron $\beta -$decay involves a
proton, an electron, an antineutrino and $\gamma -$rays, the
probability of the polarized neutron $\beta -$decay, upon
summarizing the absolute square
 $|{\cal M}|^2$ of the transition amplitude over the polarizations
of all the particles in the final state, is obviously put into the
following well-known general form
\begin{eqnarray} d{\bf W}({\bf p_{e}}, {\bf P},
{\bf p_{\nu}}, {\bf k}, {\bfgr \xi})=
(2\pi)^{4}{\delta}(M_{n}-E_{P}-{\omega}_{\nu}-\varepsilon -\omega )
{\delta}({\bf P{+}p_{e}{+}p_{\nu}{+}k}) \times \nonumber\\
\frac{1}{2M_n}\sum_{i f}{|{\cal M}_{if}|^{2}}
\frac{d{\bf P} d{\bf p_{e}} d{\bf
p_{\nu}} d{\bf k}} {(2\pi)^{12} \, 2E_{p} \, 2{\varepsilon} \,
2{\omega}_{\nu}
\, 2{\omega}}= \nonumber \\ w({\bf p_{e}}, {\bf P}, {\bf p_{\nu}}, {\bf k},
{\bfgr \xi}) \, d{\bf P} d{\bf p_{e}} d{\bf p_{\nu}} d{\bf k}
{\delta}(M_{n}-E_{P}-{\omega}_{\nu}-\varepsilon -\omega )
{\delta}({\bf P{+}p_{e}{+}p_{\nu}{+}k}) \, , \label{gw}
\end{eqnarray} where $\bfgr \xi$ stands for the polarization vector
of a resting neutron, and $ p_{e}{=}({\varepsilon},{\bf p_{e}}) ,
\; P{=}(E_{P},{\bf P}) , \;$ $p_{\nu}{=}({\omega}_{\nu},{\bf
p_{\nu}}) ,\; k{=}({\omega},{\bf k})$ are the electron, proton,
antineutrino and $\gamma -$ray four-momenta, respectively. The
familiar expression (\ref{gw}) renders the momentum distribution of
electrons, protons, antineutrinos and $\gamma
-$rays in the final state.

In the work presented, our purpose is to calculate the $\beta
-$decay probability integrated over the final proton, antineutrino and
photon momenta and summarized over the polarizations of all the
final particles,
\begin{eqnarray}
\mbox{d}{\bf W}(\varepsilon ,{\bf
p}_e , \bbox{\xi})
=\mbox{d}{\bf W}^R(\varepsilon ,{\bf p}_e , \bbox{\xi})+
\mbox{d}{\bf W}_{1\gamma}(\varepsilon ,{\bf p}_e , \bbox{\xi})+
\mbox{d}{\bf W}_{2\gamma l}(\varepsilon ,{\bf p}_e ,
 \bbox{\xi}) \; , \label{i32}
\end{eqnarray}
where $\mbox{d}{\bf W}^R , \;
\mbox{d}{\bf W}_{1\gamma} , \;
\mbox{d}{\bf W}_{2\gamma l}$ are due to $|{\cal M}^R|^2 , \;  |{\cal
M}^{(r)}_{1\gamma}|^2 , \; 2\mbox{Re}[{\cal M}^0{\cal M}_{2\gamma
l}], $ (\ref{abc}) , respectively.

Although the calculation of the distribution (\ref{i32}) turns out
to be cumbersome and labour-consuming, it runs along a plain and
unsophisticated way, as a matter of fact. So, we shall not expound
this calculation at full length, in details, but only set forth the
main stages in evaluating $ \, \mbox{d}{\bf W}(\varepsilon ,{\bf
p}_e ,\bbox{\xi}) \, $ (\ref{i32}).

As being due to $|{\cal M}^R|^2$, the quantity
\begin{eqnarray}
\mbox{d}{\bf W}^R(\varepsilon ,{\bf p}_e , \bbox{\xi})\approx
\mbox{d}{\bf W}^0(\varepsilon ,{\bf
p}_e ,
\bbox{\xi})\Bigr\{1-\frac{\alpha}{2\pi}\Bigl(2\ln{\frac{M_W}{M_p}}+
4\ln{\frac{\lambda}{m}}+\frac{9}{2}-\ln{\frac{M_p}{m}}-\nonumber\\
 - \frac{6}{s_W^2} -6\ln{\frac{M_Z}{M_S}} -
 \frac{5+2c_W^2}{s_W^4}\ln{(c_W)}
\Bigr)\Bigr\}  \label{j32}
\end{eqnarray}
is apparently proportional to the uncorrected decay probability
\begin{eqnarray}
\mbox{d}{\bf W}^0(\varepsilon ,{\bf
p}_e , \bbox{\xi})=\mbox{d}{\bf w}(\varepsilon , {\bf p}_e)
\bigl(1+3g_A^2+{\bbox{v\xi}}2g_A(1-g_A)\bigr)
 \, , \nonumber\\ \mbox{d}{\bf w}(\varepsilon , {\bf
p}_e)=\frac{G^2}{2\pi^3}\varepsilon |{\bf p}_e| k_m^2
\mbox{d}\varepsilon\frac{\mbox{d}{\bf n}}{4\pi} \, , \; \; \;
 {\bf n}={\bf p}_e{/}|{\bf p}_e| \, , \; \; \;
{\bf v}={\bf p}_e{/}\varepsilon \, , \; \; \;
k_m=M_n-M_p-\varepsilon \; .
\label{i33}
\end{eqnarray}
The contribution of the real $\gamma$-radiation $\mbox{d}{\bf
W}_{1\gamma}(\varepsilon ,{\bf p}_e , \bbox{\xi})$ \cite{i} stems
from $|{\cal M}_{1\gamma}|^2$
\begin{eqnarray}
\mbox{d}{\bf W}_{1\gamma}(\varepsilon ,{\bf p}_e)=
\mbox{d}{\bf w}(\varepsilon , {\bf p}_e)
\int\limits_{0}^{k_m}\mbox{d}k\Bigl((1+3g_A^2)
W_{0\gamma}(\varepsilon , k)+{\bf v\bfgr \xi}2g_A(1-g_A)W_{\xi
\gamma}(\varepsilon , k)\Bigr)=
\nonumber\\
 =\mbox{d}{\bf w}(\varepsilon , {\bf p}_e)
\bigl\{(1+3g_A^2)[\tilde B(\varepsilon)+\tilde
C'_0(\varepsilon)] +
 {\bf v\bfgr \xi}2g_A(1-g_A)[\tilde B(\varepsilon)+\tilde
C'_{\xi}(\varepsilon)]\bigr\} \, ,\label{j33}
\end{eqnarray}
where
\begin{eqnarray}
{\tilde B}=\frac{2\alpha}{\pi}\bigl({[\frac{1}{v}
\ln\frac{p_e+\varepsilon}{m} -1]}\cdot
\ln (\frac{2k_{m}}{\lambda})-
\frac{{\cal K}(\varepsilon)}{2v}\bigr)\, , \nonumber \\
{\tilde
C}'_{0}=\frac{2\alpha}{\pi}\Bigl\{[\frac{1}{v}
\ln\frac{p_e+\varepsilon}{m} -1]\Bigl(\frac{k_m}{3\varepsilon}
-\frac{3}{2}\Bigr) +\frac{k_m^2}{24v\varepsilon ^2}
 \ln\frac{p_e+\varepsilon}{m}\Bigr\} ,\label{j34}
\\ {\tilde C}'_{\xi}=\frac{2\alpha}{\pi}[\frac{1}{v}
\ln\frac{p_e+\varepsilon}{m} -1]\cdot \Bigl(\frac{k_m}{\varepsilon v^2
}\bigl(\frac{1}{3}+\frac{k_m}{24\varepsilon}\bigr)-\frac{3}{2}\Bigr)
\, , \nonumber \\
{\cal K}=\frac{1}{2}(F(x)-F(1/x)-\ln(1/x)\cdot \ln(\frac{1-v^2}{4}))
-v+\frac{1}{2} \ln(x) + F(v) - F(-v) \,  .
\nonumber
\end{eqnarray}

The contribution from $2\mbox{Re}{\cal M}^0 {\cal M}_{2\gamma l}$
is
\begin{eqnarray}
\mbox{d}{\bf W}_{2\gamma l}(\varepsilon , {\bf p}_e , \bbox{\xi}) =
\mbox{d}{\bf w}(\varepsilon , {\bf p}_e)\bigl\{[1+3g_A^2 + {\bf v\bfgr
\xi}2g_A(1-g_A)]B_{2\gamma}(\varepsilon) + \nonumber\\
+C_{02\gamma}(g_A ,
\varepsilon ) + {\bf v\bfgr \xi} C_{\xi 2\gamma}(g_A , \varepsilon)
\bigr\} \, .\label{j35}
\end{eqnarray}
Here
\begin{eqnarray}
B_{2\gamma}({\varepsilon})=\frac{\alpha}{2{\pi}^3} [(2{\cal
I}_1(\varepsilon)-I_{1})- I_{10}], \nonumber\\ C_{02\gamma}(g_A ,
{\varepsilon})=
\frac{\alpha}{2{\pi}^3} {\{}-I_{1}v^{2}[1+
3(g_{A})^{2}]+ 2 I_{2s}[5+12g_{A}+15 g_{A}^{2}]-2 I_{20}[2+3g_{A}+
3g_A^2] {\}},
\nonumber \\
C_{\xi 2\gamma}(g_A , {\varepsilon})=
\frac{\alpha}{2{\pi}^3} {\{}-I_{1}2g_{A}(1-
g_{A})+2 I_{2s}[3+4g_{A}-7 g_{A}^{2}]
-2I_{20}[1+g_{A}-2g_{A}^{2}] {\}},
\label{j36} \\
I_{2s}=\frac{{\pi}^2}{4}[3/2 +
2\Bigl(\ln{\frac{M_W}{M_p}}-\frac{M_W^2}{M_W^2-M_S^2}\ln{\frac{M_W}{M_S}}
\Bigr)] \; ,\nonumber\\
{\cal I}_1(\varepsilon )
=-\frac{\pi^2}{v} [\ln(x) \ln({\lambda}{/}{m}) -
\frac{1}{4}(\ln{(x)})^2+F(1/x-1)-
\frac{v\pi^2}{\tilde{v}(\varepsilon )} \, ] \, ,\nonumber\\
 {\tilde v}({\varepsilon})=\frac{1}{2}\Biggl(
\sqrt{(v+\frac{m k_m}{M_p\varepsilon} )^2 +
2v\frac{k_m}{\varepsilon} (\frac{m}{M_p})^2} +
 \sqrt{(v-\frac{m k_m}{M_p\varepsilon} )^2 -
2v\frac{k_m}{\varepsilon} (\frac{m}{M_p})^2}
 \, \, \Biggr) \, , \label{vt}
\end{eqnarray}
where the quantities $I_1 , I_{10} , I_2 , I_{20}$ are given in
(\ref{11i}). It is to recall once more that all the results are
obtained utilizing the relations (\ref{j27}). Let us behold the
last term in ${\cal I}_1(\varepsilon )$ could naturally be
associated with the contribution of the Coulomb interaction between
an electron and a proton in the final state.

Eventually, upon adding up (\ref{j32}), (\ref{j33}), (\ref{j35}),
the electron momentum distribution (\ref{i32}) in the $\beta
-$decay of a polarized neutron results to be
\begin{eqnarray}
\mbox{d}{\bf W}(\varepsilon , {\bf p}_e , \bbox{\xi}) =
\mbox{d}{\bf w}(\varepsilon , {\bf p}_e)
\bigl\{ W_0(g_A , \varepsilon) + \bbox{v\xi}
W_{\xi}(g_A , \varepsilon)\bigr\} \, ,\label{i34}\\
 W_0(g_A , \varepsilon)
= (1+3g_A^2)[1+\tilde C_0(\varepsilon)+{\cal
B}(\varepsilon)]+C_0(g_A , \varepsilon)\nonumber\\ W_{\xi}(g_A ,
\varepsilon)=2g_A (1-g_A)[1+\tilde C_{\xi}(\varepsilon) +{\cal
B}(\varepsilon)] +C_{\xi}(g_A , \varepsilon)\nonumber\\
C_0=\frac{\alpha}{2\pi}
[2\ln{(\frac{{\varepsilon}+p}{m})}v(1+3g_{A}^{2})
+\frac{33g_{A}^{2}}{4}+6g_{A}+
\frac{7}{4}+ \nonumber \\
 + \ln(\frac{M_W}{M_{p}}) (3+12g_{A}+9g_{A}^{2})-
\frac{M_W^2}{M_W^2-M_S^2}\ln\frac{M_W}{M_S}
 (5+12g_A+15g_A^2)] \, ,\nonumber\\
C_{\xi}=\frac{\alpha}{2\pi}[\frac{4g_{A}(1- g_{A})}{v}\ln(\frac{{
\varepsilon}+p}{m})+\frac{5}{4}+2g_{A}-
\frac{13}{4}{g_{A}}^{2}+ \nonumber \\
\ln(\frac{M_W}{M_{p}})3(1-{g_{A}}^2) -
\frac{M_W^2}{M_W^2-M_S^2}\ln\frac{M_W}{M_S}
(3+4g_A-7g_A^2)] \, ,\nonumber \\
 {\cal B} =
\frac{2\alpha}{\pi}{\cdot}{[\frac{1}{v}
\ln\frac{p_e+\varepsilon}{m} -1]}{\cdot}\ln(\frac{2k_{m}}{m}),
 \; \; \; \; {\tilde C}_{0}={\tilde C}'_{0}+ {\tilde C}_{1} , \;
\; \; \; {\tilde C}_{\xi}={\tilde C}'_{\xi}+{\tilde C}_{1} ,
\nonumber \\ {\tilde C}_1{=}\frac{2\alpha}{\pi}[\frac{{\cal
J}}{2v}-\frac{{\cal K}}{2v}+\frac{1}{4}\Bigl(3 \ln(M_{p}/m)-9/2 +
\frac{6}{s_W^2}{+}6\frac{M_Z^2}{M_Z^2-M_S^2}\ln\frac{M_Z}{M_S}
+\frac{5+2c_W^2}{s_W^4}\ln{(c_W)}\Bigr)] \, ,\nonumber\\ {\cal
J}(\varepsilon )=
\frac{1}{4}(\ln{(x)})^2-F(1/x-1)+\frac{\pi^2v}{\tilde v(\varepsilon)}
\, , \; \; \; \; \; \; \; \;
\nonumber
\end{eqnarray}
which can also be rewritten as
\begin{eqnarray}
\mbox{d}{\bf W}(\varepsilon , {\bf p}_e , \bbox{\xi}) =
\mbox{d}{\bf W}^{0}(\varepsilon , {\bf p}_e , \bbox{\xi})\cdot [1+{\cal
B}(\varepsilon)+\tilde C_1(\varepsilon)] +
\mbox{d}{\bf w}(\varepsilon , {\bf p}_e)\times\label{i341}\\
\times\Bigl((1+3g_A^2)\tilde C'_0(\varepsilon)+2\bbox{v\xi}g_A(1-g_A)\tilde
C'_{\xi}(\varepsilon) + C_0(g_A
,\varepsilon)+\bbox{v\xi}C_{\xi}(g_A ,\varepsilon)\Bigr) .\nonumber
\end{eqnarray}
We purposely retain the factors ${M_{W,Z}^2}{/}{(M_{W,Z}^2-M_S^2)}$
in front of $\ln{M_{W,Z}}{/}{M_S}$ in order to clarify that nothing
out-of-the-way will occur even in the case
 $M_S{\rightarrow}M_{W,Z}$ and the dependence on $M_S$ is very
 smooth.

The total decay probability $W$, reverse of the lifetime $\tau$,
and the asymmetry factor of electron momentum distribution
$A(\varepsilon )$ are acquired from (\ref{i34}) in the familiar
way:
\begin{eqnarray}
W=\frac{1}{\tau}=\frac{G^2}{2\pi^3}\int\limits_{m}^{M_n-M_p}
\mbox{d}{\varepsilon} \, \varepsilon|{\bf p}_e| k^2_m \,
W_0(g_A , \varepsilon) \; , \label{wt}\\ A(g_A ,
\varepsilon)=\frac{W_{\xi}(g_A , \varepsilon)}
{W_{0}(g_A , \varepsilon)} \; .\label{ae}
\end{eqnarray}

The radiative corrections cause the relative modification of the
total decay probability W
\begin{eqnarray} \frac{\int
\limits_{m}^{M_n-M_p} \mbox{d}{\varepsilon} \,
\varepsilon |{\bf p}_e|k^2_m W_{0}({g_A , \varepsilon})}
{(1+3g_A^2) \int
\limits_{m}^{M_n-M_p} \mbox{d}{\varepsilon} \,
\varepsilon|{\bf p}_e|k^2_m }-1
 = \delta W \, . \label{dw}
\end{eqnarray}

The uncorrected asymmetry factor of the electron angular
distribution $A_0$ is replaced by the quantity $A({\varepsilon})$
accounting for the radiative corrections,
\begin{eqnarray}
A_{0}=\frac{2g_{A}(1-g_{A})} {1+3g_{A}^{2}}\Longrightarrow
\frac{W_{\xi }(g_A , {\varepsilon})}
{W_{0 }({g_A , \varepsilon})}= A({\varepsilon} , g_A) \, .
\label{ada}
\end{eqnarray}

So, the quantities $\delta W$ (\ref{dw}) and
\begin{equation}
\frac{A(\varepsilon , g_A)-A_0}{A_0}={\delta A}(\varepsilon)
 \; \; \; \label{da}
\end{equation}
render the effect of radiative corrections on the total decay
probability $W$ (\ref{wt}) and asymmetry coefficient $A$
(\ref{ae}).

The results of numerical evaluation of $\delta W \, , \; \, \delta
A$ are discussed in the next section.

\section{Discussion of the results.}
\label{sec:level12}

Before setting forth the numerical evaluation, several valuable
features of the ultimate result (\ref{i34}) deserve to be
spotlighted.

Surely, upon adding the contributions from the processes involving
virtual and real infrared photons, the fictitious infinitesimal
photon mass $\lambda$ has disappeared from the final expression
(\ref{i34}), amenably to the received removal of the infrared
divergency \cite{jfc,ll,d,com}.

Let us behold that if we got a neutral initial particle in place of
a $d-$quark and a final particle with the charge $+1$ in place of
an $u-$quark in the expression (\ref{j26}), the coefficient $6$ in
front of $\ln{M_Z}{/}{M_S}$ in $\tilde C_1$ would be replaced by
$8$, following what was observed at the end of Sec. 8.
Subsequently, if $g_A$ were therewith equal to $1$, the subsidiary
parameter $M_S$ would be cancelled in the final result (\ref{i34}).
Being generically represented by the first diagram in (\ref{j26}),
this conceivable case might be associated with the
$neutron{\rightarrow}proton$ transition involving exchange of a
$W-$boson and a ``massive photon" between leptons and quarks, with
the weak nucleon transition current being pure left. As one can
see, the description of the neutron $\beta-$decay would not involve
the parameter $M_S$ in this case.

The form of dependence of (\ref{i34}) on the UV cut-off , i.e. on
${\ln{M_W}}{/}{M_p}$, asks for a special attention. First, it is
readily seen straight away that the portion of (\ref{i34}) multiple
to ${\ln{M_W}}{/}{M_p}$ would strictly vanish, if there were
${g_A}{=}{-g_V}{=}{-1}$, that is if the nucleon weak transition
current were pure right, $(V+A)$, instead of the actual current
(\ref{i8}), (\ref{j0}). This fact is associated with the general
theorem ascertained in Refs.\cite{35}.

As one might infer from Refs. \cite{sr,sd,h}, the amplitude ${\cal
M}$ and the probability $\mbox{d}{\bf W}$ of any semileptonic decay
ought generically to be of the form
\begin{eqnarray}
{\cal M}{\approx}{\cal M}^{0}[1+\frac{3 \alpha}{2
 \pi}\tilde{q}\ln (\frac{M_W}{M_{p}})]{\cdot} [1+{\cal
O}_{1}(\alpha )] \, ,\label{h1} \\
\mbox{d} {\bf W}{\approx} \mbox{d} {\bf W}^{0}[1+\frac{3
\alpha}{2\pi}{\cdot}2\tilde{q}\ln (\frac{M_W}{M_{p}})]{\cdot}[1+{\cal
O}_{2}(\alpha)] \, ,\label{h2}
\end{eqnarray}
up to the terms of order $\alpha$. Here, ${\cal M}^{0} \, $ and
$\mbox{d} {\bf W}^{0}$ render the uncorrected (Born) values of
${\cal M} , \mbox{d} {\bf W}$, and
\begin{equation}
 \tilde{q}=\frac{2\bar{Q}+1}{2}=-(Q_{1in}Q_{2in}+Q_{1out}Q_{2out}) \,
 ,\label{h3}
\end{equation}
where $\bar Q$ is the average charge of the isodoublet involved in
the decay \cite{sr,sd}, and $$Q_{1in}Q_{2in} \, , \; \,
Q_{1out}Q_{2out}$$ are the products of charges of incoming and
outgoing particles, respectively \cite{h}. In the case of the
neutron $\beta-$decay, i. e. for the $(n , p)$ doublet,
$\bar{Q}=1/2$, and $$Q_{1in}Q_{2in}=0 \, , \;
\, Q_{1out}Q_{2out}=-1 \, , $$ so
 that $\tilde{q}{=}1$. So, the distribution (\ref{i34}) ought to
 have taken the form
\begin{equation}
\mbox{d} {\bf W}(g_{A} , \varepsilon , \alpha ){\approx}
\mbox{d} {\bf W}^{0}(g_{A} , \varepsilon){\cdot}[1+\frac{3
\alpha}{2\pi}{\cdot}2 \ln (\frac{M_W}{M_{p}})]{\cdot}[1+{\cal
O}_{P}(\alpha)] \, ,\label{h5}
\end{equation}
with the quantity ${\cal O}_{P}(\alpha)$ independent of $g_{V}\, ,
\; \, g_{A} \, , \; \, \ln{M_W}{/}{M_p}$.
Apparently, it is not the case: the expression (\ref{i34}) can
never be reduced to the form (\ref{h5}). Yet though one might think
we encounter some puzzling mismatch, there is no real contradiction
between the assertions of Refs. \cite{sr,sd,h} and our
straightforward consistent calculation based on the electroweak
Lagrangian (\ref{l2})-(\ref{l8}) and the parameterization
(\ref{i8}), (\ref{j0}), (\ref{g0}) of the nucleon weak transition
current. To perceive the matter, we rewrite the actually used
current ${\cal J}^{\beta}_{np}$ (\ref{j0}) and the distributions
$\mbox{d} {\bf W}^0$ (\ref{i33}), $\mbox{d} {\bf W}$ (\ref{i34}) in
terms of the amplitudes
\begin{equation}
g_L =\frac{g_V +g_A}{2} \; , \; \; \; g_R =\frac{g_V -g_A}{2}\label{k2}
\end{equation}
introduced instead of the original ones ${g_V}{=}{1} , \;
 {g_A}{\neq}{g_{V}}$ :
\begin{eqnarray}
{\cal
J}^{\beta}_{np}(0)=\gamma^{\beta}[(1-\gamma^5)g_L+(1+\gamma^5)g_R]
\, , \; \; \; \label{55}\\
\mbox{d} {\bf W}^0=\mbox{d} {\bf w}\cdot [4(g_L^2+g_R^2-g_Lg_R)+
\bbox{v\xi}4g_R(g_L-g_R)] \, , \; \label{57} \\
\mbox{d} {\bf
W}\approx\mbox{d} {\bf
w}\Biggl\{4g^2_L\Biggl(1+\frac{3\alpha}{2\pi}\ln\frac
{M_W}{M_p}\Biggr)^2\bigl(1+{\cal O}_L(\alpha)\bigr)+\nonumber\\ +
4g^2_R\bigl(1+{\cal
O}_R(\alpha)\bigr)-4g_Lg_R\Biggl(1+\frac{3\alpha}{2\pi}\ln\frac{M_W}
{M_p}\Biggr)\bigl(1+{\cal O}_{RL}(\alpha)\bigr)+\label{58}\\ +
\bbox{v\xi}\Biggl(-4g^2_R\Bigl(1+{\cal
O}_R^{\xi}(\alpha)\Bigr)+4g_Lg_R\Biggl(1+\frac{3\alpha}{2\pi}\ln\frac
{M_W}{M_p}\Biggr)\bigl(1+{\cal
O}_{RL}^{\xi}(\alpha)\bigr)\Biggr)\Biggr\} \, , \;
\nonumber\\ {\cal O}_L(\alpha)={\cal
O}_0(\alpha)+\frac{\alpha}{2\pi}\Bigl(4+ 2\ln\frac{M_W}{M_S}\Bigr)
\, , \; \; \; {\cal O}_R(\alpha)={\cal
O}_0(\alpha)+\frac{\alpha}{2\pi}\Bigl(1+ 8\ln\frac{M_W}{M_S}\Bigr)
\, , \; \nonumber\\ {\cal O}_{RL}(\alpha)={\cal
O}_0(\alpha)+\frac{\alpha}{2\pi}\Bigl(\frac{13}{4}-
5\ln\frac{M_W}{M_S}\Bigr) \, , \; \; \; {\cal
O}_R^{\xi}(\alpha)={\cal
O}_{\xi}(\alpha)+\frac{\alpha}{2\pi}\Bigl(1-
2\ln\frac{M_W}{M_S}\Bigr) \, , \; \nonumber\\ {\cal
O}_{RL}^{\xi}(\alpha)={\cal
O}_{\xi}(\alpha)+\frac{\alpha}{2\pi}\Bigl(\frac{9}{4}-
5\ln\frac{M_W}{M_S}\Bigr) \, , \; \; \; {\cal O}_0(\alpha)={\tilde
C}_0+{\cal B}+\frac{\alpha}{\pi}v\ln\frac{\varepsilon +p_e}{m} \, ,
\nonumber\\ {\cal O}_{\xi}(\alpha)={\tilde C}_{\xi}+{\cal
B}+\frac{\alpha}{\pi}\frac{1}{v}\ln\frac{\varepsilon +p_e}{m} \, .
 \qquad \qquad \qquad \qquad \nonumber
\end{eqnarray}
It stands to reason that the values $g_L{\ne}1 , \, g_R{\ne}0$
reflect the mixture of the left and right hadronic currents on
account of the effect of nucleon structure. In confronting
(\ref{57}) and (\ref{58}), one grasps that the amplitude $g_L$ gets
the renormalization factor which corresponds to that in (\ref{h1})
 accordingly to Refs. \cite{d,sr,sd,h}, whereas the modification of
$g_R$ does not depend on the cut-off $\ln{M_W}{/}{M_p}$ at all, in
accordance with Ref. \cite{35} as was discussed above. If there
were the pure left hadronic current, i.e. $g_L{=}1,
\, g_R{=}0$, the relation (\ref{h5}) between the uncorrected
(\ref{57}) and corrected (\ref{58}) distributions would apparently
hold true as prescribed by Refs. \cite{d,sr,sd,h}. In the case of
the pure right hadronic current, i.e. $g_L{=}0, \, g_R{=}1$, the
final result (\ref{58}) would not depend on $\ln{M_W}{/}{M_p}$ at
all.

Inquiring carefully into the calculations carried out in Refs.
\cite{sr,sd,h}, we realize that the semileptonic decays considered
therein are actually described by the interactions which correspond
to the case $g_L {=}1 ,
\, g_R{=}0$, i.e. a pure left hadronic current. It is to emphasize that the
assertions (\ref{h1}), (\ref{h2}) of Refs. \cite{sr,sd,h} hold true
for any decays induced by a pure left ($g_L{=}1, \, g_R{=}0,
\, g_V{=}g_A{=}1$) hadronic current, in particular for the
semileptonic decays which can be reduced to the pure $\large
d{\rightarrow}u$ transitions of free quarks . Thus, Eqs.
(\ref{h1}), (\ref{h2}) are valid to describe the manifold
semileptonic decays such as
${\pi}{\rightarrow}{\mu}\bar{\nu}_{\mu}\gamma
\, , \; \; {\pi}{\rightarrow}{e}\bar{\nu}_{e}\gamma \, , \; \;
K{\rightarrow}{\mu}{\nu}_{\mu}\gamma \, , \; \;
{\tau}{\rightarrow}{\pi}{\nu}_{\tau}\gamma \, , \; \;
{\tau}{\rightarrow}K{\nu}_{\tau}\gamma $ and so on (see, for
instance, \cite{41,42} in addition to \cite{sr,sd,h}). The Eqs.
(\ref{h1}), (\ref{h2}) might although be pertinent to treat the
transitions caused by the pure axial ($g_L{=}g_{R}{=}-g_A{=}-1 , \,
g_V{=}0$) hadronic current, such as
${\Sigma}^{\pm}{\rightarrow}{\Lambda}^{0}e^{\pm}{\nu}(\bar{\nu})\gamma$,
or by the pure vector current ($g_L{=}g_R{=}g_V{=}1, g_A{=}0$),
such as the super-allowed $0^{+}{\rightarrow}0^{+}$ nuclear
transitions. But all the aforesaid is not our case, it is not
relevant for describing the neutron $\beta-$decay.

Evidently, as the total amplitude $\cal M$ (\ref{i5}) is not
multiple to ${\cal M}^{0} \, $ (\ref{i7}), the distribution
(\ref{i34}) can never be transformed to an expression multiple to
(\ref{i33}), unlike the results asserted in several calculations
\cite{g1,34,g2,g3,g4,ga,ga1,sha} which were entailed by the
original work \cite{s67} where the decay probability was reduced,
to all intents and purposes, to the ``model-independent" part
merely proportional to $\mbox{d} {\bf W}^0 \, $ (\ref{i33}), that
is explicitly not our case.

The original investigation \cite{s67} had been undertaken before
the Standard Model of elementary particle physics was brought to
completion in the nowaday form \cite{d,ao,h1,h2,b}. Then, for the
lack of the renormalizable Electroweak Weinberg-Salam Theory, there
was seen no way to treat the neutron $\beta-$decay with
self-contained allowance for the radiative corrections. The purpose
of the ingenious work \cite{s67} was to circumvent the problem of
 UV divergence and sidestep the consideration of the
electromagnetic corrections in the UV region, by appropriate
separating the whole electromagnetic corrections of order $\alpha$
into two conceivable parts, a ``model-independent" (MI) and a
``model-dependent" (MD), of different purports. The first one, MI,
was chosen and sorted out so that it should evidently be UV-finite
and could merely be obtained by multiplying the uncorrected (Born)
decay probability $\mbox{d} {\bf W}^0$ (\ref{i33}) by a single
universal function $g(\varepsilon , M_n-M_p , m)$, see Eqs. (20) in
Ref. \cite{s67}, which was calculated within the effective
4-fermion-interaction approach (\ref{1})-(\ref{l}), without taking
into consideration the electroweak and strong interactions as
prescribed by the Standard Model. In Ref. \cite{s67}, this MI part
was presumed to describe the electromagnetic effects on the neutron
$\beta-$decay. All the left-over radiative corrections were
conceived to be incorporated into the second, MD part, assuming the
electroweak and strong interactions conspire somehow to give the
finite corrections to the quantities $g_V \, , \; \, g_A \, ,
\; \, V_{ud}$ which reside in the uncorrected, Born decay
probability (\ref{i33}), see Eqs. (19), (20) in Ref. \cite{s67}.
Thus, in all the calculations, such as
\cite{g1,34,g2,g3,g4,ga,ga1,sha}, presuming the approach launched
by the work \cite{s67}, the corrected decay probability merely
shows up to be reduced to the uncorrected one multiplied by the
function $g(\varepsilon , M_n-M_p , m)$, with the whole effect of
the remained MD part absorbed into the quantities $g_V \, , \; \,
g_A \, , \; \, V_{ud}$ which thereby would get the new values $g'_V
\, , \; \, g'_A \, , \; \, V'_{ud}$ instead of the original ones : the
$CKM$ matrix element $V_{ud}$ in (\ref{l5}) and the amplitudes $g_V
\, , \; \, g_A$ specifying the nucleon weak transition current
(\ref{i8}), (\ref{j0}). Thus, the experimental data would be
described in terms of these ``new" quantities $g'_V
\, , \; \, g'_A \, , \; \, V'_{ud}$. However, any explicit and
definite, quantitative one-to-one correspondence between these two
sets of parameters, $g_V \, , \; \, g_A \, , \;
\, V_{ud}$ and $g'_V \, ,\; \, g'_A \, , \; \, V'_{ud}$,
 would never be asserted in Refs.
\cite{s67,g1,34,g2,g3,g4,ga,ga1,sha}. Yet the guide tenet is to
ascertain, as precise as possible, the very genuine values of $g_V
\, , \; \, g_A \, , \; \, V_{ud}$ from experimental data
processing. In particular, we are in need of the stringent
$|V_{ud}|$ value in order to verify strictly the validity of the
$CKM$ identity (\ref{ckm}) \cite{ckm}. So, the aforesaid
calculations \cite{s67,g1,34,g2,g3,g4,ga,ga1,sha} making use of the
very handy, but rather untenable simplifications cannot be said to
be eligible for now, in so far as an accuracy ${\sim}1{\%}$ or even
better goes.

In our treatment, the amplitude $\cal M$ (\ref{i5}), (\ref{mm})
and, subsequently, the distribution $\mbox{d} {\bf W}$ (\ref{i34})
comprise all the $\alpha-$order radiative corrections, without
disparting the Coulomb term and separating the MI and the MD parts.
Adopting $M_S{=}10 \, \mbox{GeV} \, ,\; \,
\; (M_p^2{\ll}M_S^2{\ll}M_W^2) \, $ and taking all the input
parameters in (\ref{i34}) from Ref. \cite{pdg}, we obtain the
corrections (\ref{dw}) and (\ref{da})
\begin{equation}
\delta W =8.7\% \; , \qquad \qquad \delta A =-2\% \qquad
\label{dwa}
\end{equation}
to the uncorrected $W^0$ and $A_0$ values. As a matter of fact, the
correction $\delta A$ (\ref{da}), (\ref{dwa}) is independent of
$\varepsilon$. Apparently, our results (\ref{dwa}) pronouncedly
differ from the respective MI-values
\begin{equation}
\delta W_{MI} \approx 5.4\% \; , \qquad \qquad \delta A_{MI}\approx 0
\qquad \label{mi}
\end{equation}
asserted in Refs. \cite{s67,g1,34,g2,g3,g4,ga,ga1,sha}.
Consequently, the values of $|V_{ud \, MI}|$ and $g_{A \, MI}$
ascertained from experimental data processing with utilizing
$\delta W_{MI}$, $\delta A_{MI}$ (\ref{mi}) will alter, when they
are obtained with $\delta W$, $\delta A$ (\ref{dwa}). The
modifications are of the noticeable magnitude: $\delta
g_A{\approx}0.47\%$, $\delta |V_{ud}|{=}{-1.7\%}$. For instance,
the values $g_A{=}1.2739 , \, {|V_{ud}|}{=}{0.9713} $ given in
\cite{t,a,pdg} will be modified to $g_A{\approx}1.28, \,
|V_{ud}|{\approx}0.96 $, provided the same value of the quantity
$G$ is used.

Now we are to discuss what is the precision attainable in the
actual calculations nowadays, a pivotal question that matters a
lot.

As from the first we have been calculating the radiative
corrections in the one-loop order, $O(\alpha)$, the relative
uncertainty ${\sim}{\alpha}{\sim}10^{-2}$ resides in the evaluated
radiative corrections (\ref{dwa}), from the very beginning.

We further recall that the terms of relative order
$$\frac{M_n-M_p}{M_p} \, , \; \,
\frac{M_n-M_p}{M_p}{\cdot}\ln\frac{M_n-M_p}{M_p} \; , $$
and smaller have been
neglected far and wide, with a relative error ${\lesssim}10^{-3}$
entrained thereby.

Yet a far more substantial task than the aforesaid ones is to
inquire into the ambiguities caused by entanglement of the strong
quark-quark interactions in the neutron $\beta$-decay.

The final result (\ref{i34}) involves the matching parameter $M_S
\; , \; ( M_p^2{\ll}M_S^2{\ll}M_W^2 ) \, $ posited to treat
separately quark systems with large, $k^2{\gtrsim}M_S^2$, and
comparatively small, $k^2{\lesssim}M_S^2$, momenta. The dependence
of the results $\delta W \, , \; \, \delta A$ (\ref{dwa}) on the
$M_S$ value shows up to be very faint : we have got $\delta
W{=}8.6\%$ at $M_S{=}5 \,
\mbox{GeV}$ and $\delta W{=}8.8\%$ at $M_S{=}30 \,
\mbox{GeV}$, and $\delta A$ is practically independent of $M_S$ at
all. So, the uncertainties because of the $M_S$ involvement in
(\ref{i34}) are about $0.1\%$ in $\delta W$ and practically
 zero in $\delta A$ (\ref{dwa}).

Further, $M_S$ is chosen so that $M_p^2{\ll}M_S^2$, and we took for
granted the generally accepted standpoint of the Standard Model
that the strong quark-quark interactions die out when a quark
system possesses
 momenta $k^2{\gtrsim}M_S^2{\gg}M_p$. At relatively small momenta
$k^2{\lesssim}M_S^2$, a quark system was considered to form various
baryonic states, including the nucleon.
 Let us recall all the actual calculations have been carried out
assuming Eqs.(\ref{j0})-(\ref{g0}) and retaining only the single
nucleon intermediate state (\ref{gn}) in the expressions
(\ref{j18}), (\ref{j19}), (\ref{j20}), (\ref{j29}), what counts is
the final result (\ref{i34}), (\ref{dwa}). In calculating the
radiative corrections, we did not intend to allow for nucleon
compositeness rigorously, but (in sections $\mbox{VI}$ and
$\mbox{IX}$) we only tried and estimated how those basic
calculations alter when including the nucleon excited states
(\ref{gb}) and the form factors (\ref{i8}), (\ref{a})-(\ref{ff})
into the expressions (\ref{j18}), (\ref{j20}), (\ref{j29}). As was
found out in sections $\mbox{VI}$ and $\mbox{IX}$, the different
terms in the amplitude ${\cal M}$ (\ref{i5}), (\ref{mm}) (and
subsequently in the distribution $\mbox{d} {\bf W}$ (\ref{i34}) )
 are affected by allowance for compositeness of the nucleon to a
different extent. As a matter of fact, there is no modification in
the first, prevailing term in (\ref{0}) which is determined by the
integral $I(2M_p\varepsilon , \lambda)$ (\ref{1ij}). It includes,
in particular, the Coulomb correction. The direct evaluation shows
that this major term causes the share of about $\delta
W_I{\approx}5\%$ in the whole correction $\delta W{\approx}8.7\%$
(\ref{dwa}). All the other left-over terms
 in the decay amplitude ${\cal M}$ provide the remnant portion
 ${\delta}W{-}{\delta}W_I{\approx}4\% $ of $\delta W$ and the whole
value ${\delta}A{=}{-2\%}$ (\ref{dwa}). The effect of nucleon
compositeness on these terms was estimated (in sections
$\mbox{IV}$, $\mbox{IX}$) to constitute no more than ${\sim}10\%$
to their whole value. For now, there sees no real reliable way to
calculate precisely these corrections-to-corrections in treating
the neutron $\beta-$decay. With the ascertained estimations, they
are abandoned in the actual calculation which has provided
(\ref{i34}), (\ref{dwa}).
 Consequently, in respect of all the aforesaid, the uncertainties
in the result (\ref{dwa}) prove to make up no more than
\begin{equation}
\Delta (\delta W )\approx 0.4\% \, , \qquad \qquad \Delta (\delta
A)\approx 0.2\% \; . \label{ddwa}
\end{equation}

Thus, our inferences are realized to hold true up to the accuracy
about a few tenth of per cents, never worse.

If anything, let us behold the energy released in the $\beta-$decay
of free neutrons is rather negligible as compared to the nucleon
mass, $M_n-M_p{\ll}M_p$, whereas the
 energy released in manifold semileptonic decays is comparable to
the masses of the hadrons involved in the process, or even greater
than they. That is why accounting for compositeness of the hadron
proves to play no decisive role in the neutron $\beta-$decay, but
can be of
 significant value in other semileptonic decays (see, for example,
\cite{41,42}).

In the current treatment of the radiative corrections to the
neutron $\beta-$decay, we have actually allowed for the effects of
nucleon structure by introducing only one fit-parameter $g_A$ to be
specified, simultaneously with the fundamental quantity $|V_{ud}|$,
by processing the experimental data on the lifetime \cite{t} and
electron momentum distribution \cite{a}. Evidently, the ambiguities
(\ref{ddwa}) put bounds on the accuracy which can be attained in
obtaining the $|V_{ud}| \, , \; \, g_A$ values thereby.

Thus, introducing only the usual parameters $g_V \, ,\; \, g_A \, ,
\,
\, g_{WM}
\, ,\,  \, g_{IP}$ to describe the weak nucleon transition current
 does not suffice to parameterize the whole effect of strong
interactions in treating the neutron $\beta -$decay with allowance
for the radiative corrections, in so far as the accuracy one per
cent or better goes. Nowadays, no way is thought to get rid of the
errors (\ref{ddwa}), but to parameterize ingeniously the effects of
nucleon compositeness by expedient introducing some additional
fit-parameters (besides $g_A$) to describe the radiative
corrections to various characteristics of the neutron
$\beta-$decay. These additional parameters are to be fixed by
 processing, simultaneously with the results of measurements of
$\tau$ \cite{t} and $A$ \cite{a}, the experimental data obtained in
the additional experiments, such as proposed in \cite{ba,by1,f} and
other in this line. For instance, these extra parameters might be
conceived to render generically the ``effective" mass in the
intermediate state in (\ref{j18}), (\ref{j19}), (\ref{j20}),
(\ref{j29}), (\ref{dm5}) and the ``effective" vertices (\ref{j1b}),
(\ref{i8}), (\ref{j2b})-(\ref{a1}). They are to be fixed, together
with $g_A \, , \; \, |V_{ud}| \, , \; \, M_S$, from the
simultaneous analysis of all the available experimental data, the
kinematic corrections \cite{sm} respected as well.

So we are in need of the manifold tenable experiments to measure
various characteristics of the neutron $\beta-$decay, besides
$\tau$ and $A$, with an accuracy about $0.1\%$, and even better.
Obtained such high-precision experimental data, the high accuracy,
better than ${\sim}0.1\%$, is believed to be attained within the
unified self-contained analysis of the different experimental data
amenably to the Standard Model.
\section{Acknowledgments.}
Author is thankful to E.A. Kuraev for the invaluable discussions and
encouragements.


\begin{thebibliography}{99}
\bibitem{t} W. Mampe, L. Bondarenko, V. Morosov et al., JETP Lett.
{\bf 57}, 82 (1993).\\ S. Arzumanov, L. Bondarenko, S. Chernyavski
et al., NIM A {\bf 440}, 511 (2000).\\ J. Byrne, P. G. Dawber, C.
G. Habeck et al., Eur. Lett. {\bf 33}, 187 (1996). \\ L.
Bondarenco, E. Korobkina, V. Morosov et. al., JETP Lett. {\bf 68},
691 (1998).
\bibitem{a} P. Liaud, K. Schreckenbach, R. Kossakowski et al.,
Nucl. Phys. {\bf A612}, 53 (1997).\\ K. Schreckenbach, P. Liaud, R.
Kossakowski et al., Phys. Lett. B {\bf 349}, 427 (1995). J. Reich,
H. Abele, M. A. Hofmann et al., NIM A {\bf 440},
 535 (2000).\\ H. Abele, S. B\"a{\ss}ler, D\"ubbers et al., Phys.
Lett. B {\bf 407}, 212 (1997).\\ H. Abele, M. Astruc Hoffman, S.
B\"a{\ss}ler et. al., Phis. Rev. Let. {\bf 88}, 211801-1 (2002).
\bibitem{ba} I. A. Kuznetsov, A. P. Serebrov, I. V. Stepanenko et al.,
Phys.  Rev.  Lett. {\bf 75}, 794 (1995).\\ I. A. Kuznetsov, A. P.
Serebrov, I. V. Stepanenko et al., JETP Lett.  {\bf 60}, 311 (1994).\\ A.
P. Serebrov, I. A. Kuznetsov, I. V. Stepanenko et al., JETP {\bf 113},
1963 (1998).
\bibitem{by1} J. Byrne, P. G. Dawber, S.R. Lee, NIM
 A {\bf 349}, 454 (1994).\\ P.G. Dawber, J. Byrne, M.G.D. van der
Grinten et al., NIM A {\bf 440}, 543 (2000); NIM A {\bf 440}, 548
(2000).\\ J. Byrne, P. G. Dawber, M. G. D. van der Grinten et. al.,
J. Phys. G {\bf 28}, 1325 (2002).
\bibitem{tk} L. J. Lising, S. R. Hwang. P. Mumm et. al., Phys. Rev. C
{\bf 62}, 055501 (2000).\\ K. Bodek at. al., Neutron News, {\bf 3},
29 (2000); Nucl. Ins. and Meth. A {\bf 473}, 326 (2001).\\ T.
Soldner, in Proceedings of the ``Quark-Mixing, CKM-Unitarity"
Workshop, Internationales Wissenschaftsforum Heidelberg, 19-20
September 2002, http//ckm.uni-hd.de .
\bibitem{f} W.S.Wilburn, J.S.  Kapustinsky, J.D. Bowman et al., in
{\it 1999 Division of Nucl. Phys. Fall Meeting}, October 20-23,
1999, Pacific Grove, CA, Bul. of American Phys. Soc. {\bf 44},
[BC.09] (1999).\\ M.S. Dewey, F.E. Wietfeldt, B.G. Yerozolimsky et
al., in {\it 1999 Division of Nucl. Phys. Fall Meeting}, October
20-23, 1999, Pacific Grove, CA, Bul. of American Phys. Soc. {\bf
44}, [BC.11], [BC.12] (1999).\\ A.R. Yong, S. H\"odle, C.-Y. Lin at
al., in {\it 1999 Division of Nucl. Phys. Fall Meeting}, October
20-23, 1999, Pacific Grove, CA, Bul. of American Phys. Soc. {\bf
44}, [BC.08] (1999).\\ H.P. Mumm, M.C. Browne, R.G.H. Robertson at
al., in {\it 1998 Division of Nucl. Phys. Fall Meeting}, October
28-31, 1998, Santa Fe, NM, Bul. of American Phys. Soc. {\bf 43},
[B2.05], [B2.06] (1998).\\ Y. Liao and X. Li, Phys. Lett. B {\bf
503}, 301 (2001).\\ S. Balashov, Yu. Mostovoy, Preprint of Rus.
Res. Center ``Kurchatov Institute" , IAE-5718/2, Moscow (1996).\\
B. G. Yerozolimsky, in Proceedings of ``Quark-Mixing,
CKM-Unitarity" Workshop, Internationales Wissenschaftsforum
Heidelberg, 19-20 September 2002, http://ckm.uni-hd.de.\\ Yu.
Mostovoy, Preprint of Rus. Res. Center ``Kurchatov Institute" ,
IAE-6040/2, Moscow (1997).
\bibitem{ll} V.B. Berestezky, E.M.  Lifshitz and L.P.
Pitajevsky, {\it Relativistic Quantum Field
 Theory, part I}, Nauka, Moscow (1971).\\ E.M. Lifshitz and L.P.
Pitajevsky, {\it Relativistic Quantum Field Theory, part II},
Nauka, Moscow (1971). \bibitem{d} J.F. Donoghue, E. Golowich and
B.R. Holstein, {\it Dynamics of the Standard Model}, Cambridge
University Press, Camgridge, UK (1994).
\bibitem{com} L. B. Okun, {\it Leptons and Quarks},
Nauka, Moscow (1982).\\ E.D.  Commins and P.H.  Bucksbaum, {\it Weak
Interactions of
 Leptons and Quarks}, Cam. Univ. Press, Cambridge England (1983).\\
 E.D. Commins, {\it Weak Interactions}, McGraw-Hill Book Company,
 New York (1973).
\bibitem{ckm} N. Cabibbo, Phys.  Rev.  Lett. {\bf 10},
 531 (1963).\\
 M. Kobajashi and T. Maskawa, Prog. Theor. Phys. {\bf 49}, 625
(1973).\\ D. E. Groom et. al., (PDG), 11 CKM-Quark-Mixin Matrix,
Eur. Phys. J. C {\bf 15}, 110 (2000).
\bibitem{sm} S.M.
Bilin'ky, R.M. Ryndin, Ya.A. Smorodinsky and Ho Tso-Hsin, {ZhETF} {\bf
37}, 1758 (1959).
\bibitem{jap} Y. Yokoo, S. Suzuki and M. Morita, Prog. Theor.
Phys. {\bf 50}, 1894 (1973); Sup. of Prog. Theor. Phys. ${\cal
N}$60, 37 (1976).
 \bibitem{g1} K.  Toth, K.  Szeg\"o and A.  Margaritis, {
Phys.  Rev.} D {\bf 33}, 3306 (1986).
\bibitem{34} F. Gl\"uck and K.
Toth, { Phys. Rev.} D {\bf 41}, 2160 (1990).
\bibitem{g2} F. Gl\"uck and K. T\' oth, Phys. Rev. D {\bf 46},
2090 (1992).
\bibitem{g3} F. Gl\"uck, I. Jo\' o and J. Last, Nucl. Phys.
{\bf A295}, 125 (1995).
\bibitem{g4} F. Gl\"uck, { Phys.  Rev.} D {\bf 47}, 2840 (1993).
\bibitem{ga} A. Garcia, { Phys. Rev.} D {\bf 25}, 1348 (1982); D {\bf
35}, 232 (1987).
\bibitem{ga1} A. Garcia, J. l. Garcia-Luna and G. Lopez Castro, Phys.
Lett. B {\bf 500}, 66 (2001).
\bibitem{i} G.G. Bunatian, Phys Atomic Nuc. {\bf 63}, 502 (2000);
/aps1999mar$11_{-}$005.
\bibitem{iii} G.G. Bunatian, E4-2000-19 Preprint of JINR, Dubna,
Russia, 2000; /aps2000feb$17_{-}$001\\ G.G. Bunatian, Part. and
Nucl., Lett. No.{\bf 6[103]-2000}, 63 (2000).
\bibitem{ao} K.I.Aoki et al., Suppl. Progr. Theor. Phys.
{\bf 73}, (1982) 1.
\bibitem{h1} M. B\"ohm, W. Hollik, H. Spiesberger, Fortschr. Phys. {\bf
34}, 687 (1986).
\bibitem{h2} W. Hollik, Fortschr. Phys. {\bf 38}, 165 (1990).
\bibitem{b} D. Bardin, G. Passarino,
 ``{\it The Standard Model in the Making }"
, Oxford, 1999.
\bibitem{jfc} D.R. Yenni, S.C. Frautschi and Suura, Ann. of Phys.
{\bf 13}, 379 (1961).
\bibitem{sr1} A.Sirlin, Phys.Rev. D {\bf 5}, 436 (1972).
\bibitem{h} B.R.Holstein, Phys.Lett. B {\bf 224}, 83 (1990).
\bibitem{41} R. Decker and Finkemeier, Nucl. Phys. {\bf B438}, 17
(1996).
\bibitem{23} J.D. Bjorken, J.D. Walecka, Ann. of Phys. {\bf 38}, 35
(1966).\\
 M. Gourdin and Ph. Salin, Nuovo Cim. {\bf 27}, 193 (1963).\\ H.
Munczek, Phys. Rev. {\bf 164}, 1794 (1967).
  \bibitem{ms} A. Sirlin, Phys.Rev. D {\bf 22}, 971 (1980).
\bibitem{sf} K. Mitchell, Phil. Mag. (Ser. 7), {\bf 40}, 351 (1949).
\bibitem{35} S.M.  Berman, { Phys.  Rev.} {\bf 112}, 267 (1958).\\
S.M. Berman and A. Sirlin, Ann. of Phys. {\bf 20}, 20 (1962).\\
 R.P. Feynman and M. Gell-Mann, { Phys. Rev.} {\bf 109}, 193
(1958).\\ Ya.A. Smorodinsky and Ho Tso-Hsin, { ZhETF} {\bf 38},
1007 (1960).
\bibitem{sr} A. Sirlin, { Nucl.  Phys.} {\bf B71}, 29 (1974);
{ Nucl. Phys.} {\bf B100}, 291 (1975); \\
 { Rev. Mod. Phys.} {\bf 50}, 573 (1978); { Phys. Rev.} D {\bf 22},
 971 (1980).
\bibitem{sd} A. Sirlin, { Nucl.  Phys.} {\bf B196}, 83 (1982).
\bibitem{42} W.J. Marciano, Phys. Rev. D {\bf 45}, R721 (1992).
\bibitem{sha} R.T. Shann, Nuov. Cim. A {\bf 5}, 591 (1971).
\bibitem{s67} A. Sirlin, Phys. Rev. {\bf 164}, 1767 (1967).
\bibitem{pdg} D.E. Groom et al., (PDG), EUR. Phys. J. C {\bf 15},
1 (2000).
\end{thebibliography}
\end{document}